\documentclass[a4paper,fleqn,usenatbib]{mnras}

\usepackage{newtxtext,newtxmath}

\usepackage[T1]{fontenc}
\usepackage{ae,aecompl}

\usepackage{graphicx}	
\usepackage{amsmath}	
\usepackage{amssymb}	

\title[AME in the TMC and L1527 with QUIJOTE]
{QUIJOTE Scientific results. III. Microwave spectrum of
  intensity and polarization in the Taurus molecular cloud
  Complex and L1527}

\author[F. Poidevin et al.]{F. Poidevin$^{1,2}$,\thanks{E-mail: fpoidevin@iac.es (IAC)}
 J.A. Rubi{\~n}o-Mart{\'{\i}}n,$^{1,2}$
C. Dickinson$^{3}$, 
R. G\'{e}nova-Santos$^{1,2}$, 
\newauthor 
S. Harper$^{3}$, 
R. Rebolo,$^{1,2}$, 
B. Casaponsa$^{6}$, 
A. Pel\'{a}ez-Santos,$^{1,2}$, 
R. Vignaga$^{1,2}$, 
\newauthor 
F. Guidi$^{1,2}$,   
B. Ruiz-Granados,$^{1,2}$
D. Tramonte$^{1,2}$ 
F. Vansyngel$^{1,2}$, 
M. Ashdown$^{4,5}$,  
\newauthor
D. Herranz$^{6}$, 
R. Hoyland$^{1,2}$, 
A. Lasenby$^{4,5}$, 
E. Mart\'{i}nez-Gonz\'{a}lez$^{6}$,
L. Piccirillo$^{3}$, 
\newauthor 
R. A. Watson$^{3}$ 
\\
$^{1}$ Instituto de Astrofis\'{i}ca de Canarias, 38200 La Laguna,Tenerife, Canary Islands, Spain\\
$^{2}$ Departamento de Astrof\'{\i}sica, Universidad de La Laguna (ULL), 38206 La Laguna, Tenerife, Spain\\
$^{3}$ Jodrell Bank Centre for Astrophysics, Alan Turing Building, School of Physics and Astronomy, The University of Manchester, \\
Oxford Road, Manchester, M13 9PL, U.K\\
$^{4}$ Astrophysics Group, Cavendish Laboratory, University of Cambridge, J.J. Thomson Avenue, Cambridge CB3 0HE, UK\\
$^{5}$ Kavli Institute for Cosmology, University of Cambridge, Madingley Road, Cambridge CB3 0HA, UK\\
$^{6}$ Instituto de F\'{\i}sica de Cantabria (CSIC-Universidad de Cantabria), Avda. de los Castros s/n, 39005 Santander, Spain\\
}

\date{Accepted XXX. Received YYY; in original form ZZZ}

\pubyear{2017}

\begin{document}
\label{firstpage}
\pagerange{\pageref{firstpage}--\pageref{lastpage}}
\maketitle

\begin{abstract}
We present new intensity and polarization observations of the Taurus molecular
cloud (TMC) region in the frequency range 10--20 GHz with the Multi-Frequency
Instrument (MFI) mounted on the first telescope of the QUIJOTE experiment.  
From the combination of the QUIJOTE data with the WMAP 9-yr data
release, the \textit{Planck} second data release, the DIRBE maps and ancillary
data, we detect an anomalous microwave emission (AME) component with flux density 
$S_{\rm AME, peak} = 43.0 \pm 7.9\,$Jy in the Taurus Molecular
Cloud (TMC) and $S_{\rm AME, peak} = 10.7 \pm 2.7\,$Jy in the dark
cloud nebula L1527, which is part of the TMC. 
In the TMC the diffuse AME emission peaks around a frequency of 19
GHz, compared with an emission peak about a frequency of 25 GHz in L1527.
In the TMC, the best constraint on the level of AME polarization
is obtained at the \textit{Planck} channel of 28.4\,GHz, with an upper limit 
$\pi_{\rm AME}<$4.2$\,\%$ (95$\,\%$ C. L.), which reduces 
to $\pi_{\rm AME}<$3.8$\,\%$ (95$\,\%$
C.L.) if the intensity of all the free--free, synchrotron and thermal
dust components are negligible at this frequency.
The same analysis in L1527 leads to 
$\pi_{\rm AME}<$5.3$\%$ (95$\,\%$C.L.), or $\pi_{\rm AME}<$4.5$\,\%$ 
(95$\%$C.L.) under the same assumption. We find that
in the TMC and L1527 on average about $80\%$ of the HII gas 
should be mixed with thermal dust. Our analysis shows how the 
QUIJOTE-MFI 10--20 GHz data provides key information to properly 
separate the  synchrotron, free--free and AME components. 
\end{abstract}


\section{Introduction}

Anomalous microwave emission (AME) is a component of Galactic
radiation at frequencies of $\sim$10--60\,GHz first detected about twenty
years ago  \citep[see][]{lei97,oli98}. AME strongly correlates
with far infrared (FIR) thermal dust emission almost everywhere in our Galaxy but
cannot be explained by synchrotron, free--free, thermal dust, or CMB
emission. The nature of the carriers
producing AME are still not known but theory predicts that electric dipole
emission from very small, rapidly rotating  dust grains could 
be the origin of this emission
\citep[see][]{draine98,ali09,hoang10,ysard10}. 
Spinning dust models \citep[e.g.][]{ali09} have been 
developed and spectral energy distributions (SEDs) inferred
from the observations are well described by such models.
Magnetic dipole emission is an alternative proposed mechanism
to explain AME \citep[see][]{draine99,draine13}. 
A review of some observations, mostly at 1-degree resolution, 
discussed by \citet{genova17} shows that AME is expected to 
be linearly polarized, if at all, at a very low level. 
This could be in agreement with quantum suppression of alignment in
very small grains as proposed by \citet{draine16}.
Recent studies shed doubt on polycyclic aromatic hydrocarbons (PAHs) being 
AME carriers \citep[][]{hensley16,hensley17}. On the 
other hand, high resolution observations at 11.3 $\mu$m in the MWC
1080 Nebula show that some PAHs emit linear polarization 
at a level of two per cent \citep[see][]{zhang17}. Such observations
leave room for further attempts at high resolution polarization 
measurement of AME in different astronomical environments, thus
making the origin of AME  a much debated subject.
A full review on the state of our knowledge
of AME and future avenues of research is given by \citet{dickinson18}.

The Q-U-I-JOint TEnerife (QUIJOTE) experiment is a collaborative project consisting of two
telescopes and three linear polarimeters covering the frequency
range 10--42$\,$GHz. The main science driver of this project is
to constrain or detect the B-mode anisotropy in the CMB
polarization down to a tensor-to-scalar ratio of $r=0.05$. Such a goal
is attainable only under the condition that the polarization of the
low-frequency foregrounds, mainly the synchrotron and the AME, are
well understood. The first instrument of the QUIJOTE experiment, the
Multi-Frequency Instrument (MFI), observing through bandpasses
centred at 11, 13, 17, and 19$\,$GHz  has been specifically designed for
this purpose. A more detailed
description of some of the technical aspects can be found in \citet{rubino12}.
The first scientific results obtained with the MFI 
are presented in \citet{genova15} and in  \citet{genova17}. 
Recent reviews on the status of the project are detailed in
\citet{rubino17} and in \citet{poidevin18}.

The detection and study in intensity of AME has been investigated towards
different types of regions including supernova remnants and planetary
nebulae, and several molecular
cloud regions \citep[see][]{cpp2015-25}. A follow-up of a series 
of molecular clouds in the frequency range 10--20 GHz showing potentially strong
AME candidates will be the subject of a
future paper. Here we focus on the Taurus molecular cloud (TMC)
which is the closest and one of the most studied low mass star-forming
regions in our Galaxy. The science driver that leads us to observe the
TMC with the QUIJOTE experiment 
is that the TMC is pervaded by a large-scale uniform 
magnetic field, as proved from polarization by extinction 
\citep[e.g.][]{arce98,chapman11} in the visible and near infrared (IR), 
and by emission \citep[e.g.][]{pir19} at submillimetre wavelengths, 
i.e.\ from the point of view of thermal dust grains.
In addition, the TMC is situated below the Galactic plane at galactic
declinations $b<-11^\circ$,which decreases confusion with components
associated with the Milky Way. These conditions makes this region 
an interesting laboratory in which to study AME 
intensity and polarization properties.  

\begin{figure}
\begin{center}
\vspace*{2mm}
\centering
\includegraphics[width=90mm,angle=0]{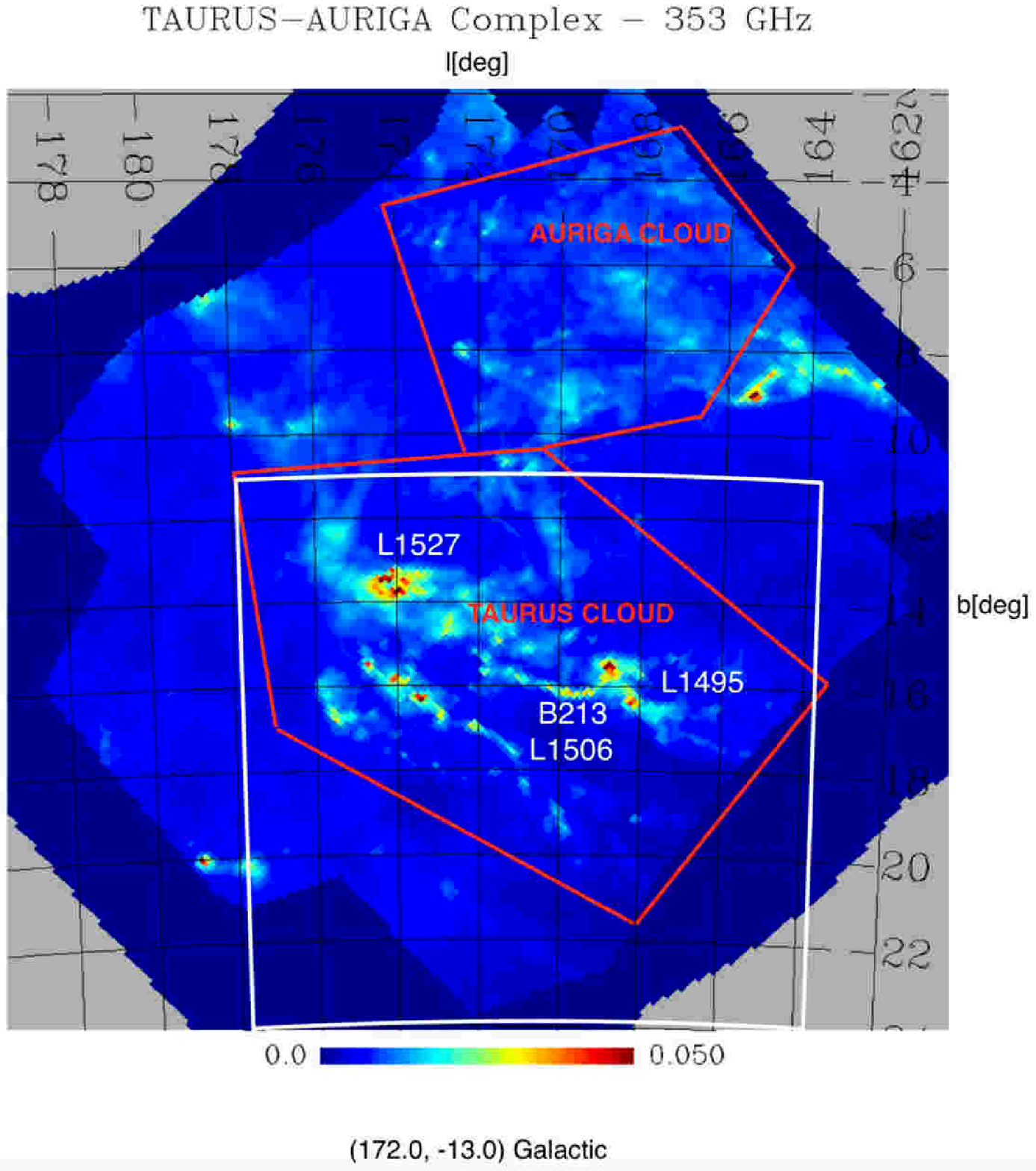}
\caption{\small Location of the Taurus molecular cloud (TMC) part of
the Taurus--Auriga complex (red boxes) in the field of view common 
to the four horns of the QUIJOTE MFI instrument displayed on
the 353 GHz \textit{Planck} map of thermal dust emission. Our analysis 
is focused on the region delineated by the white box.}
\label{fig:taurus-auriga-regions}
\end{center}
\end{figure}

The TMC is located at a mean distance of 140$\,$pc with an expected thickness
of $\sim$0.7$\,$pc in the line of sight
\citep[][]{qian15} and a total mass of about 1.5 $\times10^{4}\,M_{\odot}$
derived from CO data analysis \citep[][]{pineda2010}. The L1527 dark
cloud nebula is a region located at position
$(l,b)=(174.0^{\circ},-13.7^{\circ})$ where the brightest
bulk of thermal dust emission is detected in the TMC. 
Figure~\ref{fig:taurus-auriga-regions} shows the location of the TMC 
and  L1527 in the field of view of the QUIJOTE MFI instrument.
From $^{13}$CO observations
the TMC as a whole appears to be a complex network of filaments, the
width of which follows a broad distribution \citep[][]{panapoulou14}.  
This network of filaments is pervaded by well-structured large-scale
magnetic fields as seen from IR \citep[see][]{chapman11} and
visible \citep[see][]{arce98} polarization of starlight produced by absorption
of dust grains aligned perpendicular to the magnetic field. 
The large scale structure of the magnetic
fields inferred from these data reveals a picture consistent with the
magnetic fields traced with polarized thermal dust emission
\citep[see][]{pir20}. 
Taking advantage of all the astrophysical information provided by
the \textit{Planck} satellite an intensive study of the thermal dust properties
is provided by \citet{per25}. Thermal dust in equilibrium with the
incident radiation field is shown to cool down from 16--17$\,$K in the
diffuse regions to 13--14$\,$K in the coldest regions. Dust spectral
emissivity indices are centred at 1.78 with a standard deviation of
0.08 and a systematic error of 0.07. A study of the spatial variations of the
dust optical depth at 250 $\mu$m ($\tau_{250}$) shows an increase in
$\tau_{250}/N_{\rm H}$ by a factor of about 2 between the atomic phase
and the molecular phase, thus attesting radiative transfer effects as the
density increases.
Additional pictures of certain physical mechanisms 
have been inferred from the study of specific filaments of the TMC. 
\citet{ysard13} have shown an increase in the far IR
dust grain opacity valid along the entire length of the dense filament of L1506 that is interpreted as being due to grain growth and
therefore dynamic dust grain evolution. Striations perpendicular to
the B211/B213/L1495 filament have been detected by the \textit{Herschel}
satellite \citep[see][]{palmeirim13}. These structures tend to be aligned
with the magnetic field directions and suggest the accretion of material
on to the filament. 

In this work we present new intensity and polarization measurements 
of the emission between 10 and 20$\,$GHz 
with the first instrument of the QUIJOTE experiment towards the 
Taurus--Auriga complex (see Figure~\ref{fig:taurus-auriga-regions}). 
These measurements have been obtained
in order to probe and constrain the level of AME inside the TMC and L1527.
At Galactic latitudes $> -11^{\circ}$ the synchrotron radiation 
increases subtantially in the direction of the Galactic plane. For
that reason, in the following, unless specified otherwise, we
focus our analysis on the area observed at Galactic latitudes $< -11^{\circ}$. 
We begin by summarizing the data set
employed in Section~\ref{sec:data}. 
The \textit{Planck} component separation products are compared to the
QUIJOTE 13 GHz map and are discussed in Section~\ref{sec:planck_products}. 
In Section~\ref{sec:analysis} we analyse the morphology
  structure of the free--free maps obtained by using the H$\alpha$ map
  and compare it to the Commander free--free map. We use a TT-plot
  analysis to estimate the synchrotron component power law. We show 
how the QUIJOTE data improve the detection of AME, quantify the
amount of AME associated with the TMC and L1527, and then 
put constraints on the level of polarization of the AME in
both regions. 
The results and the various methods used to estimate the contributions
of the synchrotron, free, and AME components are compared and
discussed in Section~\ref{sec:discussion}. 
The conclusions are given in Section~\ref{sec:conclusions}.


\section{Observations and data} \label{sec:data}

\subsection{QUIJOTE data}\label{sec:quijote_data}

The data used in this work were obtained with the first instrument of
QUIJOTE, also called the multi-frequency instrument (MFI). It consists
of four horns that provide eight independent maps of sky intensity and
polarization at an angular resolution close to one degree and in four
frequency bands centred at $11$, $13$, $17$, and $19\,$GHz
(each frequency is duplicated in two different horns), each with a
2$\,$GHz bandwidth. The data from this instrument have been used for the
first time to analyse the intensity and polarization properties of the AME in
the Perseus molecular complex \citep{genova15}. A similar analysis was
conducted with data obtained towards the molecular complexes W43 and W47
and towards the supernova remnant W44 \citep{genova17}. This analysis provides 
the most stringent constraints on the level of AME polariZation associated
with a molecular cloud structure with $\pi_{\rm AME}<$0.39$\,\%$ (95$\,\%$
C. L.) from QUIJOTE 17 GHz data and $\pi_{\rm AME}<$0.22$\,\%$ (95$\,\%$
C. L.) from WMAP 41 GHz data. Here we present new data obtained towards the TMC.

\subsubsection{Observations and basic data processing}

The observations used in this work were carried out from 2015 March
4 to 2015 July 27 using the
MFI. They consisted of raster scans at constant elevation in local coordinates centred on
the TMC complex. We performed the observations by scanning the sky
in azimuth over a range of $15^\circ$ at a constant elevation
and at a velocity of $1^\circ$ in sky coordinates, stepping the
elevation by $0.1^\circ$ after each scan. Each of these observations
took around 25~min and produced a map of the sky 
capturing the transient of the TMC at the
elevation defined as a function of the time of observation. 
The MFI horns point to sky positions
separated by up to $5^\circ$, so each horn rasters a slightly
different patch. Therefore, the total sky area surveyed in each of
these observations is slightly wider. Once all the observations are 
combined together, the area common to all the horns is about 
$\sim 17^\circ\times 17^\circ$ (see Figure~\ref{fig:taurus-auriga-regions}). 

The final observing time was of 451$\,$h. All the data were first
carefully inspected by eye. Data affected by out-of-range housekeeping
values, contamination by geostationary satellites and their
near-sidelobes imprints, and  transient sources such as the sun, 
moon, and planets were automatically removed at the pipeline
level and not used in this study. All data less than $5^\circ$ from any satellite on the
plane of the sky were also automatically removed. The data showing 
glitch-type variations,
radio frequency interference (RFI), and any uncommon variations 
in the calibrated time-ordered-data (CTOD), including rapid atmospheric
variations, that might produce strong signal variations were also
removed. For this purpose a model of the sky \citep[][]{harper17} was removed from 
each of the 32 CTODs and the dispersions in groups of data of size 8
seconds were calculated for all the data. All groups of data of
size 8 seconds with a dispersion 
higher than a threshold value were flagged as bad data and not included when
producing the final maps. The remaining calibrated data not flagged at the CTOD
level were stacked as a function of azimuth in bins of one third of a degree. 
Additional flagging was applied by identifying ranges of
azimuth where the stacked data still showed high values. 
For horn 2 17$\,$GHz maps we directly identified by eye the
TOD pausing problems by looking at the stacked data and removing the
full CTODs from the analysis. After flagging, the final effective observing
time in the area common to all the horns is of about 290, 272, 281, and 241$\,$h, 
for horns 1, 2, 3, and 4 respectively. Horns 1 and 3 provide the 11 and 
13$\,$GHz bands. They are less affected by unstable channels than horns 2
and 4 but are more sensitive to external RFI signals entering through
far side lobes from geostationary satellites. Horns 2 and 4 are expected
to be less sensitive to RFI contamination. The filtering process to remove
groups of data contaminated by spurious signals is working well but
horn 4 was particularly exposed to an unidentified source of RFI during
the observations of the TMC which results in high RMS values in some
areas of the maps thereby rendering them of limited use for science. 
For this reason all the maps from horn 4 are not used in our analysis. 
As will be discussed in more detail below, the noise levels in
intensity are in the range 40--120 $\mu$Jy $ ($1$^\circ$ beam)$^{-1}$
and 3 to 5 times lower in $Q$ and $U$.

\subsubsection{Calibration}

Our calibration process is similar to the one explained in
\citet{genova17}. 
The main calibrator is Cass A, whose flux scale is assumed to follow
that derived by \citet{weiland11} with WMAP data. The secular variation of CASS A
is taken into account and corrected for by using the model proposed 
by \citet{hafez08}. 
The amplitude calibrator Cass A was observed
at least once a day except on a few days when observations had to
be stopped owing to brief intervals of poor weather. 
A total of 230 individual Cass A observations were obtained during the
period of observations of the TMC and the averaged gain
calibration factors are calculated for all the Cass A 
data obtained between 2014 April and 2015 December. The scatter in the
distribution of the measured gain factors for the 32 channels is in the range
0.3--2.2\% and is on average around $0.7\,\%$. 
An improvement with respect
to the previous calibration \citep[][]{genova15,genova17} is that 
the pipeline now takes advantage of the calibration diode installed in 
the middle of the secondary mirror in
front of the MFI. This diode is switched on for one second every 31
seconds and is used to track gain variations over time scales of
30 minutes for horns 1 and 3, and of one hour for horns 2 and 4.
The median value of the calibration factors calculated for each
channel were modulated as a function of time by these variations and
then used to calibrate each observation of Taurus. 
The errors due to the secular variations of Cass A, as well as 
the errors from the extrapolation of the Cass A
fluxes at WMAP and \textit{Planck} frequencies to the QUIJOTE frequencies
of order $4\,\%$, were added in quadrature to the gain calibration errors. 
All in all, we conservatively estimate that our absolute calibration errors are
accurate to within 5$\,\%$ for each channel. 

The reference zero position angles of the modulator for each horn of the MFI was
estimated by using Tau A (Crab Nebula) observations. Tau A was
observed at least on every day the Taurus region was observed. The polarization angle of Tau A,
estimated at 22.8$\,$GHz by \citet{weiland11}, varies by less than 5$\,\%$ 
from 22.7 to 90$\,$GHz. We assume the same to be true, or better, in the frequency
range of the MFI. We checked that the reference position angles
remained constant over time and found the scatter of the measurements
to be lower than 1$\%$. We then combined all the results from
the individual observations and derived the final values used for each
horn. The statistical accuracy of the values is of the order of a tenth of a
degree for each horn. 

\subsubsection{Map making}

The optics of the first QUIJOTE telescope provide highly symmetric 
beams (ellipticity $> 0.98$) with very low sidelobes ($\lesssim−40\,$dB) 
and polarization leakage ($\lesssim−25\,$dB). 
The MFI has been modified a few times since its first light.
The half-wave plate was blocked to position A (i.e.\ 0$^{\circ}$ in the
reference frame of the half-wave plate) from 2013 September,  and
for that reason the polarization maps of this horn are not used.
As a test bed Horn 1 was modified during 2014 April and 
all its paired channels have been correlated from that time.
Further modifications of the instrument took place
in 2015 December in order to get all the pairs of channels of 
horns 2, 3, and 4 correlated. For horns 2, 3, and 4, 
of the four channels per frequency range two channels 
were correlated and two were still uncorrelated during the Taurus
observations. In intensity the typical knee frequencies
are of order $f_{\rm k}$ $\sim$ 10--40$\,$Hz depending on the
channel. The correlated channels are those used
to calculated the $Q$ and $U$ maps because,
 for the measurement of polarization the 
subtraction results in much lower values of $f_{\rm k}$ $\sim$0.1--0.2$\,$Hz.
The method of deriving the values of the 
$I$, $Q$, and $U$ Stokes parameters is therefore the same 
as that used by \citet{genova17}, and we refer the reader to this
article for more details.

The final maps were produced with a destriper map-maker
implemented at the IAC and used to 
lower the stripes produced by the scanning strategy \citep[e.g.][]{kurki2009}.
Because of the low fraction of the sky (with respect to the full sky) covered by the observations a
gradient perpendicular to the direction of the scans was present in the
QUIJOTE intensity maps. This effect was removed a posteriori by
removing a two-dimensional plane from the final maps. 
As a consistency test, the same procedure was applied 
to the WMAP 22.8 GHz and showed a null residual between the original and
corrected map over the area of interest, thereby attesting that the origin of
the gradient observed in the QUIJOTE maps came from residual 
baseline drifts and RFI signals propagating into the maps as smoothed gradients.
Our final maps use the HEALPix format \citep[for Hierarchical Equal Area isoLatitude
Pixelization (HEALPix) of a sphere][]{gorski05} 
with N$_{\rm side}=$512 (i.e.\ a pixel size of $\lesssim$
6.9$\,$arcmin) that is sufficient enough to sample our beam. 
For further comparisons and combinations all our maps were smoothed with a kernel
corresponding to an angular full width at half maximum (FWHM) of
1$^{\circ}$. 
 

\begin{table*}
\begin{center}
\begin{tabular}{ccccccccccccc}
\hline\hline
\noalign{\smallskip}
Horn & Freq.   && \multicolumn{2}{c}{$\sigma_{I}$ ($\mu$K beam$^{-1}$)}
  && \multicolumn{2}{c}{$\sigma_{Q}$ ($\mu$K beam$^{-1}$)}    &&
                                                              \multicolumn{2}{c}{$\sigma_{U}$
                                                                 ($\mu$K
                                                                 beam$^{-1}$)} && $\sigma_{Q,U}$ (mK~s$^{1/2}$) \\
\noalign{\smallskip}
\cline{4-5}\cline{7-8}\cline{10-11}\cline{13-13}
\noalign{\smallskip}
& (GHz) && Map & NT && Map & NT && Map & NT && NT \\
\noalign{\smallskip}
\hline
\noalign{\smallskip}
     1&11 &&       45.9 &       29.6 &&        ... &       ... &&       ... &       ... &&       ...\\
     1&13 &&       38.5 &       25.2 &&        ... &       ... &&       ... &       ... &&       ...\\
     2&17 &&       98.0 &       63.3 &&       16.1 &       12.0 &&       15.1 &       12.7&&       2.8\\
     2&19 &&       115.0 &       78.5 &&       21.9 &  16.2 &&       16.6 &      14.6 &&       2.9\\
     3&11 &&       70.9 &       49.2 &&       21.4 &       14.1 &&       15.1 &       12.6 &&       2.6\\
     3&13 &&       58.2 &       37.6 &&       18.7 &       12.6 &&       16.5 &       11.8 &&       2.3\\
     4&17 &&       103.4 &       71.9 &&       18.1 &       11.9 &&       14.4 &       11.2 &&       2.2\\
     4&19 &&       122.0 &       82.4 &&       21.5 &       16.2 &&       19.3 &       19.1 &&       3.3\\
\noalign{\smallskip}
\hline\hline
\end{tabular}
\end{center}
\normalsize
\medskip
\caption{RMS per 1$^\circ$ beam in intensity and polarization, calculated on the QUIJOTE 
maps in a circular aperture of diameter 1$^{\circ}$ around Galactic coordinates $(l,b)=(167 ^{\circ},-13 ^{\circ})$.
For each Stokes parameter ($I,Q,U$) we show the RMS estimated from the
original maps and from half the difference of the two null test (NT) maps
as a function of Julian date. The RMS obtained from the 
original maps should be representative of the combined background 
and instrumental noise uncertainties, whereas the RMS calculated from the 
combination of the null test maps should be indicative of the level of
instrumental noise only. In the last column we display the instrument 
instantaneous sensitivities in polarization obtained by normalizing 
the averaged $Q$ and $U$ noise estimates calculated from the null test 
maps by the averaged integration time per beam. 
 }
\label{tab:time}
\end{table*}


\begin{figure*}
\begin{center}
\vspace*{2mm}
\centering
\includegraphics[width=125mm,angle=-90]{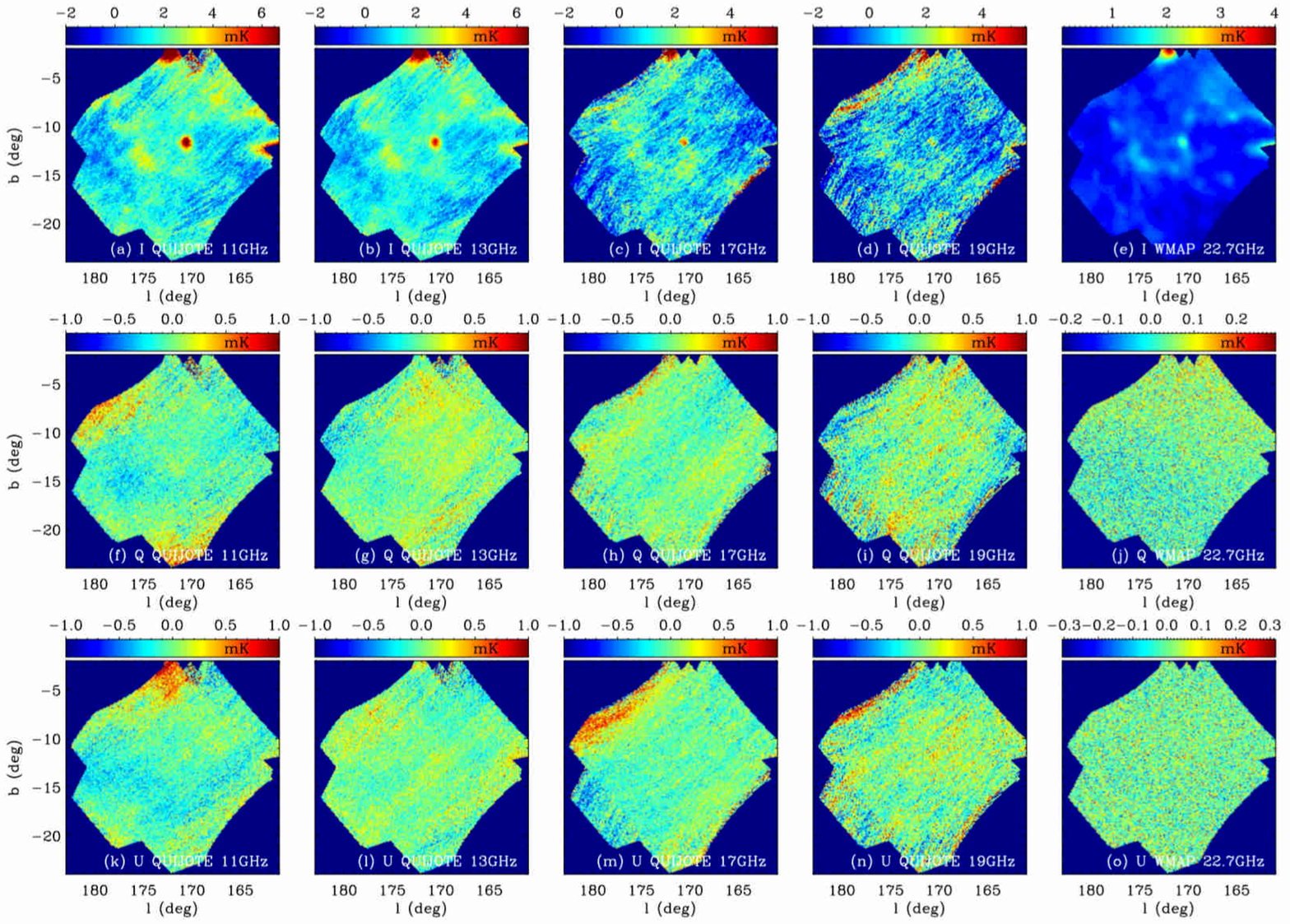}
\caption{\small Top row: intensity maps of the Taurus region observed with
  Horn 3 of the MFI at a frequency of 11$\,$GHz (a), 13$\,$GHz
  (b), with Horn 2 at a frequency of 17$\,$GHz (c)
  and 19$\,$GHz (d), and with WMAP at 22.7$\,$GHz (e). 
Second row: Stokes parameter $Q$ maps observed  with
  Horn 3 of the MFI at a frequency of 11$\,$GHz (f), 13$\,$GHz
  (g), 17 GHz (h), 19 GHz (i) and with WMAP at 22.7$\,$GHz (j). 
Third row: Stokes parameter $U$ maps observed  with
  Horn 3 of the MFI at a frequency of 11$\,$GHz (k), 13$\,$GHz
  (l), 17 GHz (m), 19 GHz (n), and with WMAP at 22.7$\,$GHz (o)}
\label{fig:quijote_wmap}
\end{center}
\end{figure*}

\subsection{QUIJOTE maps of the Taurus Molecular Clouds}

The intensity maps from horn 1 (11$\,$GHz and 13$\,$GHz), horn 2 (17$\,$GHz and
19$\,$GHz) and horn 3 (11$\,$GHz and 13$\,$GHz) are used for our analysis. 
As mentioned previously, the intensity maps obtained with horn 4 lack sensitivity
because many data had to be flagged out because of the presence of 
RFIs during the observations. The polarization maps from horn 
3 are also used for our analysis. The polarization maps from horn 4 are discarded
because of the problem with the intensity maps. The polarization maps from horn 1
are also not used because of the lack of definition to reconstruct
Stokes $Q$ and $U$ parameters. 
The sensitivities of all the $I$, $Q$, and $U$ maps from the correlated
channels are given in Table~\ref{tab:time}. The sensitivities of the 
$I$ destriped maps are in the sensitivity range $\approx$
40--120$\,\mu$K beam$^{-1}$ 
to a resolution of 1$^{\circ}$.
Given the low cut-off of the $1/f$ noise the sensitivies of the $Q$
and $U$ maps are three to five times better than those from the intensity maps. 
Null tests have been implemented that show consistency 
between the different data sets (see Table~\ref{tab:time} and
additional Tables in Appendix~\ref{sec:addjktests}) 
and ensure low level contamination by RFIs above the
detector nominal sensitivities in data from horns 1, 2, and 3.
The $I$, $Q$, and $U$ maps obtained with the MFI at 11$\,$GHz and 
13$\,$GHz with horn 3, and at 17$\,$GHz and 19$\,$GHz with horn 2
are shown in Figure~\ref{fig:quijote_wmap} and are compared with the 
maps obtained by WMAP at 22.7$\,$GHz. 


\begin{table*}
\begin{center}
\begin{tabular}{ccccccc}
\hline\hline
\noalign{\smallskip}
Frequency & Wavelength & Telescope$/$   & Angular resolution & Stokes
                                                               & Units
  & References\\
\noalign{\smallskip}
[GHz]& [mm] & survey & [$^{\prime}$] & Parameters  & & \\
\noalign{\smallskip}
\hline
\noalign{\smallskip}
  0.408 & 735.42& JB$/$Eff$/$Parkes & $\approx$ 60&     $I$ &   [K$_{\rm  RJ}$]    &\citet{haslam82}\\
&&&&&& \citet{remazeilles15}  \\
  1.420 & 211.30& Stockert/Villa-Elisa & 36 &     $I,Q,U$ &
                                                                    [K$_{\rm RJ}$]&\citet{reich82}\\ 
&&&&&&\citet{reich86}\\
&&&&&&\citet{reich01}   \\
&&&&&& \citet{wolleben2006} \\
  22.8 & 13.16& WMAP 9-yr & $\approx$ 49 &     $I,Q,U$ &  [mK$_{\rm CMB}$]&\citet{bennett13}     \\
  28.5 & 10.53& $Planck$ LFI & $32.60$ &     $I,Q,U$ &   [K$_{\rm CMB}$]&\citet{cpp2015-1}    \\
  33.0 & 9.09& WMAP 9-yr & $\approx 40$&     $I,Q,U$ &     [mK$_{\rm CMB}$]&\citet{bennett13}    \\ 
  40.7 & 7.37& WMAP 9-yr & $\approx 31$ &     $I,Q,U$ &    [mK$_{\rm CMB}$]&\citet{bennett13}   \\
  44.1 & 6.80& $Planck$ LFI &$27.92$ &     $I,Q,U$ &   [K$_{\rm  CMB}$]&\citet{cpp2015-1}   \\
  60.7 & 4.94& WMAP 9-yr & $\approx 21$&     $I,Q,U$ &   [mK$_{\rm CMB}$]&\citet{bennett13}    \\
  70.3& 4.27& $Planck$ LFI & $13.01$&     $I,Q,U$ &  [K$_{\rm CMB}$]&\citet{cpp2015-1}    \\
  93.5 & 3.21& WMAP 9-yr & $\approx 13$&     $I$ &     [mK$_{\rm CMB}$]&\citet{bennett13}  \\
  100 & 3.00& $Planck$ HFI &$9.37$ &     $I$ & [K$_{\rm CMB}$]&\citet{cpp2015-1}   \\
  143& 2.10& $Planck$ HFI &$7.04$ &     $I$ & [K$_{\rm CMB}$]&\citet{cpp2015-1}   \\
  217 & 1.38& $Planck$ HFI & $4.68$&     $I$ & [K$_{\rm CMB}$]&\citet{cpp2015-1}   \\
  353 & 0.85& $Planck$ HFI & $4.43$&     $I,Q,U$ & [K$_{\rm CMB}$]&\citet{cpp2015-1}      \\
  545 & 0.55& $Planck$ HFI & $3.80$&     $I$ &  [MJy sr$^{-1}$]&\citet{cpp2015-1}     \\
  857& 0.35& $Planck$ HFI & $3.67$&     $I$ &  [MJy sr$^{-1}$]&\citet{cpp2015-1}     \\
  1249 & 0.24& COBE-DIRBE & $\approx 40$ &     $I$ &  [MJy sr$^{-1}$]&\citet{hauser98}      \\
  2141 & 0.14& COBE-DIRBE & $\approx 40$&     $I$ &  [MJy sr$^{-1}$]&\citet{hauser98}      \\
  2997 & 0.10& COBE-DIRBE & $\approx 40$&     $I$ &  [MJy sr$^{-1}$]&\citet{hauser98}       \\
\noalign{\smallskip}
\hline\hline
\end{tabular}
\end{center}
\normalsize
\medskip
\caption{List of surveys and maps used in our analysis.
}
\label{tab:surveydata}
\end{table*}

\subsection{Ancillary and survey data}\label{sec:ancillary_&_survey_data}

The turnover of the AME spectral energy distribution is on average expected to be detected at
frequencies lower than 30$\,$GHz, including the frequency range 10--20$\,$GHz
in which the QUIJOTE MFI observes. In
order to fully sample the spectrum and the turnover of the AME SED the \textit{Planck} (28.5,
44.1, and 70.3$\,$GHz, 2015 release) and WMAP 9-yr (22.8, 33.0, 40.7, and 60.7$\,$GHz) low
frequency maps are added to our data. The WMAP 9-yr and \textit{Planck} HFI maps
at frequencies higher than 90 GHz and the low frequency DIRBE maps (100, 140, and
240$\,\mu$m) are combined together to characterize the contribution of
the thermal dust that dominates at frequencies higher than $\approx$100$\,$GHz.
On the other side of the spectrum, at very low radio frequency,
ancillary data are used to characterize the respective contributions
of the synchrotron and free--free emissions. 

All the data used in our analyses and the surveys they have been
obtained from are
summarized in Table~\ref{tab:surveydata}. We use the most up to date
version of these maps regridded into the HEALPix format
\citep{gorski05} at $N_{\rm side}$ = 512. All the maps were smoothed to a common resolution of
1$^\circ$. Some of the maps show significant baseline, or offset,
uncertainties, but we circumvent this problem by using T-T plots \citep{turtle1962},
or by removing these offsets by subtracting a local background in each
map when conducting SED analyses. In the
following section we discuss some of these data in more detail.

\subsubsection{Low frequency ancillary data}\label{sec:low_ancillary_data}

At low frequencies we use the all-sky 408 MHz map of \citet{remazeilles15},
and the \citet{reich86} $I$ map and \cite{wolleben2006} $Q$ and $U$ maps at 1.420$\,$GHz. 
The map provided by \citet{remazeilles15}  is an
improved version of the total-power radio survey map of \citet{haslam82}. Various
corrections have been applied to improve the original four individual
maps by using filtering techniques. The most notable corrections include
the reduction to a level $\ll 1$ K of large scale striations associated
with $1/f$ noise in the scan direction and the removal of the brightest 
extragalactic sources ($\gtrsim 2\,$Jy). 
In order to take into account the power that is introduced through 
the sidelobes in the 1.420$\,$GHz map we apply 
the correction factor of 1.55 derived by \citet{reich88} to compensate for 
the initial calibration of the map that was referred to the full-beam
solid angle. 
We assume a 10$\%$ uncertainty in the radio data at these two frequencies.

\subsubsection{WMAP data}\label{sec:wmap_data}

In the microwave range, at frequencies 23, 33, 41, 61, and 94$\,$GHz, we
use the $I$, $Q$, and $U$ 
maps from the 9-year data release of the \textit{WMAP} satellite \citep{bennett13}.
All the maps were retrieved from the  LAMBDA database.\footnote{Legacy Archive for
  Microwave Background Data Analysis, {\tt  http://lambda.gsfc.nasa.gov/}.}
Towards the TMC the maps are not dominated by instrumental noise and we
assume a 3.2$\%$ overall calibration uncertainty.

\subsubsection{$Planck$ HFI and LFI data}\label{sec:planck_data}

Additional $I$, $Q$, and $U$  maps are available in the microwave range at frequencies
28, 44, and 70$\,$GHz. They were obtained with the Low-Frequency Instrument
(LFI) on board of the \textit{Planck} satellite \citep{cpp2015-1}.  
We use the second public release version of these maps as provided by the 
\textit{Planck} Legacy Archive (PLA\footnote{\textit{Planck} Legacy Archive (PLA){\tt
    http://pla.esac.esa.int/pla/}.}). The LFI maps were corrected for the
bandpass mismatch, which produces intensity to polarization leakage, 
by using the correction maps also provided at $N_{\rm side}$=256 in the PLA database.  

In the submillimetre (submm) range we use the second data release version
of the intensity maps obtained with the High-Frequency Instrument (HFI) on board 
the \textit{Planck} satellite \citep{cpp2015-1} at frequencies centred at
100, 143, 217, 353, 545, and 857$\,$GHz. The type 1 CO maps \citep{cpp13}, 
were used to correct the 100, 217, and 353~GHz intensity maps for 
contamination introduced by the CO rotational transition lines
(1-0), (2-1) and (3-2), respectively. Stokes parameter $Q$ and $U$
maps were obtained with the HFI at 100, 143, 217, and 353$\,$GHz. The
second release version of the \textit{Planck} maps show a non-negligible 
leakage problem in the $Q$ and $U$ maps at 100, 143, and 217$\,$GHz;
therefore, in our analysis we only use the 353$\,$GHz $Q$ and $U$ maps.
We assume an overall calibration uncertainty of 3.4 $\%$ in the
HFI and LFI data at frequencies lower than 217$\,$GHz, but a value of 10$\%$ at 
353$\,$GHz, 545$\,$GHz, and 857$\,$GHz.

\subsubsection{High frequency ancillary data}\label{sec:high_ancillary_data}

In the FIR range we use the Zodi-Subtracted Mission Average
(ZSMA) \textit{COBE}-DIRBE maps \citep{hauser98} at 240~$\mu$m (1249$\,$GHz), 
140$\,\mu$m (2141$\,$GHz), and 100$\,\mu$m (2997$\,$GHz).
We assume an 11.9$\%$ overall calibration uncertainty in the data at
these frequencies.
 
\subsubsection{Colour corrections}\label{sec:colour_corrections}

Map colour corrections were applied to all the surveys except for those
at frequencies 0.408 GHz and 1.420 GHz, for which the narrow
bandwidths of the detectors make the corrections negligible. The colour
correction dependence on the fitted models was taken into account by
implementing an iterative process \citep[e.g.][]{pr13_9}. In each iteration the fitted model
on the QUIJOTE, \textit{WMAP} and \textit{Planck} maps is integrated over the bandpasses
retrieved from the QUIJOTE calibration database, the LAMBDA archive, and the PLA respectively. 
In the case of the DIRBE survey we use the colour correction tables
provided by the LAMBDA archive. The convergence is generally fast and is
reached after the second or third iteration. The magnitude of the colour
corrections is 
$\lesssim$1.3$\,\%$ for the QUIJOTE maps, 
$\lesssim$3.1$\,\%$ for the \textit{WMAP} maps, and
$\lesssim$12.2$\,\%$ for the \textit{Planck} and DIRBE maps. 

\section{Planck component separation products} \label{sec:planck_products}

Before analysing the set of maps obtained from direct observations of the sky, we consider various 
templates providing estimates of several physical components. 
These maps were produced with the Commander component separation
algorithm \citep{cpp2015-25} and are by-products
of various maps listed in Table~\ref{tab:surveydata}. 
Among other foreground maps we use the emission measure and electron
temperature templates that provide information relative to the level
of free--free emission. Hints on the level of emission of 
AME and the level of synchrotron emission expected towards the Taurus
region are given in the AME and synchrotron Commander templates respectively.  
In the frequency range where thermal dust emission is the major
component we use the optical depth estimates provided by the template at 545$\,$GHz.
All the maps were retrieved from the PLA.

\subsection{ Commander maps} \label{sec:first_glance}

The maps discussed in this section are the products of the
Commander component separation algorithm \citep{cpp2015-10}. 
This tool uses Monte Carlo/Markov chains in a Bayesian framework to
explore a parameterized description of the foregrounds. In the data model,
each foreground is described by a template map at a given frequency
and a spatially dependent SED. The parametric model of the SED 
depends on the foreground under consideration. It has been used on 
combinations of the \cite{haslam82} 408$\,$MHz, \textit{WMAP}, and
\textit{Planck} maps to investigate the range of spectral indices and
morphologies at frequencies around 20--50$\,$GHz. The products of 
this analysis are a CMB-nulled map at 22.8$\,$GHz. 
One of the CMB maps obtained from the analysis of the diffuse
  foregrounds is the SMICA map, which was obtained using a non-parametric method in the spherical harmonic domain
  \citep[see][]{cardoso2008,cpp2015-9}, where the foregrounds were modelled as a small number of
templates. SMICA was the method that performed best on the 
simulated temperature data. We use the SMICA CMB map to subtract the CMB component directly at the map level.
The GALPROP model
\citep[][]{strong2011,orlando2013} is used to estimate the synchrotron emission and produce full-sky maps of the free--free
electron temperature and emission measure, and two AME component maps
at 22.8$\,$GHz, and 41.0$\,$GHz respectively. The use of GALPROP is based on physically
motivated calculation since it requires less parameter tuning than
other methods, this to the cost of losing flexibility in tracing real
spatial variations. As a consequence, some cross-talk between the
synchrotron, free--free, and AME is expected. A thermal dust emission
map at a central frequency of 545$\,$GHz is also provided by the analysis.


\begin{figure}
\centering
\hspace*{-1.3cm}
\vspace{-1.cm}
\includegraphics[width=107mm]{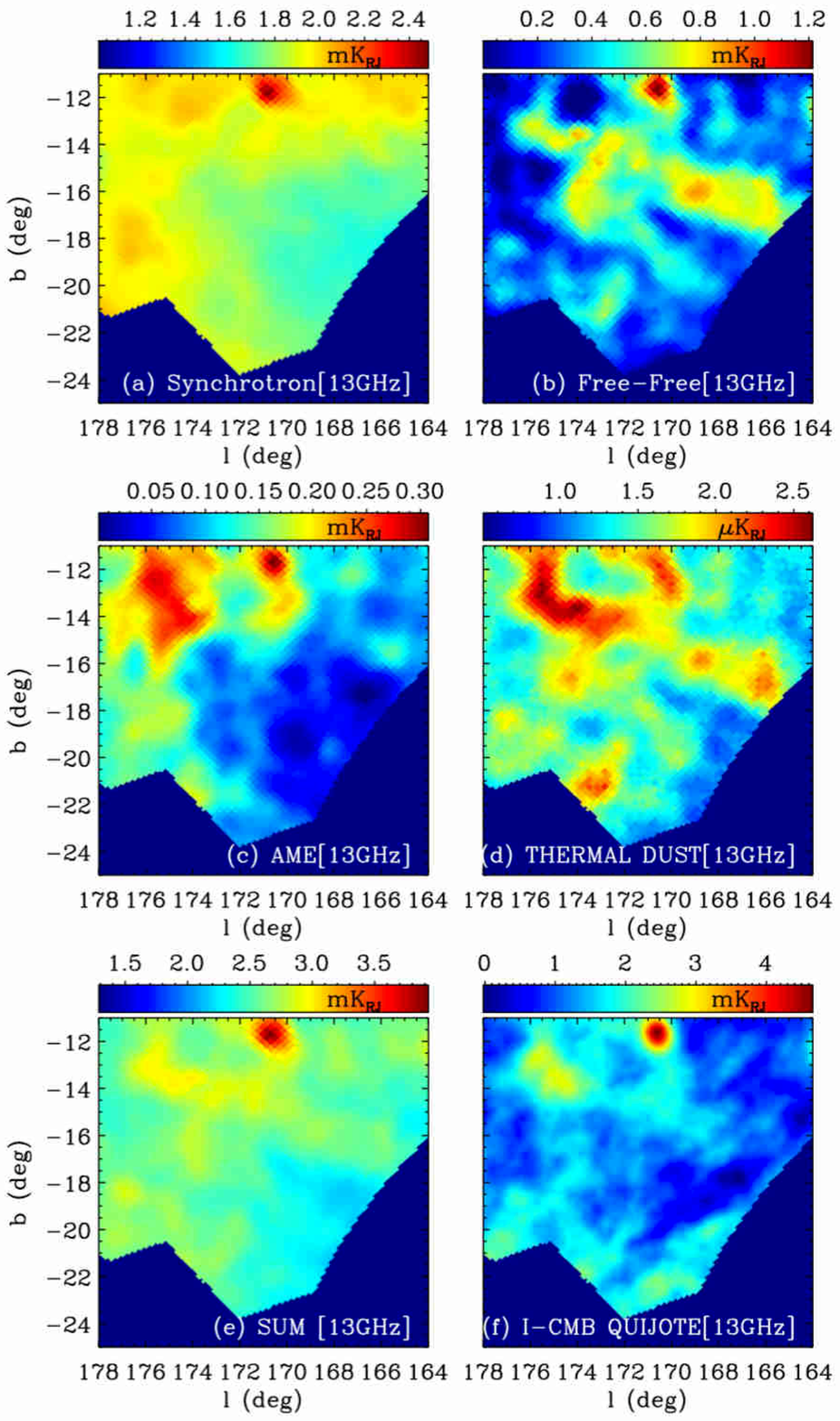}\\
\vspace{+0.1cm}
\includegraphics[width=80mm]{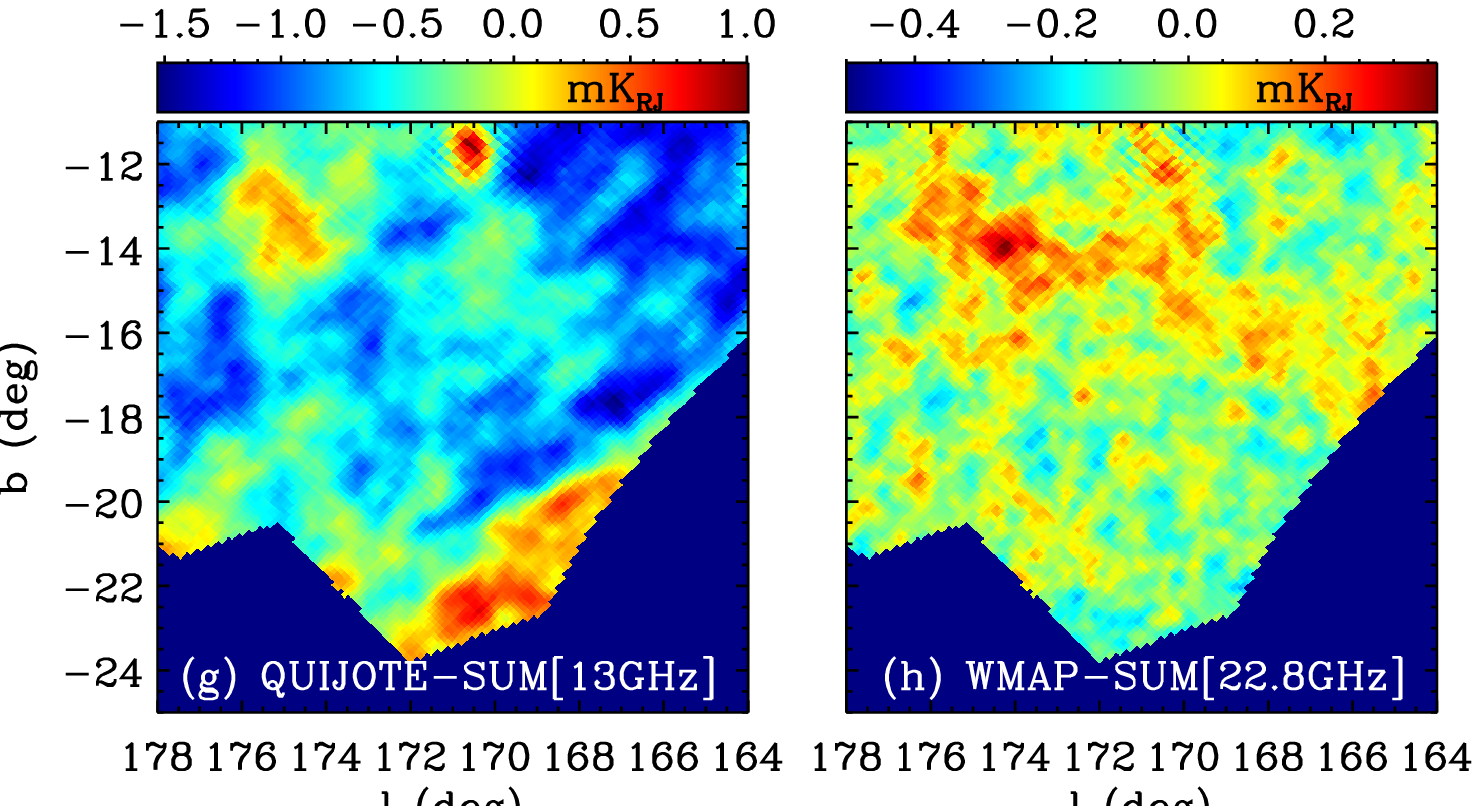}
\caption{\small Commander template maps extrapolated to 13$\,$GHz. 
Top Left to bottom right: 
a) Synchrotron intensity map from the
  Commander component separation tool \citep{cpp2015-25} in
  mK$_{\rm RJ}$ at a reference frequency of 13$\,$GHz; 
b) free--free intensity map at 13 GHz calculated with the
Commander
electron temperature and Emission Measure maps; 
c) Commander AME total intensity map at 13$\,$GHz; 
d) thermal dust intensity map in $\mu$K$_{\rm RJ}$ from Commander extrapolated
from the reference frequency  545$\,$GHz to a frequency 13$\,$GHz; 
e) sum map of the synchrotron, free--free, AME, and thermal dust 
Commander intensity maps at a reference frequency of 13$\,$GHz; 
f) QUIJOTE-MFI 13 GHz SMICA-CMB corrected map; g) QUIJOTE-MFI 13 GHz
SMICA-CMB corrected map - Commander 13 GHz Summed maps; and h) WMAP 22.8 GHz
SMICA-CMB corrected map - Commander 22.8 GHz Summed maps.}
\label{fig:foreground_ame_maps}
\end{figure}

\subsection{AME morphology of the TMC at 13 GHz from
  Commander } \label{sec:quijote_morphology}

We first estimate the contributions of the free--free, synchrotron, and thermal dust
from the Commander component maps at a central frequency of 
13$\,$GHz. Assuming a power law relation \citep{platania98} with a 
constant index of $\beta_{\rm synch}=$2.8 (the choice of this
  value will be justified in Section 4) an average estimate
of the level of synchrotron emission at 13$\,$GHz can be obtained by assuming that
the Haslam map at 408$\,$MHz mainly probes the synchrotron
radiation. The free--free intensity map is estimated at 13$\,$GHz following the
equations in Table 4 from \citet{cpp2015-10}  by using the
Commander free--free electron temperature map and emission
measure map. The distribution of the Commander  
total AME intensity map estimated towards the TMC
is obtained by summing the contribution of the two AME component maps
provided by the Commander component separation analysis
at a central frequency of 13$\,$GHz. The thermal dust intensity map extrapolated
from a frequency  545$\,$GHz to 13$\,$GHz is obtained by 
following the equations given in Table 4 from \citet{cpp2015-10}.
All these maps, as well as their sum, are shown in
Figure~\ref{fig:foreground_ame_maps}
for comparison with the 13$\,$GHz MFI CMB-corrected intensity 
map also displayed in the Figure. The CMB correction was
done by subtracting the SMICA-CMB map from the QUIJOTE 13$\,$GHz map.

From the Commander component analysis one can see 
that the synchrotron (panel a) and free--free 
(panel b) components could respectively contribute
about 50$\%$ and about 25$\%$ or more 
to the 13$\,$GHz MFI CMB-corrected intensity map. 
As expected, the thermal dust contribution (panel d) should be fully negligible at
this frequency, whereas the AME intensity (panel c) is expected to contribute about 
10$\%$ of the total intensity. In terms of morphology one can see that the 
AME is expected to be encountered almost everywhere except
in the south-west corner of the Commander map. 
At first sight the morphology of the sum of the Commander
intensity maps (panel e) and  the QUIJOTE CMB-corrected intensity map (panel f) look
slightly different. In particular the L1495 filament (see 
Figure~\ref{fig:taurus-auriga-regions}) seems to be captured
by the Commander  free--free component map but
is barely seen in the QUIJOTE CMB-corrected intensity map. 
This discrepancy and others are easier to visualize once
  one looks to the difference between the QUIJOTE CMB-subtracted map
  and the Commander summed maps at 13 GHz. This is shown in panel g.
  Even though the absolute baseline level of the
  QUIJOTE maps is not known, one can see that the difference map shows
  strong morphological differences that are not only noise-like.
A possible explanation of such differences is that sky maps at the
QUIJOTE-MFI frequencies were not available when the Commander 
analysis was performed. The Commander analysis was performed using the
408 MHz map combined with a set of maps with frequencies higher than
22 GHz. For that reason, it most probably offers a reasonable component
separation model at frequencies higher than 22 GHz, 
but its predictability at frequencies lower than 22 GHz could be limited. 
As a simple test, we repeated at a frequency of 22.8 GHz 
the analysis we did at 13 GHz. The structural features obtained at
22.8 GHz (not shown here) are very similar to those seen at 13 GHz
(Figure~\ref{fig:foreground_ame_maps}, panels a to e).
Indeed, the difference map obtained at 22.8 GHz and shown in panel h
is more consistent with noise than that obtained at 13 GHz (panel
g). Some residuals of the structure of the molecular clouds 
can still be seen, though, that may be indicative of cross-talk between the
free--free, AME, and synchrotron components at 22.8 GHz. 
Comparisons between the Commander products and the component
separation analysis obtained from spectral energy distributions
will be discussed further in Section 5.

\subsection{Extra-Galactic sources and masks} \label{sec:masks}

\begin{table*}
\begin{center}
\begin{tabular}{ccccllll}
\hline\hline
\noalign{\smallskip}
Source name & Glon $[^{\circ}]$ & Glat  $[^{\circ}]$ & 
FWHM $[^{\prime}]$&30GHz flux [mJy]& Other name$^{(a)}$&Source type$^{(a)}$\\
\noalign{\smallskip}
\hline
\noalign{\smallskip}
PCCS2 030 G163.79-11.99&       163.789&      -11.986&      33.08&774.27$\pm$      26.14&1RXS J041435.6+341842 & X-ray source\\
PCCS2 030 G168.02-19.65&       168.023&      -19.653&      33.38&1415.45$\pm$      38.46&QSO B0400+258 & Quasar\\
PCCS2 030 G169.00-22.47&       169.002&      -22.473&      33.22&724.52$\pm$      47.36&IRAS 03542+2311 & Infra-Red source\\
PCCS2 030 G170.54-11.64&       170.544&      -11.636&      35.89&3867.93$\pm$      23.38&CXOSEXSI J043659.1+294231 & X-ray source\\
PCCS2 030 G171.34-14.38&       171.338&      -14.378&      33.85&1169.57$\pm$      23.18&2MASS J04294819+2717322 & Infra-Red source\\
PCCS2 030 G176.90-18.56&       176.901&      -18.556&      32.77&577.59$\pm$      15.43&2MASS J04313041+2035383 & T Tau-type Star\\
\noalign{\smallskip}
\hline\hline
\end{tabular}
\end{center}
\normalsize
\medskip
\caption{\small Name and Galactic positions of the 
Galactic and extra-Galactic sources located in the 
field of  view (FOV) of the
Taurus--Auriga molecular cloud complex
(Figure~\ref{fig:taurus-auriga-regions}) 
identified with the second
\textit{Planck} Catalog of Compact Sources (CCS) \citep{cpp2015-26}. 
The FWHMs are the effective FWHMs from Gaussian fits to the sources.
Note: $^{a}$ information retrieved from the Simbad 
database ({\tt http://simbad.u-strasbg.fr/simbad//}). 
The other name is the one of the closest sources at a distance less than 5$^{\prime}$ from the
Galactic coordinates given in columns 2 and 3.}
\label{tab:pccs}
\end{table*}

Extra-Galactic and Galactic compact sources strongly emitting at 
30$\,$GHz and lying in the field of view (FOV) of the
TMC complex could bias our analysis. In order to identify and mask 
these sources we use the 
\textit{Planck} Catalog of Compact Sources (CCS) \citep{cpp2015-26}. 
Names identifying sources and their coordinate positions in the
Galactic reference frame are given in Table~\ref{tab:pccs}.
The Gaussian effective FWHM and the 30$\,$GHz fluxes provided by the CCS 
catalogue are also displayed. 
Additional information retrieved from the Simbad database is given in
the last two columns. The information provided by the CCS was crossed-checked
with a search of the NASA/IPAC Extragalactic Database 
(NED,\footnote{{\tt  https://ned.ipac.caltech.edu/}.}) showing that PCCS2
030 G170.54-11.64 is the well studied object also known as 3C123. 
A closer look at the observations of this source suggested a large scale
structure around 3C123 with two hotspots about one degree away 
\citep{reich76}. From comparisons with the extensions seen in the Haslam and QUIJOTE maps 
we saw that the angle formed by the large scale hotspots in the map
displayed by \citet{reich76} are not aligned with the extensions seen
in the Haslam and QUIJOTE maps, which look rather circular. Actually, 
a closer look at the NRAO VLA Sky Survey (NVSS \footnote{{\tt
    http://www.cv.nrao.edu/nvss/}.}) shows some fairly 
bright extra-Galactic sources (compact) at these locations, therefore
the extension seen in \citet{reich76} is not that of giant
  radio structures. Given the fair amount of extinction in the
visible (of order 3 magnitudes) towards 3C123,
we therefore concluded that this source is indeed a 
bright compact ($\approx$40$\,$arcsec) extra-Galactic
radio source sitting on top of an extended Galactic dust cloud. 
This result agrees with the 1400$\,$MHz data analysis of 3C123 by \cite{hanisch1984}.
In the following, in order to exclude the extra-Galactic or Galactic
compact sources in our analysis of the
series of 1$^{\circ}$ resolution maps we apply masks   
with discs of radius $90^{\prime}$ for PCCS2 030 G170.54-11.64 
and  $45^{\prime}$ for all the other sources.
Some of the masked sources are encountered towards the TMC, as can be seen in 
the maps shown in Figure~\ref{fig:map_ff_synch408}.

\subsection{Correlation plots in the TMC region} \label{sec:correlation_plots}

\begin{figure}
\begin{center}
\vspace*{2mm}
\centering
\includegraphics[width=85mm,angle=0]{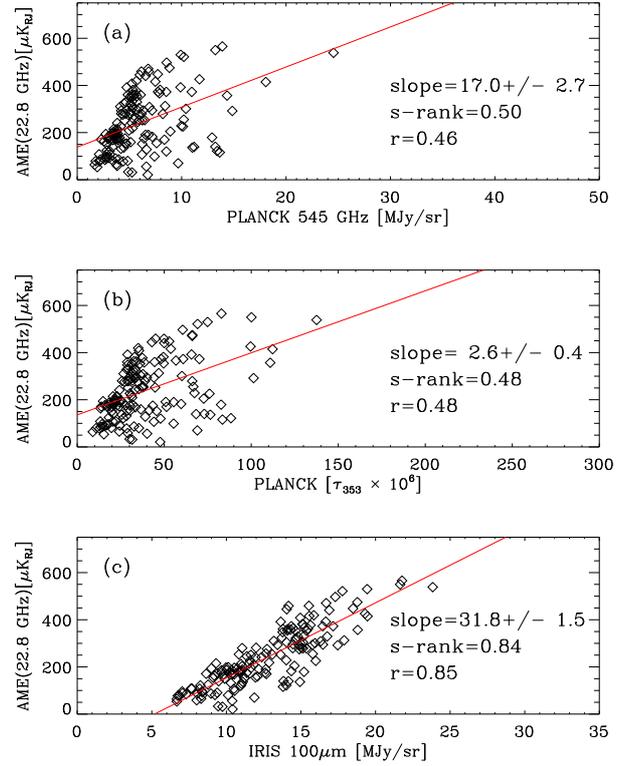}
\caption{\small TT-plots of the Commander AME intensity
  towards the TMC region (delineated in white in
Figure~\ref{fig:taurus-auriga-regions}) 
at 22.8$\,$GHz with thermal dust amplitude at 545$\,$GHz 
(a), the optical depth at 353 GHz, $\tau_{353}$ (b), 
and the IRAS map at 100$\,\mu$m (c)
from 1$^{\circ}$ resolution maps at $N_{\rm side}=64$, where the
sources displayed in Table~\ref{tab:pccs}  have been masked.   
The slopes are shown for comparison with the results
provided by \citet{cpp2015-25} towards other molecular cloud
regions. The Spearman rank correlation ($s$-rank) and
Pearson correlation coefficient ($r$) are also shown. The best fit
is shown as a red line.}
\label{fig:correl_ame}
\end{center}
\end{figure}

Diffuse AME detection is discussed in \citet{macellari11} and is 
shown to be clearly detected over a large fraction of the sky away from the Galactic
plane in the \textit{WMAP} $K$ and $K_{a}$. 
More specific diffuse AME regions are discussed in \citet{cpp2015-25} including the
molecular cloud regions R CrA, $\rho$ Oph and Musca.
In this work the Commander map AME emissivities at 22.8$\,$GHz 
relative to the thermal dust amplitude at 545$\,$GHz, the IRAS map at
100$\,\mu$m, and the optical depth at 353$\,$GHz, $\tau_{353}$ are discussed. On the galactic scale
correlations are found to be higher between AME at 22.8$\,$GHz 
relative to the thermal dust amplitude at 545$\,$GHz and the optical depth at
353$\,$GHz, $\tau_{353}$ than between AME at 22.8$\,$GHz 
relative to the \textit{IRAS} map at 100$\,\mu$m \citep[see Figure 12 in][]{cpp2015-25}. 
For comparison,
we applied a similar analysis to the 1$^{\circ}$-resolution Taurus Commander  
maps with a pixelization at $N_{\rm side}=64$ including the masks discussed
above in the region delineated with a white border in Figure~\ref{fig:taurus-auriga-regions}. 
For each map this pixelization is used to avoid any correlation between pixels.
Our results are shown in the plots in
Figure~\ref{fig:correl_ame}. The best correlation
is that for the AME emissivities  at 22.8 GHz relative to the \textit{IRAS} map at
100$\,\mu$m with a slope $= 31.9 \pm 1.6\,\mu$K$_{\rm RJ}\,$(MJy sr$^{-1}$)$^{-1}$ and 
a correlation factor $r =0.85$. In comparison, lower correlation
factors are found between the AME emissivities at 22.8 GHz relative 
to the thermal dust amplitude at 545$\,$GHz ($r =0.46$) and the optical 
depth at 353$\,$GHz, $\tau_{353} \times 10^6$ ($r =0.48$). 
This correlation between the Commander AME product and 
the 100 $\mu$m map suggests that for the TMC the AME carriers could be
very small grains, consistently with previous analyses on other
regions \citep[see][and references therein]{dickinson18}.
For the TMC the correlation factor between the AME
emissivities at 22.8$\,$GHz relative to the thermal dust amplitude at
545$\,$GHz are lower than those found for the Pegasus Plume ($r=0.68$), Musca
($r=0.63$), and $\lambda$ Orionis ($r=0.85$) regions \citep[see Figure 14 in][]{cpp2015-25}. 
On the other hand, the AME emissivity to thermal dust
optical depth ratio is quite low and of order 2.6 $\times 10^{-6}\,\mu$K$_{\rm RJ}$, similar
to estimates obtained towards other molecular cloud regions such as $\rho$
Oph and Perseus \citep[see Table 2 in][]{cpp2015-25}.

\section{Data analysis} \label{sec:analysis}

Before investigating any SEDs of the
intensities mapped towards the Taurus
region we take advantage of the low frequency map and  the
publicly available Full-Sky H$\alpha$ template map \citep{finkbeiner03,dickinson03}
to assess the level of free--free emission and the power law spectrum 
associated with the synchrotron emission. 
Such information will be useful for comparison with the
component separation that will be obtained from the SED analysis.
The H$\alpha$ map was retrieved from the LAMBDA\footnote{LAMBDA Data-Products, {\tt
    https://lambda.gsfc.nasa.gov/product/}} database.  
For the SED analyses we will use the SMICA-CMB map retrieved from 
the PLA to directly remove the CMB component at the map level. 

\subsection{Free--free components} \label{sec:ff_comp}

Estimates of the free--free component towards the TMC
complex can be assessed following the method proposed by
\citet{dickinson03}  \citep[see also][]{cpp2015-25}. The temperature
of the free--free component is given by
\noindent
\begin{equation}
\frac{T_{\rm b}^{\rm ff}}{I_{\rm H\alpha}}=1512 T_{4}^{0.517}
10^{0.029/T_{4}} \nu^{-2.0}_{\rm GHz} g_{\rm ff} ,
\end{equation}
where $T_{4}$ is the electron temperature in units of 10$^{4}\,$K,
$\nu_{\rm GHz}$ is the frequency in GHz, and $g_{\rm ff}$ is the Gaunt
factor, which takes into account quantum mechanical effects. 
As noted in \citet{cpp2015-25} a pre-factor of 1.08 to account for the
He II contribution that adds to the free--free continuum is included. In our
calculations we also use the approximation to the Gaunt factor provided by 
\citet{draine11} that is accurate to within about 1$\,\%$ in the GHz
frequency range at frequencies much higher than the `plasma
frequency', $\nu_{\rm P}$, of order a few MHz for a plasma with an
electron temperature 
$T_{\rm e} \approx\,$7000 K and electron density $n_{\rm e} \sim
10^{-2}\,\rm cm^{-3}$:
\noindent
\begin{equation}
g_{\rm ff}(\nu ,T_{e}) = \rm{ln} \left\{ \rm{exp} \left[ 5.960 -
    \frac{\sqrt 3}{\pi} ln(\nu_{\rm{GHz}} \it{T}\rm_{4}^{-3/2})\right]\right\}.
\end{equation}

To run the calculations we use the Full-Sky H$\alpha$ template for microwave
foreground prediction provided by \citet{finkbeiner03} and the electron
temperature, $T_{\rm e}$, template map from the Commander
component separation analysis to give an almost uniform 
temperature $T_{\rm e} = 7000\,$K over the whole area. 

\begin{figure}
\begin{center}
\vspace*{2mm}
\centering
\includegraphics[width=85mm]{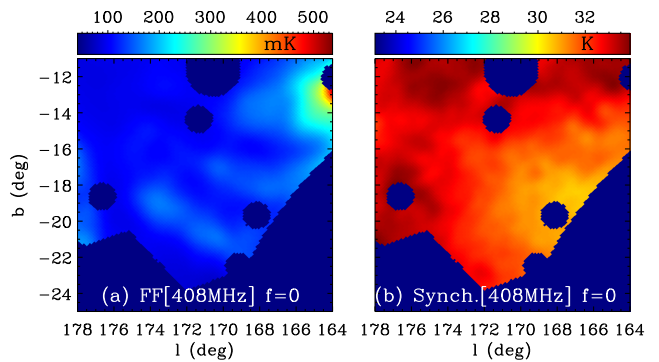}
\caption{\small
a) Estimate of the free--free emission at 408 MHz obtained from the
H$\alpha$ emission map under the hypothesis that none of the
H$\alpha$ emission is absorbed by the thermal dust ($f=0.0$). 
b) Maps at 408 MHz of the synchrotron residual 
obtained after subtraction of the free--free
component map shown on left, assuming 
a synchrotron power law index of $-2.80$. 
All maps are 1$^{\circ}$ resolution HEALPix maps at $N_{\rm side}=512$.
}
\label{fig:map_ff_synch408}
\end{center}
\end{figure}

\begin{figure}
\begin{center}
\vspace*{2mm}
\centering
\vspace{-2.5cm}
\hspace*{-2.6cm}
\includegraphics[width=135mm]{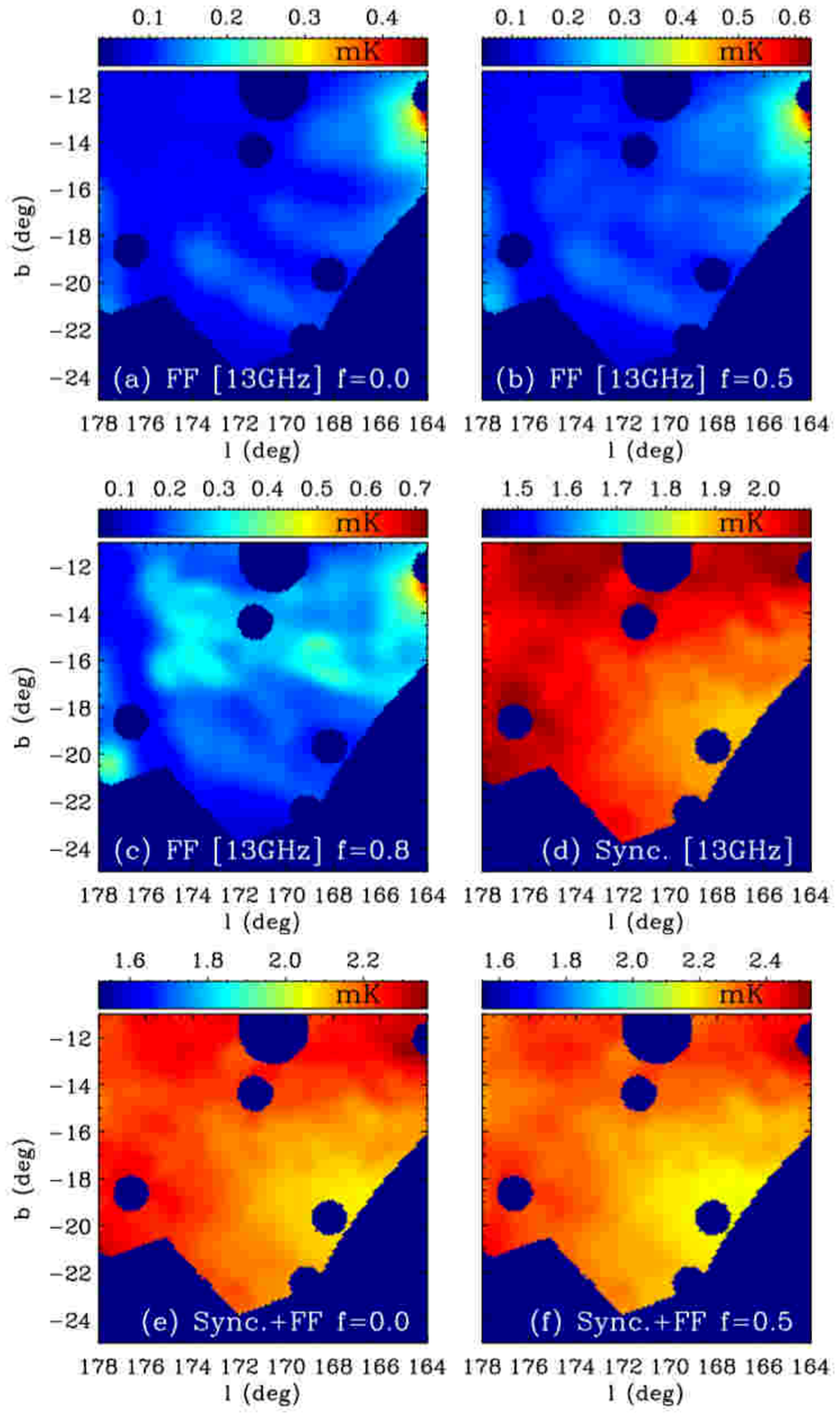}\\
\vspace{-2.3cm}
\hspace*{-3.9cm}
\includegraphics[width=46mm]{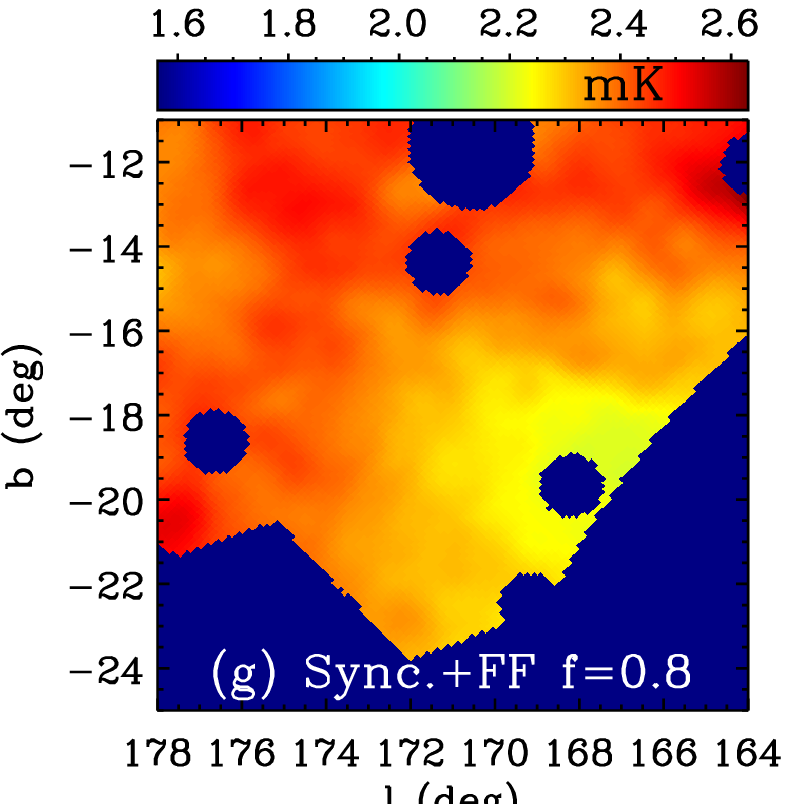}
\caption{\small
All maps are 1$^{\circ}$-resolution HEALPix maps at $N_{\rm side}=512$.
a) free--free map at 13 GHz, estimated from the
H$\alpha$ emission map under the hypothesis that none of the
H$\alpha$ emission is absorbed by the thermal dust ($f=0.0$); 
b) 13 GHz free--free map assuming a dust mixing factor $f=0.5$;
c) 13 GHz free--free map assuming a dust mixing factor
$f=0.8$;
d) synchrotron map at 13 GHz assuming 
a synchrotron power law index of $-2.80$; 
e) coadded estimated maps of the synchrotron and
free--free components at 13 GHz assuming none of the H$\alpha$ emission 
is absorbed by the thermal dust ($f=0.0$);
f) same as panel e but
assuming a dust mixing factor $f=0.5$; and
g) same as panel e but assuming
a dust mixing factor $f=0.8$. 
}
\label{fig:map_ff_synch13}
\end{center}
\end{figure}

We first make the assumption that none of the H$\alpha$ emission
is absorbed by the dust (i.e.\ we assume a dust mixing factor $f=0.0$). 
The map showing our results is shown in
Figure~\ref{fig:map_ff_synch408}. At 408 MHz the free--free brightness
temperature estimates obtained towards the TMC region are very low, 
with values showing variations below 200 mK and representing only 
$\sim 1\,\%$ of the 408 MHz total intensity map. Once the calculations are 
repeated for a central frequency of 13$\,$GHz (see
Figure~\ref{fig:map_ff_synch13}, panel a) the free--free 
brightness temperature estimates obtained towards the 
TMC region show variations lower than 0.10$\,$mK, i.e.\ about 5 to 8
times lower than the free--free component obtained from the 
Commander  analysis as shown in
Figure~\ref{fig:foreground_ame_maps}, panel b.  

In order to take into account the possibility that a fraction of the
H$\alpha$ emission is absorbed by thermal dust emission we
refer to \citet{dickinson03} and \citet{harper15}.
The expected emission $I_{\rm em}$ is related to the observed
emission $I_{\rm obs}$ by the H$\alpha$  absorption model:

 \noindent
\begin{equation} \label{equation_if}
I_{\rm obs}= \frac{I_{\rm em}}{\tau} \int_{0}^{f_{\tau}}
e^{\tau^{'}-\tau} d\tau^{'} + (1-f_{\tau}) I_{\rm em}.
\end{equation}

\begin{figure}
\begin{center}
\vspace*{2mm}
\centering
\includegraphics[width=65mm]{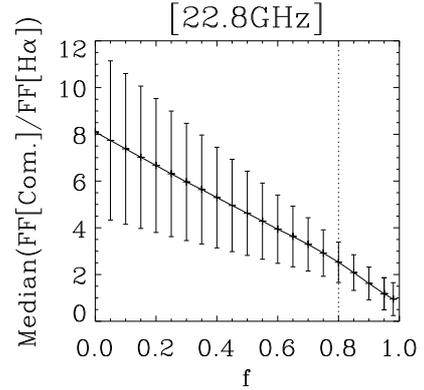}
\caption{\small
Variations of the median of the 22.8 GHz free--free ratio maps (free--free
Commander map / H$\alpha$ calculated free--free maps) as a function of the
dust mixing factor $f$. The standard deviations of the free--free
ratio maps are superimposed on the median values and illustrate the amount
of spatial variation between the pairs of maps. The dispersion
between the ratio of the maps derived by
each method is minimized and about uniform for dust mixing factors
$\ge 0.8$. The limit $f=0.8$ is shown by the dashed vertical line.
}
\label{fig:ff-difference-maps}
\end{center}
\end{figure}

This model assumes that a fraction, $f_{\tau}$, of the HII gas and
dust are mixed and in thermal equilibrium, and that the remaining
fraction of the HII gas lies in front of the TMC thermal dust such
that its H$\alpha$  emission is unabsorbed. 
 The H$\alpha$ optical depth, $\tau$, is related to the H$\alpha$
absorption by $\tau = A(\rm H\alpha)/(2.5 log_{10}(e))$, where
  $A(\rm H\alpha)=0.81 A\rm v$. Very accurate column density 
and extinction maps of the TMC have been provided by the 2MASS survey 
from analysis of near-infrared colour excesses 
\citep[see][]{pineda2010,lombardi2010}. Because of limits inherent to
the technique these maps tend to underestimate the extinction towards 
high column densities \citep[see][]{pir11,per25}, which are regions we are interested by. 
In addition, these maps do not fully cover the region we are studying;
therefore, in the following we use the PLA thermal dust optical
depth at 353 GHz, $\tau_{353}$, to estimate visible extinctions towards the TMC.
Assuming that the column density and thermal dust optical
depth at 353 GHz are proportional and related by $N_{\rm H} = 1.6 \times 10^{26}
\tau_{353}$ \citep[][]{cpp2015-19}, and that the extinction,
$E(B-V)$, is such that $N_{\rm H}/E(B-V) \approx 6.94 \times 10^{21}\,$cm$^{-2}$ 
\citep[][]{pir11}, we have, $A(\rm H\alpha) \approx 18674.4 \times R(V) \times
\tau_{353}$, where the reddening parameter, $R(V)$, is expected to lie in
the range of magnitudes $[$3.5, 5.5$]$ \citep[][]{vrba84,vrba85}.
The 13 GHz maps obtained by considering an averaged value $R(V)=4\,$magnitudes towards the
TMC are shown in Figure~\ref{fig:map_ff_synch13}  for dust mixing 
fractions $f=0.5$ and $f=0.8$. The level of free--free intensity
increases non-linearly with the dust mixing fraction, $f$, up to a
factor 4 towards the highest column density regions of the TMC for a
dust mixing factor, $f=0.8$. 
We also note that the structure of the free--free emission
looks more similar to that obtained by Commander
when considering a dust mixing factor $f \approx0.8$ than when
considering lower values for $f$ since the effect of high dust mixing 
fractions reveals the TMC morphology seen in the thermal dust maps. 
In section \ref{sec:quijote_morphology} we show that at
a central frequency of 22.8 GHz the Commander component separation 
returns self-consistent results towards the TMC region. 
In order to quantify how the
morphology of the 22.8 GHz Commander free--free map compares 
with the morphology of the H{$\alpha$} 
estimated free--free maps we made maps of the ratio between the two
sets of maps obtained for different values of $f$. We then calculated
the median values and the standard deviations of the set of maps. 
The results are shown in Figure~\ref{fig:ff-difference-maps}. 
One can see that the median of the ratio of the two maps is almost always
$>1$ and decreases with $f$,   
and that the dispersion of the ratio is minimized and uniform
for values of $f \ge 0.8$. Our calculations of the free--free maps with the
H$\alpha$ template rely on quite a simple extinction model
and also assumes a uniform dust mixing factor over a large area. In addition, the
Commander separation free--free map is expected to be subject to 
cross-talk with
other physical components (AME, synchrotron). All in all though, 
the morphology structure of the maps start to be similar for
values of $f \ge 08$ (as is also observed at a frequency of 13 GHz, see
Figure~\ref{fig:foreground_ame_maps}, panel b and 
\ref{fig:map_ff_synch13}, panel c). 
Keeping in mind all these caveats, in the following we will be conservative and 
consider a dust mixing factor $f \approx 0.8$, i.e.\ that 
about 80$\%$ of the thermal dust is mixed with the H$\alpha$ gas.


\subsection{Synchrotron frequency spectrum}

The free--free component towards the Taurus region shows a very  low
intensity with respect to the total intensities observed
at 408$\,$MHz and 1.420$\,$GHz. Therefore, one should expect synchrotron emission to be
the major mechanism to produce the intensity observed at these two
frequencies. This point is supported by the level of synchrotron emission
displayed in Figure~\ref{fig:map_ff_synch408} after the
contribution of the estimated free--free emission has been removed from
the maps. One should therefore expect a constant ratio, $r$, 
between the pixels, $i$, of the two emission maps such that
\noindent
\begin{equation}
r=\frac{T_{2}(i)}{T_{1}(i)} \times \left(\frac{\nu_{2}}{\nu_{1}}\right)^{\beta_{\rm synch}}
\end{equation}
where $\beta_{\rm synch}$ is the synchrotron power spectrum index
\citep{platania98} in the range of frequencies considered.

\begin{figure}
\begin{center}
\vspace*{2mm}
\centering
\includegraphics[width=65mm,angle=90]{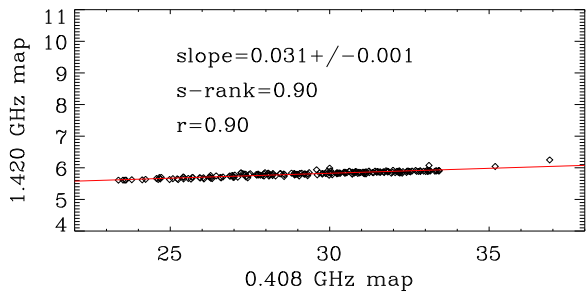}
\caption{\small Correlation plot between the emission temperatures
  at 0.408$\,$GHz and 1.420$\,$GHz towards the TMC
(region delineated in white in Figure~\ref{fig:taurus-auriga-regions})
  from 1$^{\circ}$ resolution HEALPix maps at $N_{\rm side}=$64. The
  uncertainties, assumed to be of order 10$\,\%$, are not shown for clarity.}
\label{fig:tt-408-1420}
\end{center}
\end{figure}

We conduct a $T$--$T$ plot analysis of the 408$\,$MHz and 1.420$\,$GHz maps in order 
to get estimates of the power law index $\beta_{\rm synch}$. For the map at 1.420$\,$GHz we
consider a correction factor equal to 1.55 to correct for the effect
of the near sidelobes of the beam. Figure~\ref{fig:tt-408-1420}
shows the temperature distribution at the two frequencies against
each other from the HEALPix maps at $N_{\rm side}=64$ covering the TMC
region. One can see a very tight linear relation from the fit shown in red over the data.
We repeated a similar analysis for different sub-regions of the large
region displayed in Figure~\ref{fig:taurus-auriga-regions}.
The values of the power law index are given in Table~\ref{tab:beta} 
for different cases such that all the emission is assumed to be
synchrotron and free--free emission, and for the case in which the
free--free component estimated from the H$\alpha$ map for a dust
mixing factor,$f=0.8$,  is removed from the maps. 

As expected from the very low emission of the free--free compared to the
synchrotron emission discussed above, we find similar values for  
the power law index $\beta_{\rm synch}$ to within the uncertainties, with
a tendency towards a flatter index when the free--free is not subtracted. 
Once the map was split into two regions, however, we found that the
synchrotron power spectral index is higher in the half-area map closer to the Galactic
plane than in the half-area map further from the Galactic plane. 
Within the uncertainties, though, the values of $\beta_{\rm synch}$ obtained from the two
half-maps are consistent with each other, as well as with the value of $\beta_{\rm synch}$
obtained towards the whole area and towards the TMC region. 
This means that, to within the uncertainties, a single power law
with a fixed $\beta_{\rm synch} \sim -2.80$ over the whole area 
should be sufficient to characterize the synchrotron contribution
towards the TMC. 


\begin{table}
\begin{center}
\begin{tabular}{llc}
\hline\hline
\noalign{\smallskip}
Region & Free-free$^{(b)}$ &  $\beta_{\rm synch}$ $\pm$ $\sigma(\beta_{\rm synch})$\\
\noalign{\smallskip}
\hline
\noalign{\smallskip}
  Full region & included    & -2.77 $\pm$     0.22   \\
  Full region & removed  ($f=0.8$)  &  -2.81   $\pm$  0.26   \\
  $-11^{\circ} <$b$< -2^{\circ} $ & removed ($f=0.8$)   &  -2.71  $\pm$  0.26   \\
  $-24^{\circ} <$b$< -11^{\circ} $ & removed ($f=0.8$)  &   -2.80  $\pm$  0.26  \\
TMC region$^{(a)}$ & included   &  -2.80  $\pm$ 0.26   \\
TMC region$^{(a)}$ & removed ($f=0.8$)  &  -2.80  $\pm$  0.26  \\
\noalign{\smallskip}
\hline\hline
\end{tabular}
\end{center}
\normalsize
\medskip
\caption{\small Power law index from the $T$--$T$ plots between the maps at 408
  MHz and 1.420$\,$GHz for several regions considered for cases of free--free
  emission estimates included in or removed from the maps. The
  definitions of the regions refer to Figure~\ref{fig:taurus-auriga-regions}
Notes:$^{(a)}$ TMC region as delineated in white in
Figure~\ref{fig:taurus-auriga-regions}. $^{(b)}$ 80$\%$ of the HII
gas is assumed to be mixed with thermal dust.}
\label{tab:beta}
\end{table}

\subsection{Total intensity emission} \label{sec:sed}

A method that can be used to do a component separation
analysis of the various components in emission contributing to the
total emission of the TMC is to perform an SED analysis.
This method consists of calculating the total emission of a given
source at each frequency. Once the SED has been
calculated one can use modelling to assess the fraction of the
total intensity associated with the different components (synchrotron, 
free--free, thermal dust, and AME) at all frequencies. SED modelling analysis has
been widely used in the literature \citep[e.g.][]{watson2005, per20,pir15,genova15,genova17}.

\subsubsection{Intensity flux densities of the TMC and L1527} \label{sec:flux_densities}


\begin{figure*}
\begin{center}
\vspace*{2mm}
\centering
\includegraphics[width=120mm,angle=-90]{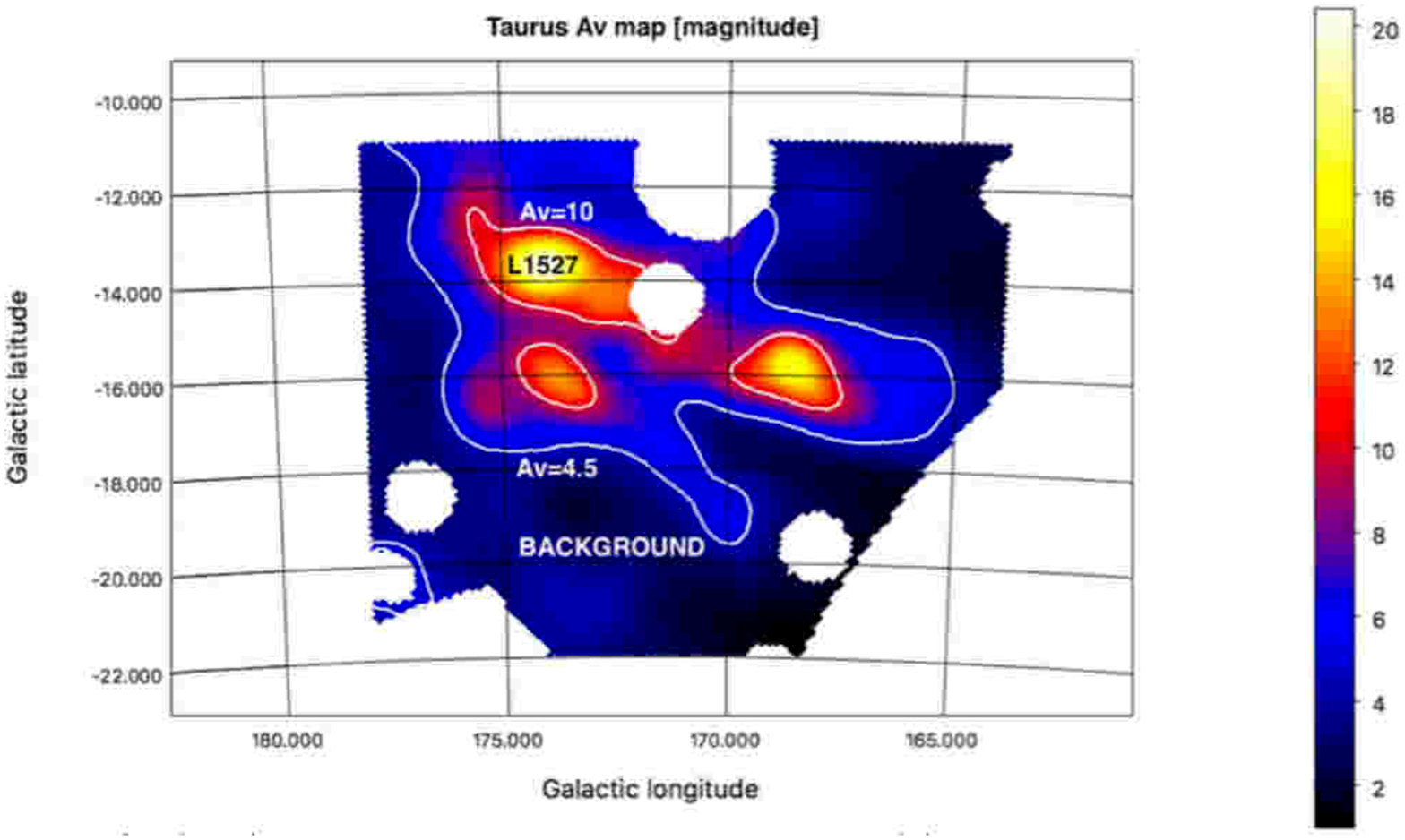}
\caption{\small TMC region defined by isolating all pixels such that $A_{V} > 4.5\,$ magnitudes (lowest contour).
The region used to define the background is the region such that
$A_{V} < 4.5\,$magnitudes. The L1527 region is defined by isolating all
pixels such that $A_{V} > 10\,$magnitudes as shown by the contour
plot (highest contour) centred on position ($l$,$b$)=(174$^\circ$,-13.5$^\circ$).}
\label{fig:ccs_bg_cloud}
\end{center}
\end{figure*}

In order to estimate the fraction of the total intensity emission
associated with the TMC that would be AME in nature we use aperture
photometry. For this purpose the units are first converted from 
CMB thermodynamic units (K$_{\rm CMB}$) to RJ units
(K$_{\rm RJ}$) at the central frequency, then the maps are converted to units of 
Jy pixel$^{-1}$ using $S = 2kT_{\rm RJ}\Omega\nu^{2}/c^{2}$, where $\Omega$ is the HEALPix pixel solid
angle. The pixels are then summed in the aperture covering the region
of interest to obtain an integrated flux density. An estimate of the background is
subtracted using a median estimator of pixels lying on the 
region defined as the background region. 
The regions we integrate the signal over are shown in
Figure~\ref{fig:ccs_bg_cloud}. The TMC is the area
inside the contour line defined by isolating all pixels such 
that $A_{V} > 4.5$ magnitudes by using the $A_{V}$ 
template from the \textit{Planck} Legacy Archive (PLA). 
The L1527 dark cloud nebula is centred at 
position $(l,b)=(174.0^{\circ},-13.7^{\circ})$ in the TMC. 
The bulk of thermal dust emission seen at this location in the 353
GHz map over an area of about 3 square degrees is also shown in the PLA
A$_{V}$ template map displayed in Figure~\ref{fig:ccs_bg_cloud}. 
This region was isolated by taking all
the pixels such that $A_{V} > 10$ magnitudes around the central
position of L1527 in the PLA template.
The median values are calculated from the background region 
(see Figure~\ref{fig:ccs_bg_cloud}) defined by isolating all pixels such 
that $A_{V} < 4.5$ magnitudes. Details on some tests 
conducted to assess whether the choice of the background region is
relevant for our study are given in Appendix~\ref{sec:assess_bg}.

\begin{figure}
\begin{center}
\vspace*{2mm}
\centering
\includegraphics[width=70mm, angle =0]{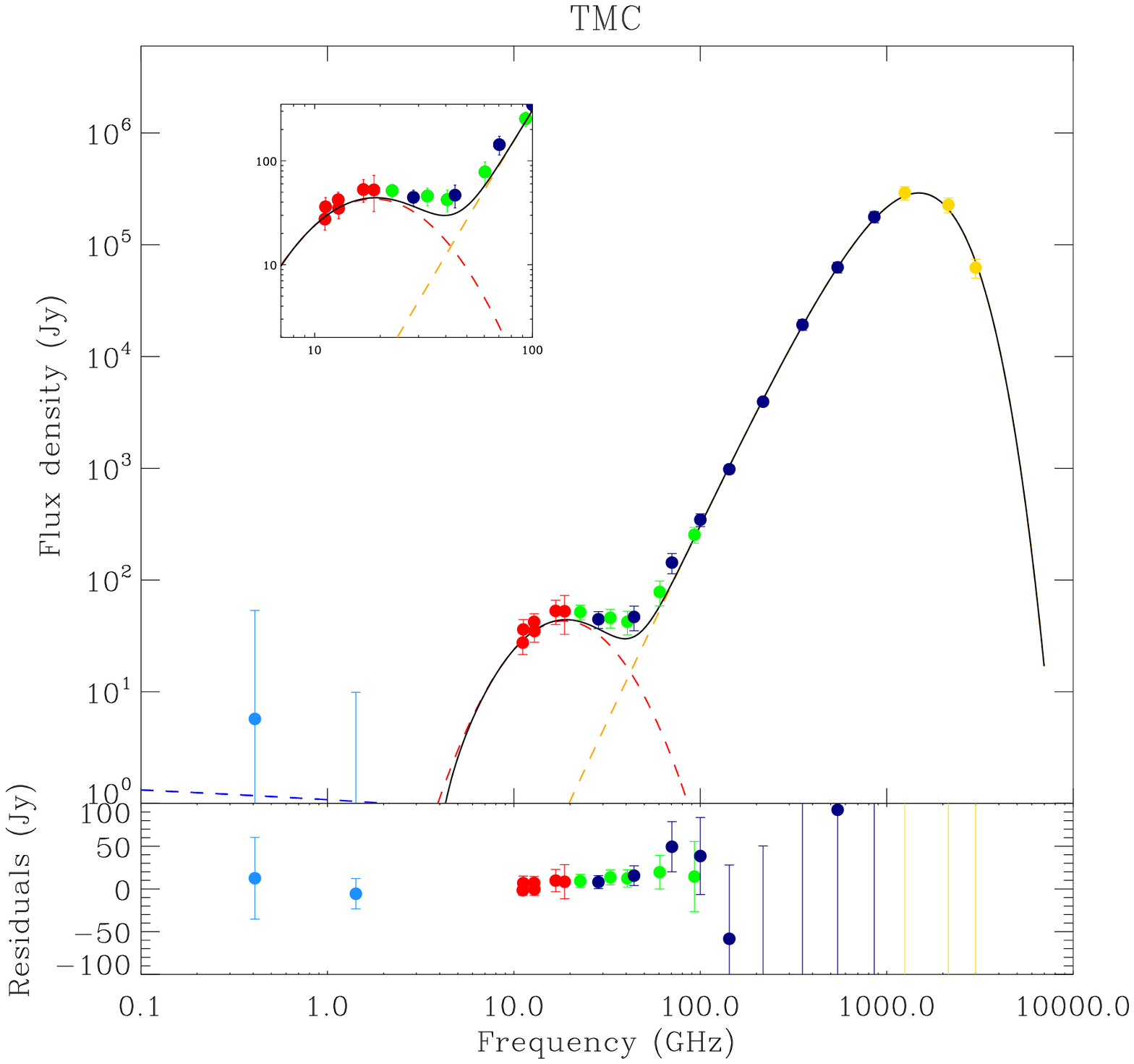}
\caption{\small SED of the TMC region as defined in
Figure~\ref{fig:ccs_bg_cloud}. The QUIJOTE intensity flux densities
are shown with red filled circles, and the \textit{WMAP}, \textit{Planck}, and
DIRBE intensity flux densities are shown with green, blue, and
yellow filled circles respectively. The low frequency points are 
shown in pale blue. The result of the multi-component fit is 
illustrated by the continuous black curve. The fit to the AME component
is shown with the dashed red line. The fit to the free--free component is shown with
the dashed blue line.
}
\label{fig:sed-int}
\end{center}
\end{figure}

\begin{figure}
\begin{center}
\vspace*{2mm}
\centering
\includegraphics[width=70mm, angle =0]{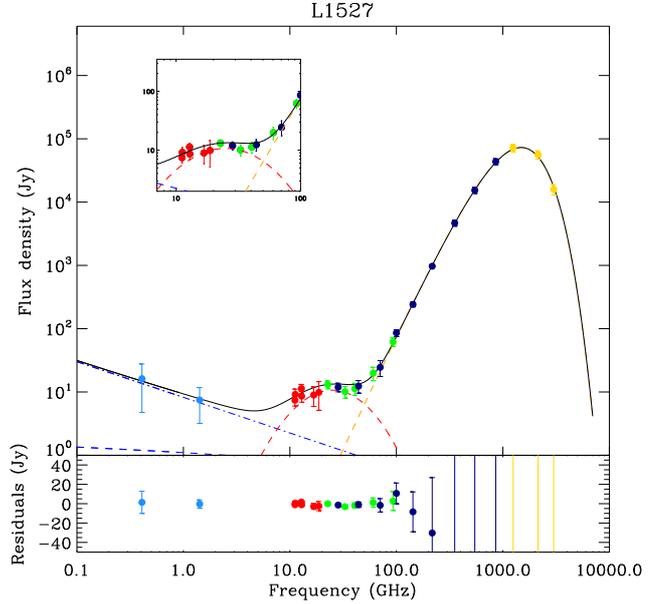}
\caption{\small SED of the L1527 dark cloud as defined in
  Figure~\ref{fig:ccs_bg_cloud}. The filled circles colour code is the
  same as in Figure~\ref{fig:sed-int}. The fit to the AME component
is shown with the dashed red line. The fit to the free--free component is shown with
the dashed blue line, and the fit to the synchrotron component is
shown with the point-dashed blue line.
}
\label{fig:sed-int-ldn1527}
\end{center}
\end{figure}

For the TMC the SEDs obtained with the masks discussed above can be seen in
Figure~\ref{fig:sed-int}, where the QUIJOTE points are shown in red, the
WMAP points in green, the \textit{Planck} points in blue, and
the DIRBE points in yellow. The low frequency points are shown in pale
blue. The intensity flux densities are listed 
in Table \ref{tab:sed-results} along with the intensity flux
densities obtained towards the L1527 dark nebula, and are plotted as a
function of frequency in Figure~\ref{fig:sed-int-ldn1527}.
At first sight one can see that the low frequency fluxes are very low in
the SED of the TMC, whereas for L1527 the low frequency points
suggest clear positive component detections. In both SEDs a 
peak in the frequency range 10--60$\,$GHz that should
be associated with an AME component is clearly seen.


\begin{table*}
\begin{center}
\begin{tabular}{ccccccc}
\hline\hline
\noalign{\smallskip}
& TMC & TMC & L1527&L1527 &&\\
\noalign{\smallskip}
\hline
Frequency & Fluxes& Background & Fluxes &   Background  &Calibration error& Survey\\ 
\noalign{\smallskip}
[GHz]        &  [Jy]     & [Jy]               &[Jy]        &   [Jy]  & [$\%$]  & \\ 
\noalign{\smallskip}
\hline
\noalign{\smallskip}
   0.4    &       5.7$\pm$   47.8    &    2277.4   &      16.2$\pm$   11.5    &     244.6&10&JB/Eff/Parkes\\
   1.4    &      -7.9$\pm$   17.7    &    5404.0  &       7.5$\pm$    4.3    &     580.5 &10& Stockert/Villa-Elisa\\
  11.2$^{(a)}$    &      27.5$\pm$    6.0    &     -27.8   &       7.4$\pm$    1.4    &      -3.0 &5& QUIJOTE-MFI\\
  11.2$^{(b)}$    &      36.1$\pm$    8.2    &     -10.7    &       9.2$\pm$    1.9    &      -1.2 &5&QUIJOTE-MFI\\
  12.8$^{(a)}$    &      42.2$\pm$    7.7    &     -17.7  &      11.3$\pm$    1.9    &      -1.9  &5&QUIJOTE-MFI\\
  12.9$^{(b)}$    &      34.9$\pm$    7.2    &     -30.1  &       8.6$\pm$    1.7    &      -3.2 &5&QUIJOTE-MFI\\
  16.7$^{(c)}$    &      52.9$\pm$   13.0    &      -0.4  &       8.9$\pm$    3.1    &       0.0&5&QUIJOTE-MFI\\
  18.7$^{(c)}$    &      52.5$\pm$   19.9    &      -10.8  &       9.9$\pm$    4.7    &      -1.2 &5&QUIJOTE-MFI\\
  22.7    &      51.6$\pm$    7.8    &     129.8   &      13.1$\pm$    1.9    &      13.9&3.2&WMAP\\
  28.4    &      44.5$\pm$    7.7    &     108.4  &      11.9$\pm$    1.9    &      11.6 &3.4& $Planck$\\
  32.9    &      45.9$\pm$    8.8    &     100.2  &      10.1$\pm$    2.1    &      10.8  &3.2& WMAP\\
  40.6    &      42.3$\pm$   10.1    &      90.3  &      11.3$\pm$    2.4    &       9.7 &3.2& WMAP\\
  44.1    &      46.7$\pm$   11.6    &      97.1  &      12.4$\pm$    2.8    &      10.4 &3.4&$Planck$\\
  60.5    &      78.2$\pm$   19.6    &      93.3  &      19.8$\pm$    4.7    &      10.0 &3.2&WMAP\\
  70.4    &     143.1$\pm$   29.4    &     122.0 &      24.5$\pm$    6.9    &      13.1  &3.4& $Planck$\\
  93.0    &     254.3$\pm$   41.0    &     248.8 &      62.1$\pm$    9.8    &      26.7  &3.2& WMAP\\
 100.0    &     346.2$\pm$   45.1    &     352.7 &      85.9$\pm$   10.8    &      37.9  &3.4& $Planck$\\
 143.0    &     980.5$\pm$   86.3    &    1036.6  &     241.7$\pm$   20.6    &     111.3 &3.4& $Planck$\\
 217.0    &    (3.94$\pm$  0.24) $\times 10^3$   &    4.57 $\times 10^3$  &     970.3$\pm$   57.3    &     490.9 &3.4&$Planck$\\
 353.0    &   (1.92$\pm$ 0.21) $\times 10^4$   &   2.12 $\times 10^4$  &    (4.67$\pm$  0.51)$\times 10^3$    &    2.27$\times 10^3$ &10.0&$Planck$\\
 545.0    &   (6.28$\pm$ 0.70) $\times 10^4$   &   6.88 $\times 10^4$   &   (1.54$\pm$ 0.17)  $\times 10^3$  &    7.39$\times 10^3$ &10.0&$Planck$\\
 857.0    &  (1.78$\pm$0.20) $\times 10^5$    &  1.92$\times 10^5$  &   (4.35$\pm$ 0.49) $\times 10^4$   &   2.07$\times 10^4$ &10.0& $Planck$\\
1249.1    &  (2.92$\pm$0.40) $\times 10^5$    &  3.45$\times 10^5$   &   (7.16$\pm$ 0.96) $\times 10^4$   &   3.71$\times 10^4$&11.9&COBE-DIRBE\\
2141.4    &  (2.27$\pm$0.35) $\times 10^5$    &  3.77$\times 10^5$  &   (5.61$\pm$ 0.84)  $\times 10^4$  &   4.05$\times 10^4$ &11.9& COBE-DIRBE\\
2997.9    &   (6.24$\pm$0.12) $\times 10^5$    &  1.64$\times 10^5$ &   (1.59$\pm$ 0.29) $\times 10^4$   &   1.77$\times 10^4$  &11.9& COBE-DIRBE\\
\noalign{\smallskip}
\hline\hline
\end{tabular}
\end{center}
\normalsize
\medskip
\caption{Intensity flux densities in each frequency of the TMC and
  L1527 regions for the areas in Figure
  \ref{fig:ccs_bg_cloud} respectively. The values
  are shown with filled circles on the SED plot in
  Figure~\ref{fig:sed-int} for the TMC and in
  Figure~\ref{fig:sed-int-ldn1527} for L1527. 
Notes: ${(a)}$: horn 1, ${(b)}$: horn 3,  ${(c)}$: horn 2. 
 }
\label{tab:sed-results}
\end{table*}

\subsubsection{Intensity SED modelling} \label{sec:sed_modelling}

As discussed previously, the CMB component was removed from
several maps by using the SMICA-CMB maps. This was done on all the maps
between 10$\,$GHz and 353$\,$GHz. The contamination level of the CMB is then 
fully negligible to within the uncertainties at frequencies higher
than 70$\,$GHz but had to be taken into account at lower frequencies.
Therefore, we consider a total of four components (synchrotron, free--free, AME,
and thermal dust) to model the intensity SED from a multi-component
fit. Following our conclusions concerning the assessment of the background
region (see Appendix~\ref{sec:assess_bg}), 
we assume that the AME associated with the molecular cloud 
material can be modelled with a single component of spinning dust 
\citep{draine98}. For this we use the \citet{bonaldi07} model, a
phenomenological model consisting of a parabola in logarithmic space
(log($S_{\nu}$)--log($\nu$)). The model is described by three free
parameters: the slope at a frequency of 60 GHz, $m_{60}$, which is a
function of the width of the parabola, the central frequency,
$\nu_{\rm AME, peak}$, and the amplitude of the parabola, $S_{\rm
AME, peak}$. The other components are modelled as follows: 
the synchrotron component is fitted by a single power law of index
assumed to be constant, $\alpha_{\rm synch, int}$, and by the intensity
of the synchrotron at a frequency of 1$\,$GHz, $S_{\rm synch, 1GHz}$.
The free--free spectrum
shape is fixed  using 
equations (2), (3), (4), and (5) of \citet{per20}. 
Our best guess for the electron temperature is the median
value of the Commander template with, $T_{\rm e} \simeq
7000\,$K. The only remaining free parameter associated with the
free--free component is the free--free amplitude, which can
be parameterized by the effective Emission Measure ($EM$).
The thermal dust emission component is modelled by a single-component
modified blackbody relation of the form, $\tau_{250} (\nu/1200 \rm
GHz)^{\beta_{\rm dust}}\it B_{\nu}(T_{\rm dust})$, where $\tau_{250}$ is the
averaged dust optical depth at 250$\,\mu$m, $\beta_{\rm dust}$ is the averaged thermal
dust emissivity and, $T_{\rm dust}$ is the thermal dust averaged temperature.
The fit procedure includes priors on some of the parameters and
consists of a minimization process using non-linear least-squares
fitting in Interactive Data Language (IDL) with MPFIT \citep[][]{markwardt2009}.
 
For the TMC the output parameters from the four multi-component fits shown in 
Figure~\ref{fig:sed-int} are listed in
Table~\ref{tab:sed-parameters}, top.
They show quite a low free--free component, which is dominated by the
AME component at the frequencies of the MFI. The averaged thermal dust temperature
and spectral dust emissivity associated with the Taurus cloud region are found
to be in the lower range of parameter values discussed by \citet{per25},
which indicates that the background region should be removing the
warmer thermal dust grain components and isolating the thermal dust
components at higher densities inside the TMC. On the other hand the
synchrotron total flux is found to be very low at 1 GHz and consistent
with no detection to within the uncertainties, still a low emission component
could remain in the maps. This happens because the
emission of the synchrotron in the region defined to estimate 
the background contribution almost dominates the total emission 
of the synchrotron component towards the TMC (see Figure~\ref{fig:map_ff_synch408}). 

For L1527 the estimates 
of the four components from the multi-component fit in intensity are
given in Table~\ref{tab:sed-parameters} (top). The thermal 
dust temperature and thermal dust emissivity in L1527 and 
the TMC are similar to within the uncertainties, but the optical depth
at 250 microns is found, on average, to be lower in L1527 than 
towards the TMC. For the TMC the area includes the regions
such that A$_{\rm V}>$4.5 magnitudes and therefore also all the regions 
such that A$_{\rm V}>$10 magnitudes while the area of L1527 covers a smaller 
patch of the sky such that A$_{\rm V}>$10 magnitudes.  In addition, at the fit
level, there is room left for degeneracy between the averaged value of 
$\beta_{\rm dust}$ and $\tau_{250}$, and we expect this effect to be higher when 
integrating and fitting a grey body over a larger area.  All these
elements put together we think this is why the averaged value of 
$\tau_{250}$ over the whole TMC is about a factor 4 higher than toward 
L1527 alone. The averaged values of the thermal dust parameters 
from the fits displayed in Table~\ref{tab:sed-parameters} should be considered as indicative 
values only and one interested in details toward some specific areas 
of the TMC should for example refer to the \textit{Planck} templates of thermal dust.
The free--free components 
show an effective $EM$ of the same order as that towards the
TMC but is better constrained, which now also allows the fitting procedure to
constrain the synchrotron component parameters with clear detection. 
This analysis also reconciles the variations of the synchrotron as
a function of frequency since the power law indices in temperature
$\beta_{\rm synch} = -2.80 \pm 0.26$, and in flux  
$\alpha_{\rm synch, int}= -0.56 \pm 0.32$, are such that $\beta_{\rm
  synch}-2 = \alpha_{\rm synch, int}$  to within the uncertainties.
The total flux of the AME component is found to be about 3 to 4 times
smaller than in the TMC, but a clear detection is still obtained thanks
to the data provided by the QUIJOTE maps.


\begin{table}
\begin{center}
\begin{tabular}{lcc}
\hline\hline
\noalign{\smallskip}
& TMC & L1527\\
\noalign{\smallskip}
\hline
SED in & Output fit & Output fit\\ 
intensity&parameters&parameters\\
\noalign{\smallskip}
\hline
\noalign{\smallskip}
$EM$ [cm$^{-6}$ pc]                              & 9.9 $\pm$ 31.7  &  10.1 $\pm$ 6.0\\ 
$S_{\rm FF, 1 \rm GHz}$  [Jy]                 &  1.1 $\pm$ 3.5   &  1.1 $\pm$ 0.7 \\ 
$S_{\rm synch, 1 \rm GHz}$  [Jy]   &  -4.2 $\pm$ 12.0  &   8.2 $\pm$ 2.6 \\ 
$\alpha_{\rm synch, int}$             &  -0.71 $\pm$ 3.42  &   -0.56 $\pm$ 0.32  \\ 
$m_{60}$                                           &   3.6 $\pm$ 5.0  &  1.7 $\pm$ 2.7 \\
$\nu_{\rm AME, peak}$ [GHz]        &  18.7 $\pm$ 7.1   &    25.4 $\pm$ 18.5     \\ 
$S_{\rm AME, peak}$ [Jy]                & 43.0 $\pm$ 7.9    &   10.7 $\pm$ 2.7    \\ 
$T_{\rm dust}$ [K]                               & 15.6 $\pm$ 0.4    &   15.5 $\pm$ 0.4  \\ 
$\beta_{\rm dust}$ [K]                         &  1.61 $\pm$ 0.05   &   1.64 $\pm$ 0.05 \\ 
 $\tau_{250} \times 10^{-3}$        & 4.2 $\pm$ 0.5     &    1.1 $\pm$ 0.1  \\ 
$\chi^2_{\rm red}$                             &        1.27      &      0.63      \\
\hline\hline
\noalign{\smallskip} 
SED in  & Output fit & Output fit\\ 
 polarization & parameters & parameters \\
\noalign{\smallskip}
\hline
$S_{\rm synch, 1 \rm GHz}$  [Jy]   &  43.60 $\pm$ 14.46 & 4.25 $\pm$ 1.47 \\ 
$\alpha_{\rm synch, pol}$           &    -1.13 $\pm$ 0.23 &   -1.02 $\pm$ 0.20 \\ 
$\pi_{\rm dust}$ [$\%$]                               &   6.54 $\pm$ 1.49  &  7.73 $\pm$ 1.91 \\ 
$\chi^2_{\rm red}$                             &        1.36 &   0.72 \\
\hline
$\gamma_{0} (^\circ)$ &   1.28 $\pm$ 0.05 &  1.31 $\pm$ 0.05\\
$RM$ [rad m$^-2$] & 6.56 $\pm$ 1.04 & 5.92 $\pm$ 1.04\\
$\chi^2_{\rm red}$                             &   0.52  &  1.55\\
\noalign{\smallskip}
\hline\hline
\end{tabular}
\end{center}
\normalsize
\medskip
\caption{\small TMC and L1527 SEDs multi-component output fit parameters in
  intensity (top table) and polarization (bottom tables). 
 }
\label{tab:sed-parameters}
\end{table}

\subsection{Limits on the level of linear polarization} \label{sec:separation_comp_anal_pol}


\begin{figure*}
\begin{center}
\vspace*{2mm}
\centering
\includegraphics[width=75mm]{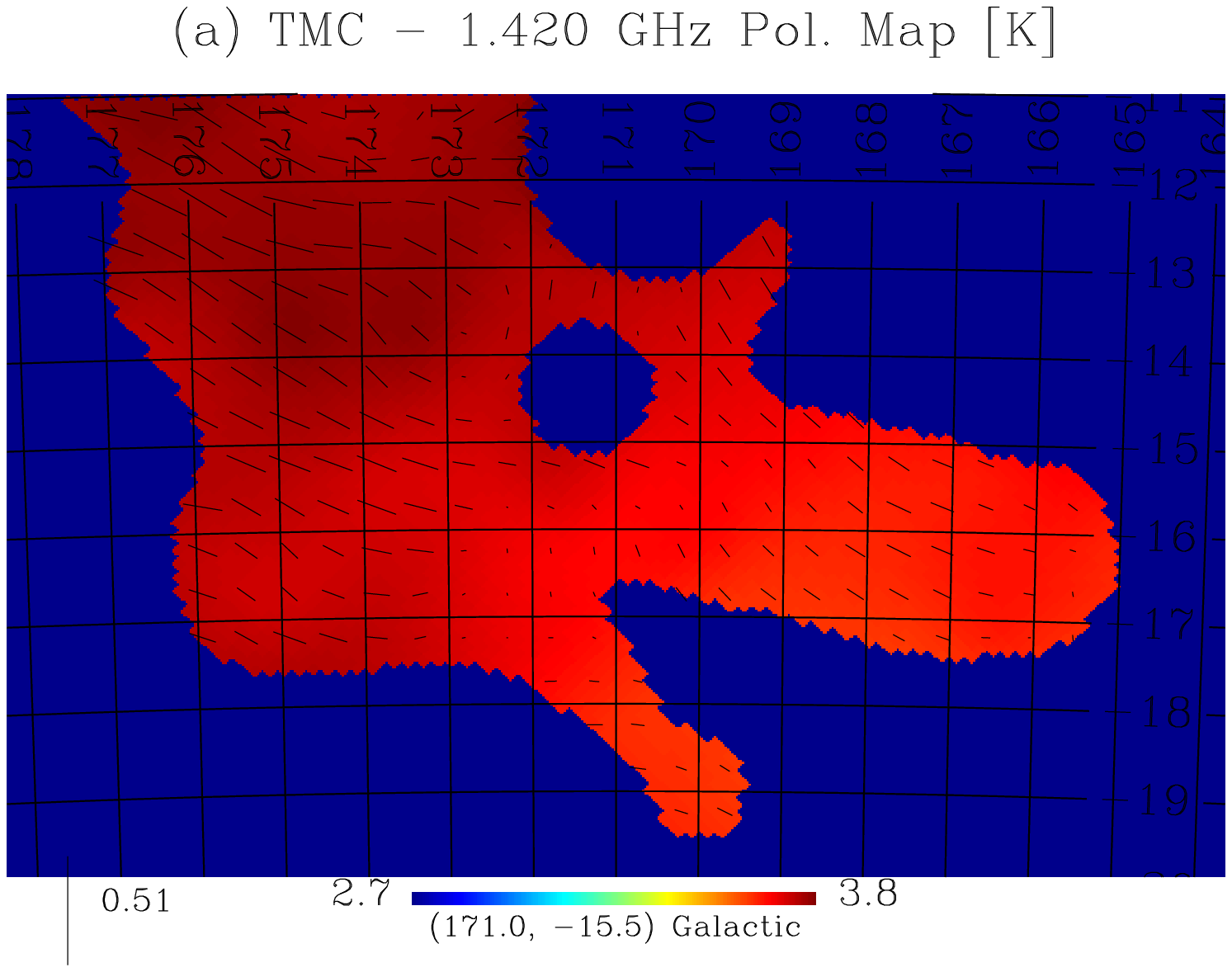}
\includegraphics[width=75mm]{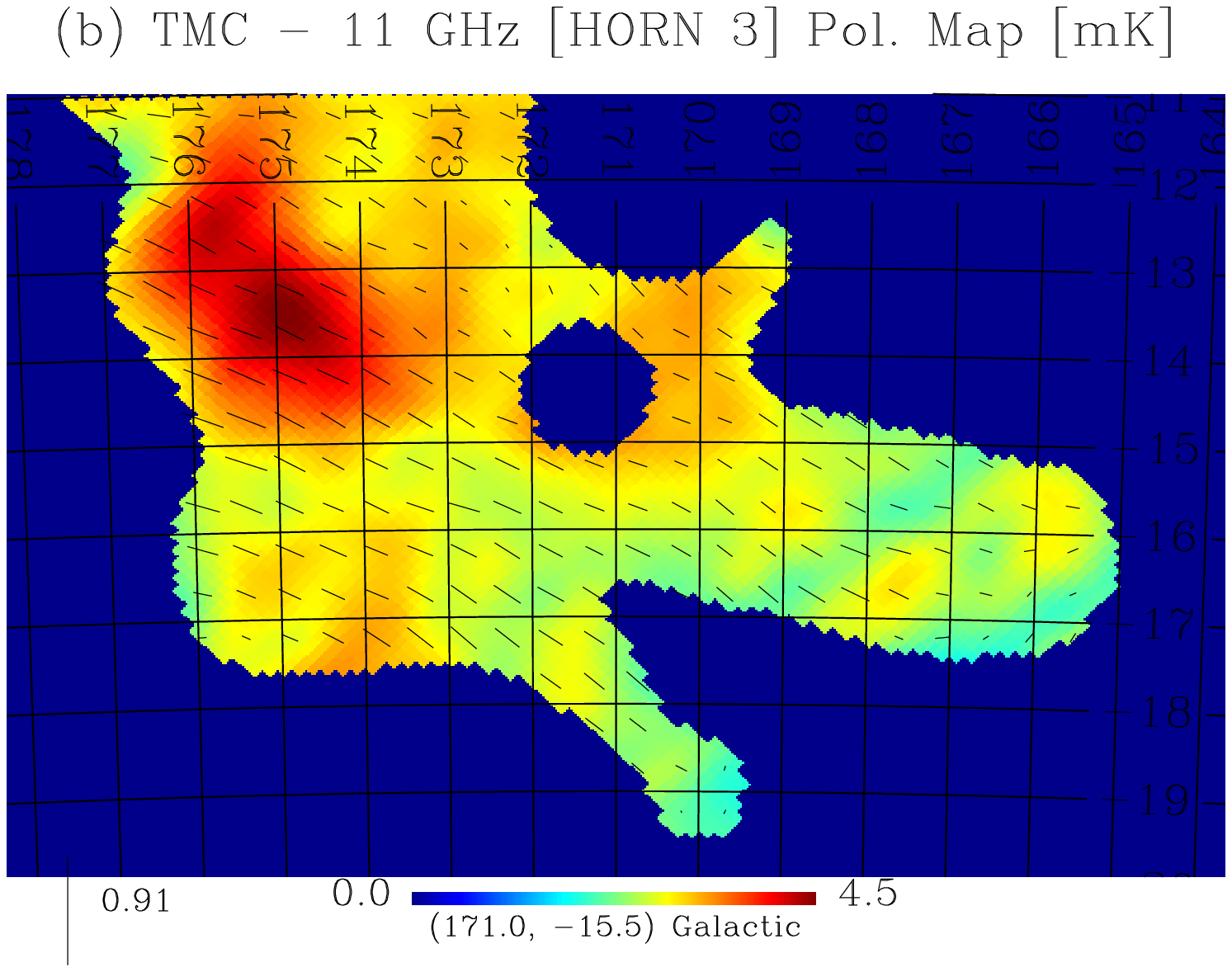}
\includegraphics[width=75mm]{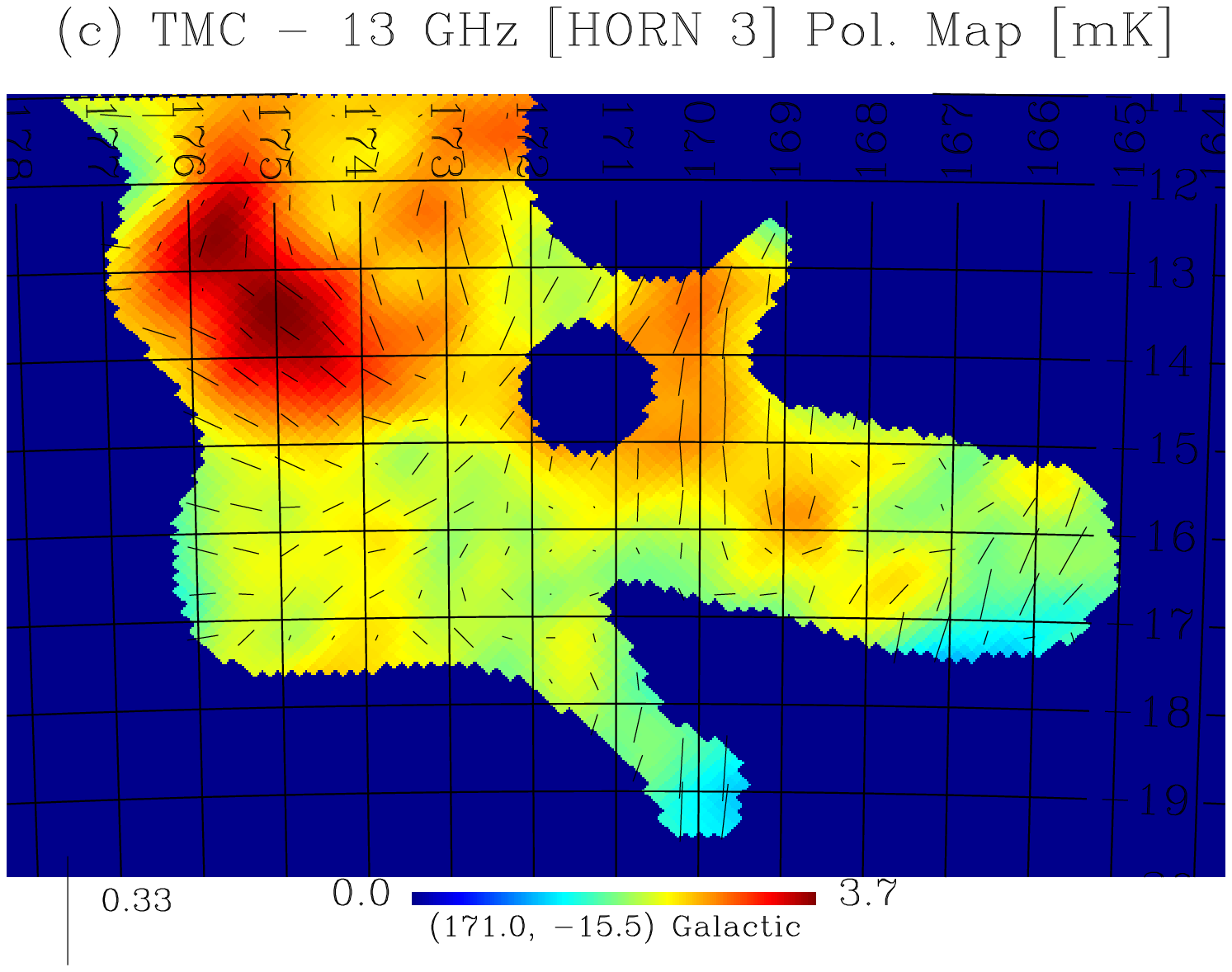}
\includegraphics[width=75mm]{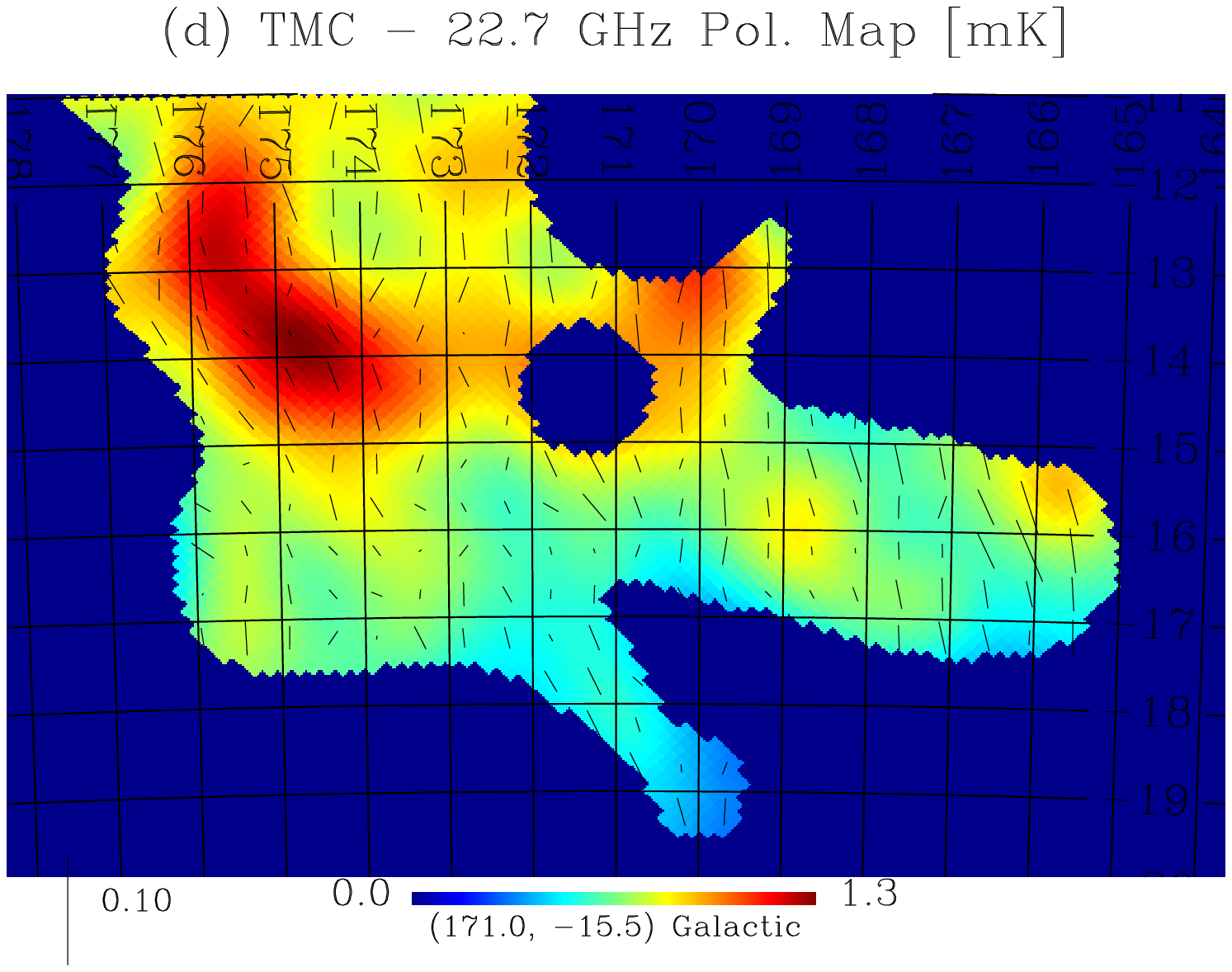}
\includegraphics[width=75mm]{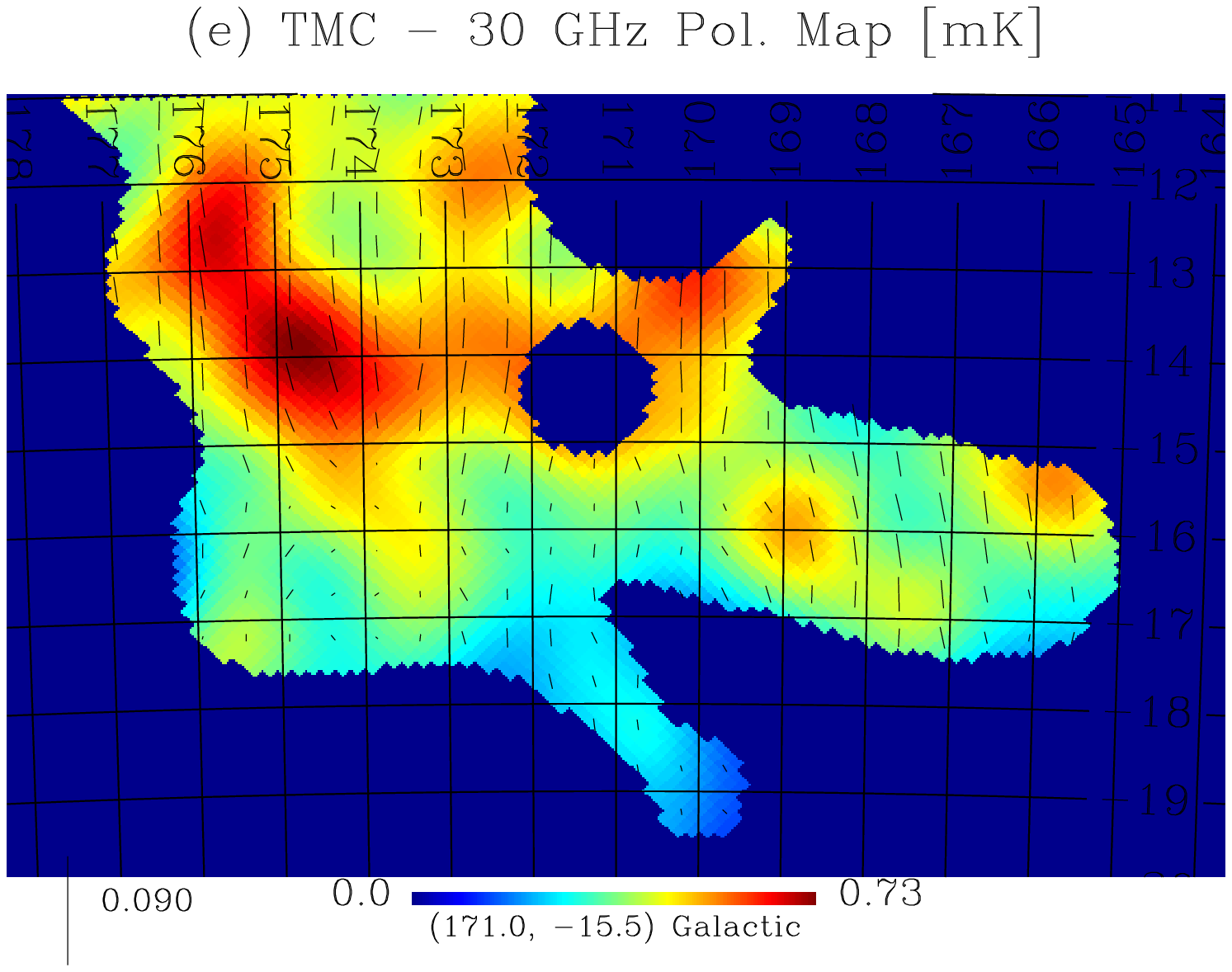}
\includegraphics[width=75mm]{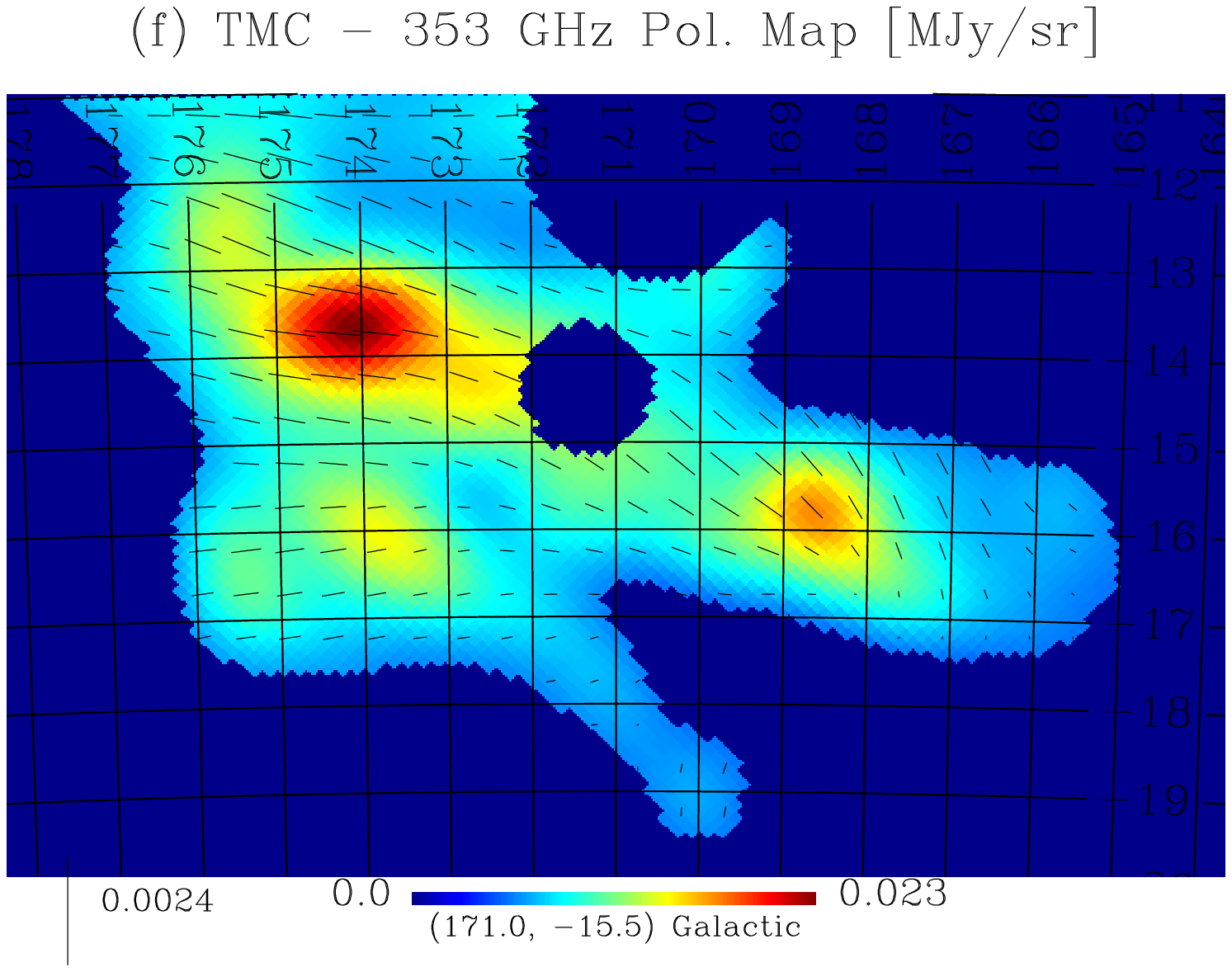}
\caption{\small Polarization maps of the TMC with polarization pseudo-vector
orientations overplotted. The polarization angles are all displayed following the cosmological 
convention \citep[polarization angles counted positively clockwise
from north to west in Galactic coordinates, see][]{gorski05}.
(a): 1.42$\,$GHz polarization map. 
(b): QUIJOTE Horn 3, 11$\,$GHz polarization map. 
(c): QUIJOTE Horn 3, 13$\,$GHz polarization map.
(d): \textit{WMAP} 22.7$\,$GHz polarization map. 
(e): \textit{Planck} 30$\,$GHz polarisation map.
(f): \textit{Planck} thermal dust polarization map at 353$\,$GHz.
}
\label{fig:pol_maps}
\end{center}
\end{figure*}


\begin{figure*}
\begin{center}
\vspace*{2mm}
\centering
\includegraphics[width=45mm, angle =0]{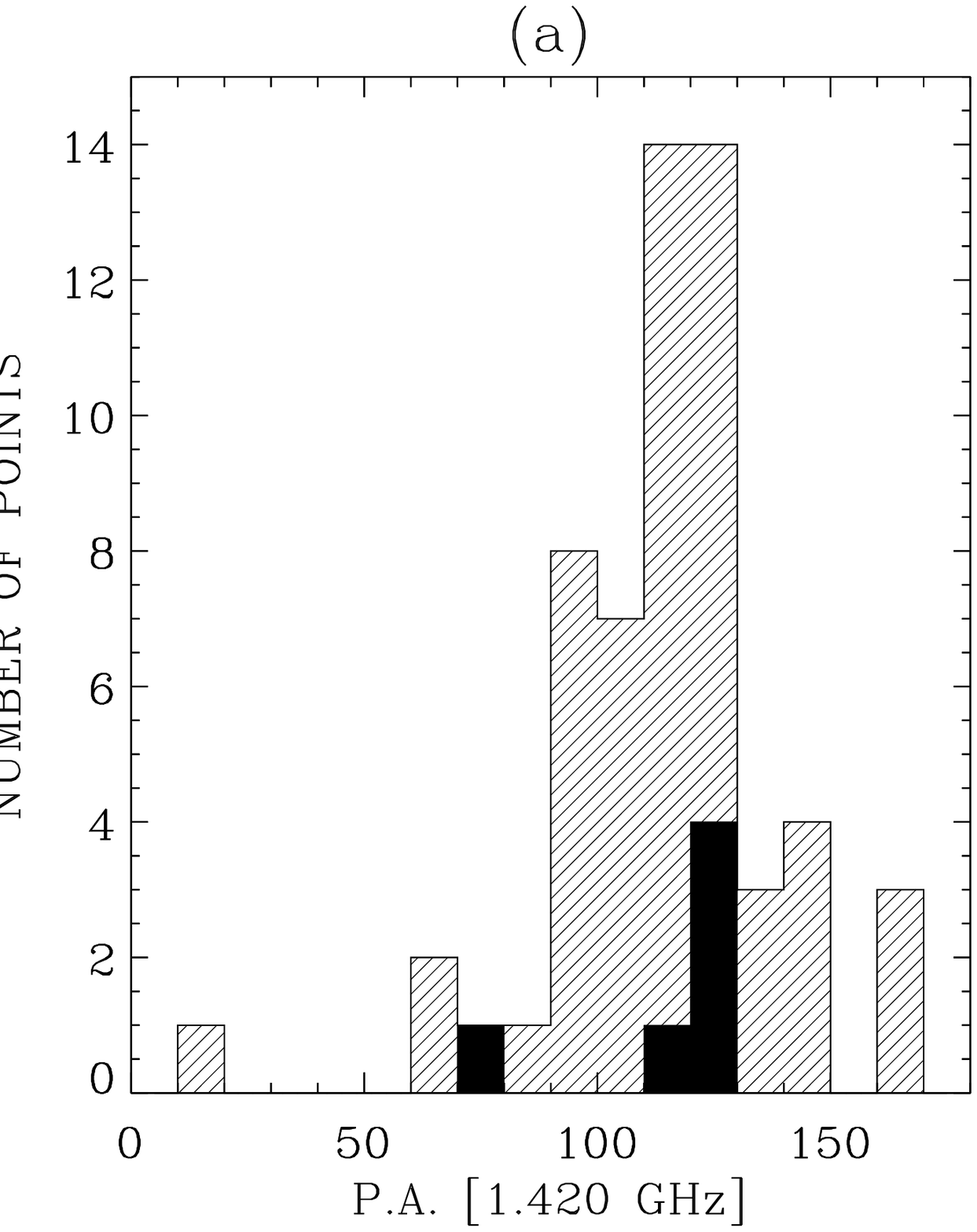}
\includegraphics[width=45mm, angle =0]{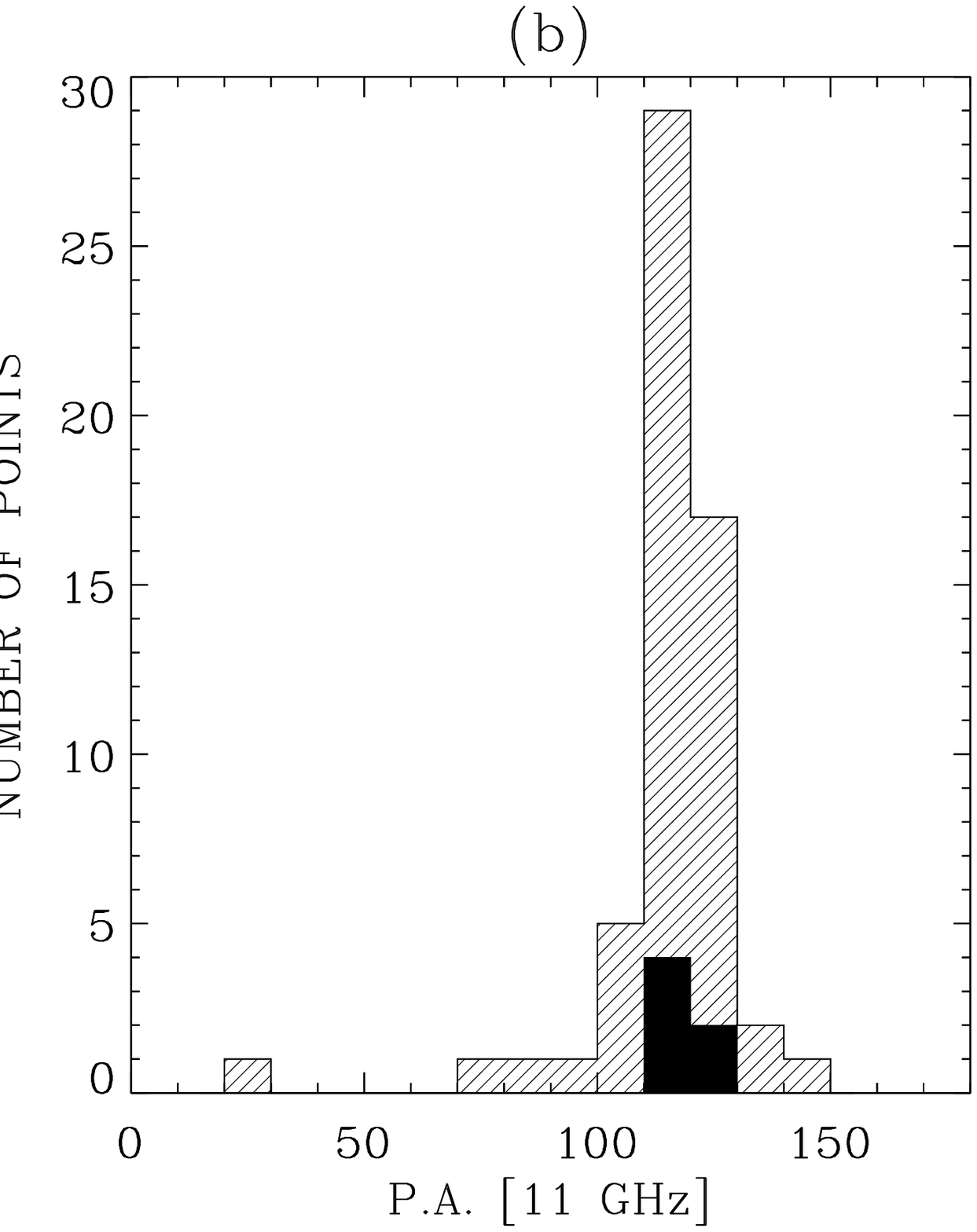}
\includegraphics[width=45mm, angle =0]{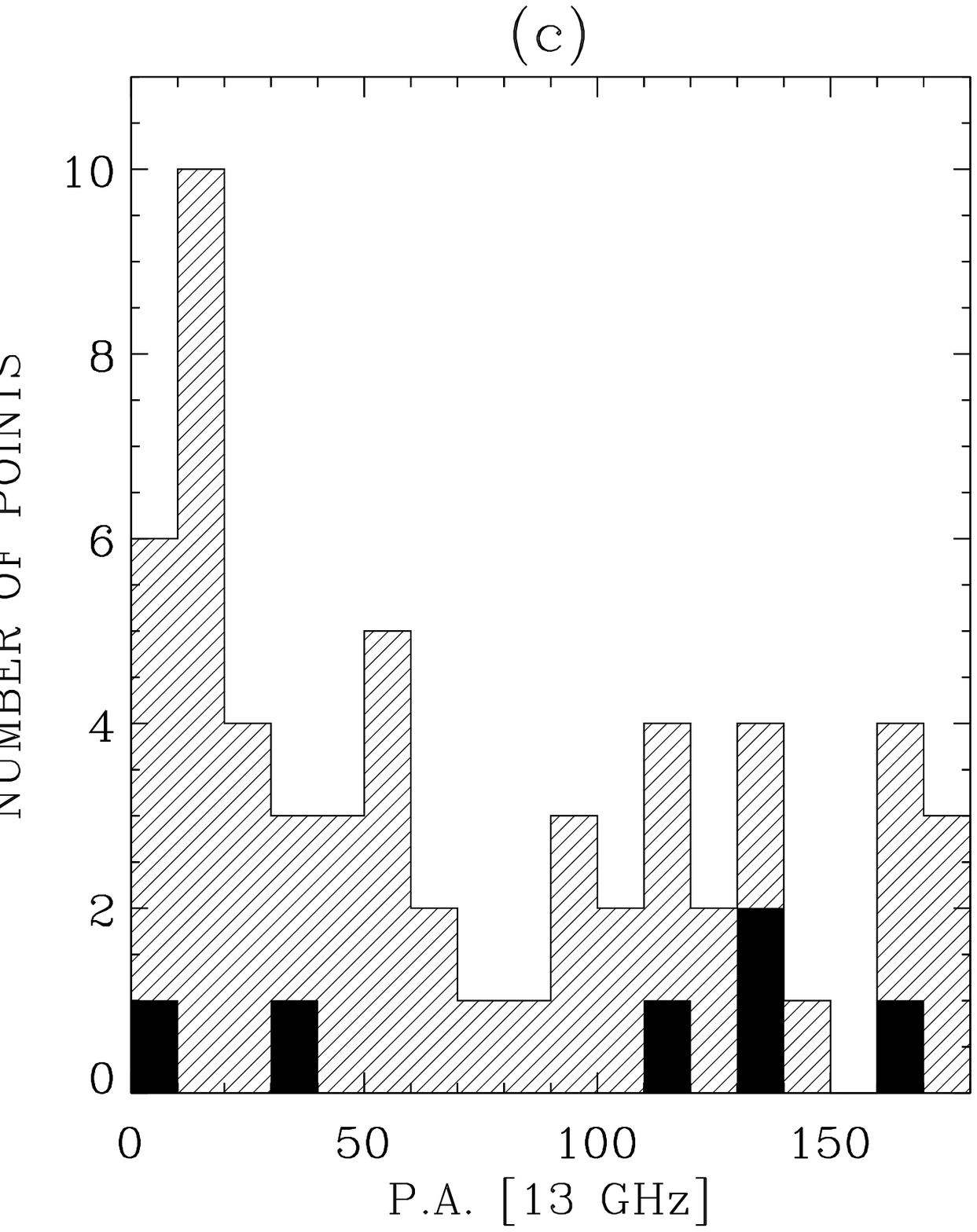}
\includegraphics[width=45mm, angle =0]{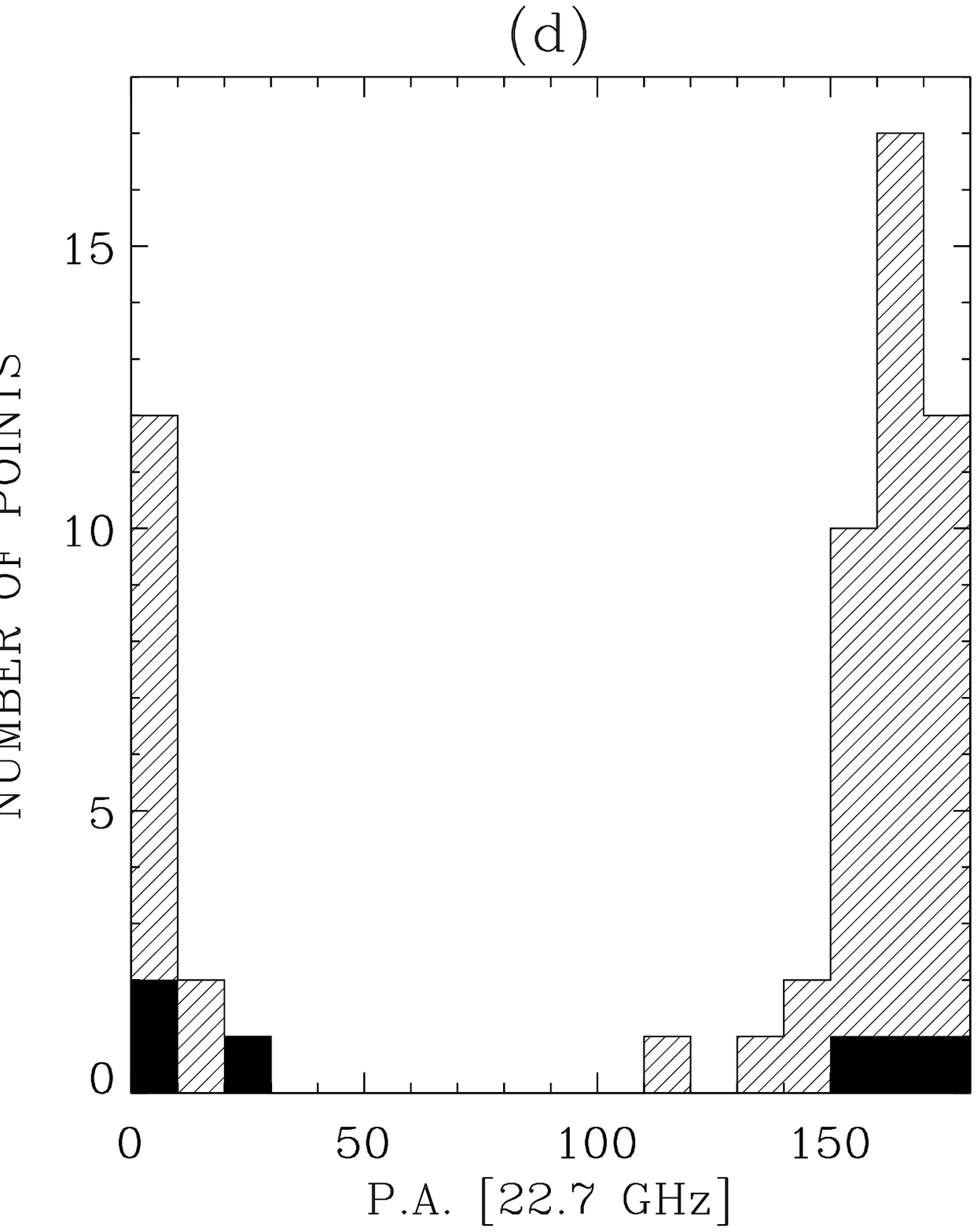}
\includegraphics[width=45mm, angle =0]{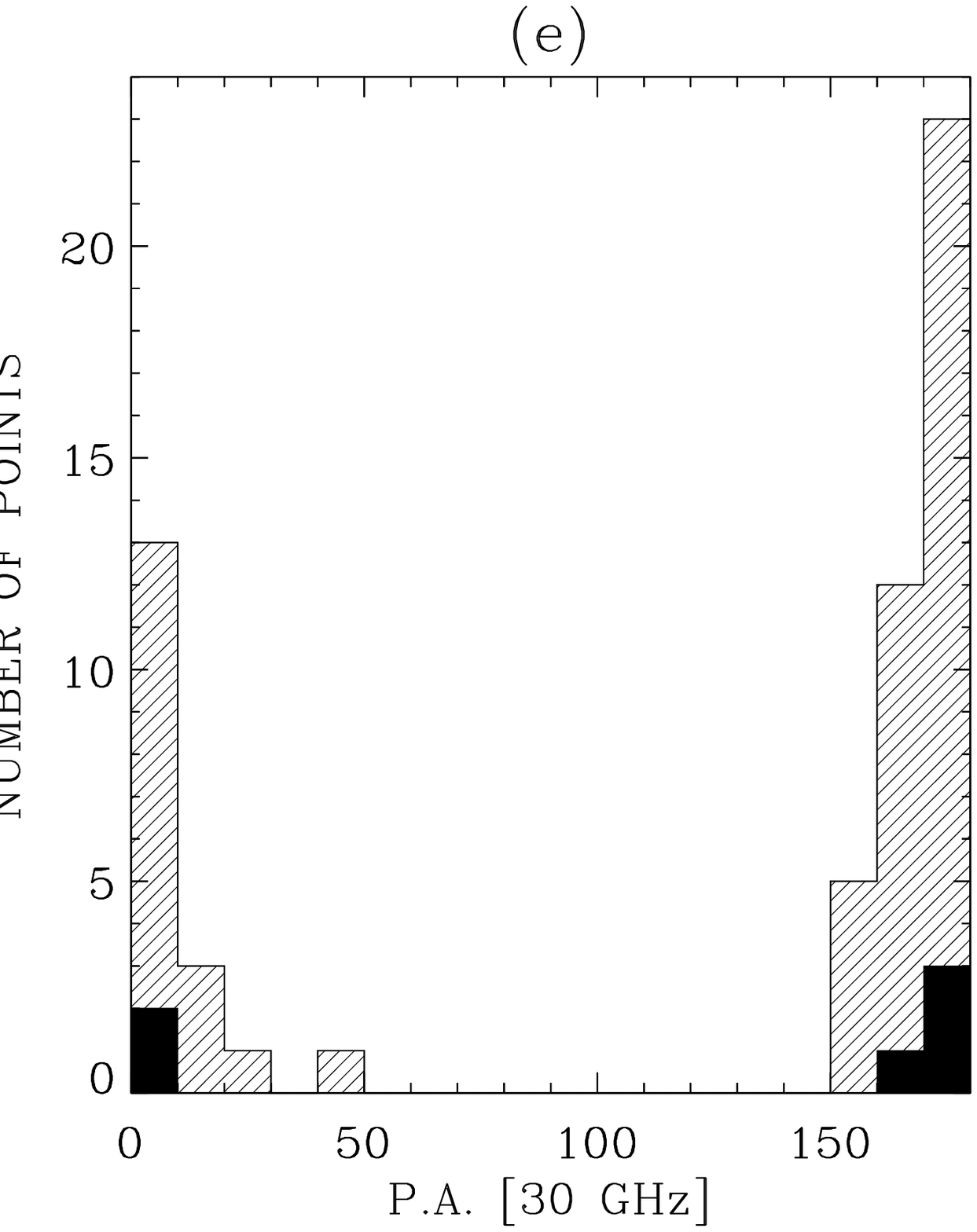}
\includegraphics[width=45mm, angle =0]{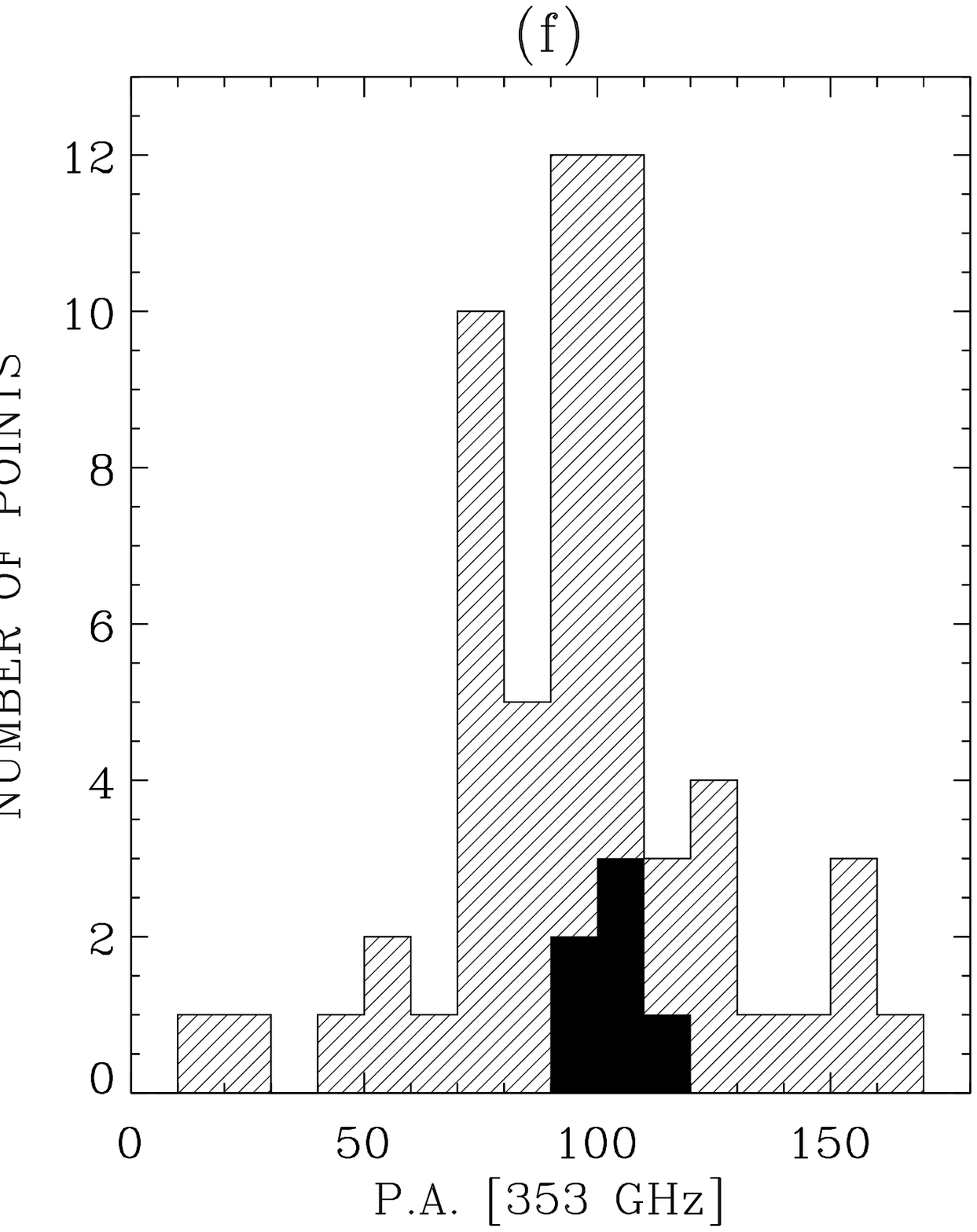}
\caption{\small Histograms of the polarization angles shown in
  Figure\ref{fig:pol_maps}. The distribution of the polarization
  angles measured towards the TMC is shown in grey, and that associated with L1527 is shown in black. 
}
\label{fig:histo_maps}
\end{center}
\end{figure*}

\subsubsection{Polarization maps of the TMC}\label{sec:polarisation}

The structures of the magnetic field pervading the TMC have been
well studied from visible wavelengths to radio frequencies \citep[e.g.][]{arce98, chapman11,pir35_2016, wolleben04}.
To get insight into these structures we first show in Figure~\ref{fig:pol_maps} the
polarization maps of the TMC at 1.420$\,$GHz (top row, left), 11$\,$GHz (top
row, middle), 13$\,$GHz (top row, right), 22.7$\,$GHz (bottom row,
left), 30$\,$GHz (bottom row, middle), and 353 GHz (bottom row, right). 
The polarization angles are all displayed following the cosmological 
convention \citep[polarization angles counted positively clockwise
from north to west in Galactic coordinates, see][]{gorski05}. 
A comparison of the maps at 1.420$\,$GHz, 11 GHz, and 
353$\,$GHz shows very similar and uniform polarization
patterns over a large fraction of the area of the TMC. 
This can be seen in the histogram in panels a, b,
  and f in Figure \ref{fig:histo_maps}.
In these maps each pseudo-vector
orientation should be perpendicular to the averaged magnetic field component
projected on to the plane of the sky, $<$\it{\^{B}}\rm$_{\rm pos}$$>$.
The data in the frequency range
1.420--10 GHz provide information about the magnetic fields producing
polarized synchrotron since the free--free emission is expected to be
unpolarized, whereas the data in the frequency range 60--3000 GHz
probe the polarized emission of thermal dust. The data in the
frequency range 10--60 GHz could provide combined information about
the synchrotron, thermal dust, and AME component if this last component
is polarized. The distribution of the polarization angles looks overall
less uniform in the polarization maps at 13$\,$GHz, 22.7$\,$GHz, and
30$\,$GHz, as can be seen with the broad distributions
  of the histogram in panels b, c, and d of Figure \ref{fig:pol_maps}. 
This could be due to averaging effects over some areas of the maps if the magnetic field
orientations change over these areas, i.e.\ if Faraday depolarization
occurs in the frequency range 10--40$\,$GHz. This is not
likely given the low level of free--free estimated in the TMC
but Faraday depolarization has been studied at frequencies $<$ 2$\,$GHz by
\citet{wolleben04} in regions to the north of the TMC and was interpreted
as produced by a Faraday screen at the boundaries of molecular clouds.
Their analysis was possible because all their maps included
absolute calibration. We do not have access to all their maps since
only the 1.420$\,$GHz map is publicly available. In addition, the baseline
level of the QUIJOTE maps is not absolutely calibrated therefore we cannot directly
repeat their analysis in the frequency range 10--40$\,$GHz. 
A second explanation is that, in this frequency range,  
the AME component is polarized, which could produce
perturbations in the polarization angle patterns. A third possibility
is that some of the regions of the polarization maps are dominated by
noise in the frequency range 10--30$\,$GHz. 
A summary of the information provided by the polarization
  angle distributions is given in Table \ref{tab:histo-pol-results}, where the 
median and standard deviations of the histograms shown 
in Figure \ref{fig:histo_maps} are shown.
In order to obtain further insights into whether the TMC is subject to
Faraday depolarization or its polarization structure perturbed by the
AME component and to decrease the impact of possibly noisy regions   
in the maps, in the following we conduct an 
analysis of the SEDs in polarization towards the TMC and L1527.


\begin{table}
\begin{center}
\begin{tabular}{lcc}
\hline\hline
\noalign{\smallskip}
& TMC & L1527\\
\noalign{\smallskip}
\hline
Frequency & Median(P.A.)$\pm \sigma$(P.A.)& Median(P.A.)$\pm \sigma$(P.A.) \\
$[\rm GHz]$&  [$^{\circ}$]&[$^{\circ}$]\\
\noalign{\smallskip}
\hline
\noalign{\smallskip}
1.420 &       117.4$\pm$       25.1&        123.6$\pm$       18.8\\
11 &       117.6$\pm$       16.1&        119.4$\pm$       2.9\\
13 &       10.5$\pm$       44.8&        160.8$\pm$       38.0\\
22.8 &       169.5$\pm$       15.3&        1.5$\pm$       15.5\\
30 &       173.9$\pm$       31.7&        164.6$\pm$       42.7\\
353 &       97.9$\pm$       29.6&        101.7$\pm$       5.8\\
\noalign{\smallskip}
\hline\hline
\end{tabular}
\end{center}
\normalsize
\medskip
\caption{\small Median and standard deviations of the histograms of
  the polarization angles (P.A.) shown in
  Figure~\ref{fig:sed-pol}. The calculations take into account that
  the P.A. distributions wrap on themselves modulo 180$^{\circ}$.
 }
\label{tab:histo-pol-results}
\end{table}

\subsubsection{Constraints on AME polarisation in the TMC}\label{sec:tmc_polarisation}


\begin{table*}
\begin{center}
\begin{tabular}{lccccc}
\hline\hline
\noalign{\smallskip}
$\nu$ & $Q \pm \sigma_{Q}$& $U \pm \sigma_{U}$ & $P^{\rm db} \pm \sigma_{P^{\rm db}}$ & $\pi^{\rm db} \pm \sigma_{\pi^{\rm db}}$ & $\gamma \pm \sigma_{\gamma}$   \\
$[\rm GHz]$ & $[\rm Jy]$& $[\rm Jy]$& $[\rm Jy]$ & $[\%]$ & $[^{\circ}]$  \\
\noalign{\smallskip}
\hline
\noalign{\smallskip}
   1.4    &    -28.37$\pm$      1.28    &     -0.02$\pm$      0.01    &     28.66$\pm$      0.77    &    ... $\pm$ ...    &     90$\pm$      1   \\
  11.2    &     -4.10$\pm$      1.73    &     -2.85$\pm$      1.17    &      4.70$\pm$      2.91    &     17.87$\pm$     13.48    &     73$\pm$     16   \\
  12.9    &     -1.42$\pm$      1.36    &      1.48$\pm$      1.06    &      1.46$\pm$      2.23    &      0.00$\pm$      6.60    &    113$\pm$     41   \\
  22.7    &     -0.38$\pm$      0.73    &     -0.97$\pm$      0.65    &      0.48$\pm$      1.16    &      0.00$\pm$      2.34    &     56$\pm$     34   \\
  28.4    &      0.37$\pm$      0.56    &     -0.65$\pm$      0.45    &      0.26$\pm$      0.84    &      0.00$\pm$      1.92    &     30$\pm$     25   \\
  40.6    &     -1.01$\pm$      1.11    &     -1.37$\pm$      1.13    &      0.82$\pm$      1.93    &      0.00$\pm$      4.87    &     63$\pm$     44   \\
  44.1    &     -1.34$\pm$      1.05    &     -1.41$\pm$      1.00    &      1.50$\pm$      2.08    &      0.00$\pm$      4.85    &     67$\pm$     40   \\
  60.5    &     -6.77$\pm$      2.00    &     -0.97$\pm$      2.45    &      6.62$\pm$      4.77    &      8.95$\pm$      7.50    &     86$\pm$     78   \\
  70.4    &     -4.39$\pm$      1.85    &     -3.83$\pm$      1.57    &      5.38$\pm$      3.45    &      4.00$\pm$      3.03    &     69$\pm$     16   \\
  93.0    &    -13.62$\pm$      6.60    &    -16.95$\pm$      7.34    &     21.12$\pm$     15.11    &      8.29$\pm$      6.45    &     64$\pm$     18   \\
 353.0    &  -1302.55$\pm$    196.99    &   -807.09$\pm$    123.46    &   1679.52$\pm$    347.92    &      8.08$\pm$      2.41    &     74$\pm$      3   \\
\noalign{\smallskip}
\hline\hline
\end{tabular}
\end{center}
\normalsize
\medskip
\caption{\small Flux density estimates of $Q$ (column 2) and $U$
  (column 3) for the TMC region as a function of the frequency
  (column 1). The quantities derived after debiasing are the 
polarized flux densities $P^{\rm db}$ (column 4), 
the fraction of polarization of the total intensity $\pi^{\rm db}$
(column 5), and the polarization angle $\gamma$ (6). 
These last three quantities are plotted as a function of frequency in Figure~\ref{fig:sed-pol}.
 }
\label{tab:sed-pol-results}
\end{table*}


\begin{figure}
\begin{center}
\vspace*{2mm}
\centering
\includegraphics[width=75mm, angle =0]{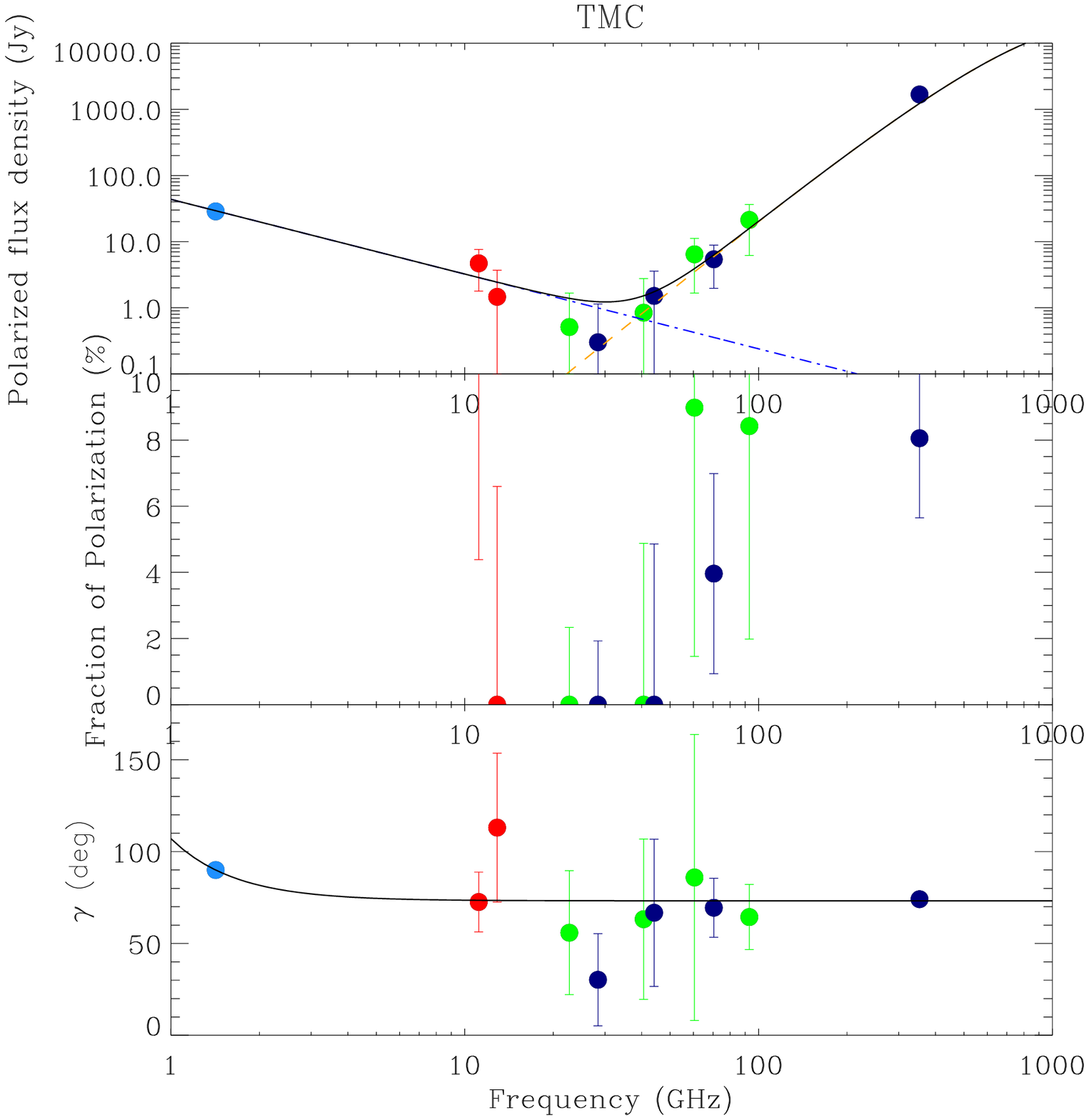}
\caption{\small Top: Polarised flux density of the TMC region as
defined in Figure~\ref{fig:ccs_bg_cloud} plotted as a function of
frequency. Middle: Fraction of polarisation of the total intensity 
plotted as a function of frequency.
Bottom: Polarisation angles estimates plotted as a function of frequency. 
Fits to the data are shown with continuous dark
lines (see text for details).
}
\label{fig:sed-pol}
\end{center}
\end{figure}

A further step to retrieve information about the various components is
to calculate the polarized flux densities in the TMC. This is done by
applying the same method as that used to estimate the flux
densities in intensity.
For this the fluxes of $Q$ and $U$ are first calculated using the
same cloud and background regions as those used to calculate the
fluxes in intensity (see Figure~\ref{fig:ccs_bg_cloud}). Since the 
values of $Q$ and $U$ can be positive or negative and $P$ is not
linear with $Q$ and $U$ the calculation of $P=\sqrt{Q^2+U^2}$ 
and $\pi=\sqrt{Q^2+U^2}/I$, can be positively biased. Various 
methods to debias $P$ exist in the literature, some of them 
taking into account priors on the distribution of $P$ \citep[e.g.][and
references therein]{alina2016}. In the following 
we use Monte Carlo simulations. For this we take the quantities 
$P$, $\pi$, and $\gamma$ and for each of these quantities we add a normal
term drawn from a normal distribution with a width characteristic of
the uncertainty on each quantity, repeat this process many times, and 
then measure the standard deviation with each quantity. 

The debiased polarized flux densities $P^{\rm db}$, the fraction of polarization of the
total intensity $\pi^{\rm db}$, and the polarization angle $\gamma$ from 
the SED analysis are all plotted as a function of frequency in 
Figure~\ref{fig:sed-pol}, from top to bottom respectively. All the values and their
uncertainties are listed in Table~\ref{tab:sed-pol-results}. 
After the data are completely processed and debiased we note that in the MFI frequency range
only data from horn 3 remain. The estimates of $P$ and $\pi$ from
the data from horn 2 are dominated by their uncertainties and for
that reason could not be debiased, therefore they are not included in the analysis.

Regarding the polarized synchrotron component one can see from
Figure~\ref{fig:sed-pol} (top plot) that at 1.420$\,$GHz 
the polarized flux is not at all consistent with zero ($\approx 35
\sigma$ detection). A residual in the 1.420$\,$GHz $Q$, and $U$ 
maps could come from the background subtraction dominating over the
flux in the cloud region or vice versa. A closer look at the data
shows that the background does not dominate the cloud region, and
that therefore the polarized flux should be associated with the TMC. 
At 1.420$\,$GHz the intensity flux is about $-8 \pm 18\,$Jy (Table~\ref{tab:sed-results}), and
the polarized flux density is $28.7 \pm 0.8\,$Jy
(Table~\ref{tab:sed-pol-results}). Within the uncertainties a total
intensity flux that would be detected to 2 sigma would be of $\approx
-8 + 2 \times 18 = 28\,$Jy and about equal to the total polarized flux density (which
would mean an approximately 100$\,\%$ polarized synchrotron component). A
detection of order 3 sigma would correspond to $\approx 46\,$Jy and imply that
about 60$\%$ of the synchrotron is polarized, assuming the free--free
contribution is negligible as expected from the
estimates shown in Figure~\ref{fig:map_ff_synch408}.

In order to get more insight into the synchrotron component we first assume that the SED
in polarization shown at the top in Figure~\ref{fig:sed-pol} can
be modelled by only two components: a polarized synchrotron
component and a polarized thermal dust component of about similar
polarization angle. The polarized synchrotron is
fitted by a power law whose parameters can be compared to those of the
power law obtained from the multi-component fit in intensity. The thermal dust
parameters from the thermal dust component fit in intensity are used as input
parameters with an additional free parameter, $\pi_{\rm dust}$,
the fraction of polarization of the thermal dust component, assumed to
be constant in the frequency range 20--353$\,$GHz. This hypothesis
is driven by the shapes of the polarized flux density and fraction of
polarization in both regions for frequencies higher than 60 GHz, taking
into account the uncertainties. 
Some observational studies
have shown that a flat polarization fraction spectrum can be expected
in low and intermediate column density molecular cloud regions such as the TMC
\citep[][]{ashton2018,gandilo2016}, whereas high column density molecular
cloud regions are not expected to show similar flat polarization 
spectra \citep[see][]{vaillancourt2012}. The output fit
parameters are displayed at the bottom of
Table~\ref{tab:sed-parameters}. A fraction of polarization, $\pi_{\rm
  dust} = 6.54 \pm 1.49\,\%$, provides a good fit to the polarized
component associated with the thermal dust. In the frequency range 1--20 GHz the polarized synchrotron is well fitted by a single power law
of index $\alpha_{\rm synch, pol} =  -1.13 \pm
0.23$. Such a power law index in flux (equivalent to a power index
$\approx -3.13$ in brightness temperature) is steeper than the power law
indices in temperature estimated from the $T$--$T$ plot analysis of the 
0.408$\,$GHz and 1.420$\,$GHz maps presented in Table \ref{tab:beta}.
Assuming that the power law indices in intensity and polarization are
of the same order, this means that one should expect 
$\alpha_{\rm synch, pol}  \approx -0.80$, which is not the
case here. On a large scale towards regions north of 
the TMC complex \citet{wolleben04} report from the
analysis of observations at 1.408$\,$GHz, 1.660$\,$GHz, 
and 1.713$\,$GHz that the diffuse
synchrotron temperature component in polarization 
is well described by a power law relation of index $\beta_{\rm synch}=
-2.7$, a value not consistent with that obtained from the SED in
polarization, but consistent to within the uncertainties 
with the values derived from our $T$--$T$ plot analysis. 
On the other hand Faraday depolarization is also
inferred from their analysis as due to Faraday screens with excessive 
rotation measures at the boundaries of molecular clouds \citep[note that
these results were not confirmed by \citet{crutcher2008} from their
analysis of OH observations, which failed to
detect the 20 $\,\mu$G magnetic field necessary to explain the
strong Faraday rotations estimated by][]{wolleben04}. 
In order to fit this effect we use the model $\gamma = \gamma_{0} + RM \times
\lambda^{2}$, where $\gamma_{0} $ is the polarization direction at 
$\lambda = 0$ ($\nu \rightarrow  \infty$ in our case) and $RM$ represents the rotation measure.
The fit parameters are listed in Table~\ref{tab:sed-parameters} and
show a rotation measure $RM=6.56 \pm 1.04\,$rad m$^{-2}$ occurring at
frequencies lower than the QUIJOTE frequencies. Our rotation measure
is lower than those reported by \citet{wolleben04} of order
-18$\,$rad m$^{-2}$ to -30$\,$rad m$^{-2}$ and suggest this 
effect to be low in the TMC at the resolution of the observations. 

Another possible scenario to account for the variations in the
polarized intensity as a function of frequency would be to assume
that, in addition to the thermal dust and the synchrotron components, 
the AME component is polarized. 
To test this possibility we show in Figure~\ref{fig:sed-pol} 
(middle) the variations of the fraction of polarization of the total 
intensity as a function of frequency. The total intensities used to
calculate $\pi^{\rm db}$ are those from
the SED in intensity. The flux density in intensity at 1.420$\,$GHz is
negative, which is why no fraction of polarization can be inferred at this
frequency. In the frequency range 20--50$\,$GHz 
the AME component peaks at a frequency
$\nu_{\rm AME, peak} \approx 19\,$GHz and the polarization fraction
are consistent with zero, and only upper limits from 
the uncertainties can be derived on the maximum of polarization. 
The best data to put constraints on the level of polarization of the
AME component comes from the point at 28.4$\,$GHz. Assuming that the
intensity and polarization components of the other components 
are negligible at this frequency, the
data show that the fraction of AME polarization  
should be $\pi_{\rm AME}<$3.8$\,\%$ with 95$\,\%$ confidence Level (C.L.).
Assuming now the extreme case in which the polarized fluxes $Q$ and
$U$ at 28.4$\,$GHz are produced only by the AME component, and 
in which only the intensity of the AME component is used to
calculate the fraction of polarization
(i.e.\ $I_{\rm AME} \approx I_{\rm total}-I_{\rm TD}$ since from the
fit $I_{\rm Sync} \approx 0$ whether the synchrotron is assumed to be 0$\,\%$ or
100$\,\%$ polarised at 28.4$\,$GHz), a simple cross product gives a
little less stringent constraint with a higher upper limit of 4.2$\,\%$ (95$\,\%$C.L.). 

Finally we show in the bottom plot in Figure~\ref{fig:sed-pol} the 
distribution of the polarization angles associated with the TMC. 
One can see some variations with respect to the fit, in particular 
the polarization angle estimated at 28.4$\,$GHz ($\gamma
\sim 30^{\circ}$) departs from the fit, as does
the polarization angle at 12.9$\,$GHz  ($\gamma \sim 113^{\circ}$), and
the two angles are almost perpendicular to each other even though the
uncertainties in both measurements are quite large. 
Polarized AME components could be the explanation but the large difference between the two angles
could also be explained by dilution effects from the integration 
over a large area like the one covered by the TMC, or 
simply because some regions in each map are noise dominated.
This second hypothesis may be more likely given the variations of the
polarization angles observed in some of the polarization maps in
Figure \ref{fig:pol_maps}. This can also be seen with the variations
with frequency of the polarization angle dispersion values in
Table \ref{tab:histo-pol-results}.


\begin{table*}
\begin{center}
\begin{tabular}{lccccc}
\hline\hline
\noalign{\smallskip}
$\nu$ & $Q \pm \sigma_{Q}$& $U \pm \sigma_{U}$ & $P^{\rm db} \pm \sigma_{P^{\rm db}}$ & $\pi^{\rm db} \pm \sigma_{\pi^{\rm db}}$ & $\gamma \pm \sigma_{\gamma}$   \\
$[\rm GHz]$ & $[\rm Jy]$& $[\rm Jy]$& $[\rm Jy]$ & $[\%]$ & $[^{\circ}]$  \\
\noalign{\smallskip}
\hline
\noalign{\smallskip}
   1.4    &     -3.00$\pm$      0.30    &      -0.01$\pm$      0.00    &      3.03$\pm$      0.08    &     51.87$\pm$     41.40    &    80$\pm$      1   \\
  11.2    &     -0.50$\pm$      0.41    &     -0.47$\pm$      0.28    &      0.57$\pm$      0.70    &      0.00$\pm$     10.62    &     68$\pm$     34   \\
  28.4    &      0.19$\pm$      0.13    &     -0.09$\pm$      0.11    &      0.15$\pm$      0.23    &      0.00$\pm$      1.98    &     13$\pm$     21   \\
  32.9    &     -0.36$\pm$      0.20    &     -0.20$\pm$      0.23    &      0.33$\pm$      0.44    &      0.00$\pm$      4.74    &     75$\pm$     59   \\
  40.6    &      0.08$\pm$      0.27    &     -0.42$\pm$      0.27    &      0.24$\pm$      0.48    &      0.00$\pm$      4.36    &     39$\pm$     26   \\
  60.5    &     -1.10$\pm$      0.48    &     -0.35$\pm$      0.59    &      0.97$\pm$      1.15    &      5.18$\pm$      6.24    &     81$\pm$     70   \\
  70.4    &     -1.07$\pm$      0.44    &     -0.38$\pm$      0.37    &      1.03$\pm$      0.84    &      4.67$\pm$      4.43    &     80$\pm$     58   \\
  93.0    &     -2.48$\pm$      1.57    &     -3.66$\pm$      1.74    &      4.18$\pm$      3.66    &      6.52$\pm$      6.19    &     62$\pm$     22   \\
 353.0    &   -305.82$\pm$     46.56    &   -156.57$\pm$     27.16    &    376.61$\pm$     79.38    &      7.23$\pm$      2.20    &     76.44$\pm$      2.75   \\

\noalign{\smallskip}
\hline\hline
\end{tabular}
\end{center}
\normalsize
\medskip
\caption{\small Flux density estimates of $Q$ (column 2) and $U$
  (column 3), for the L1527 dark cloud as a function of the frequency
  (column 1). The quantities derived after debiasing are the 
polarized flux densities $P^{\rm db}$ (column 4), 
the fraction of polarization of the total intensity $\pi^{\rm db}$
(column 5), and the polarization angle $\gamma$ (6). 
These last three quantities are plotted as a function of frequency in Figure~\ref{fig:sed-pol-ldn1527}.
 }
\label{tab:sed-pol-ldn1527-results}
\end{table*}

\subsubsection{Constraints on AME polarization in L1527}\label{sec:ldn1527_polarisation}

\begin{figure}
\begin{center}
\vspace*{2mm}
\centering
\includegraphics[width=75mm, angle =0]{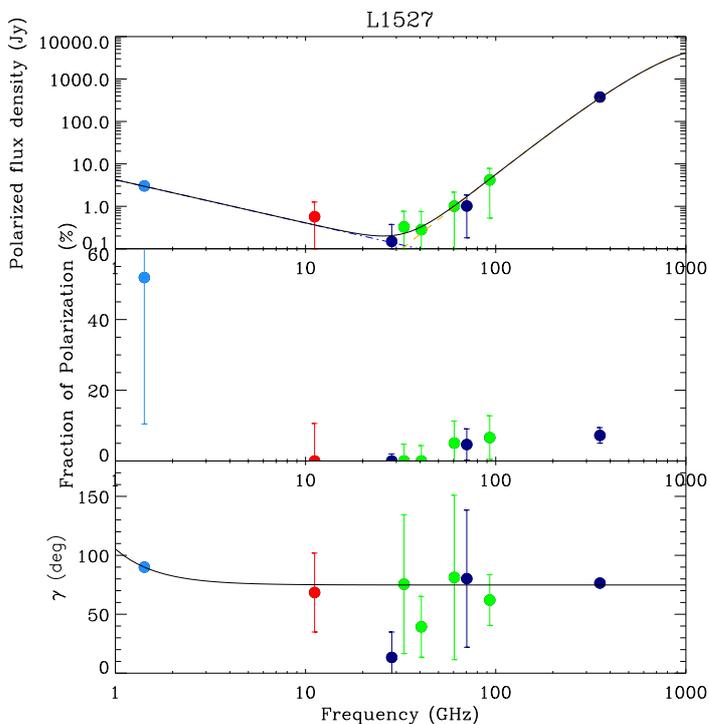}
\caption{\small Top: polarized flux density of the L1527 dark cloud as
  defined in Figure~\ref{fig:ccs_bg_cloud} plotted as a function of
  frequency. 
Middle: fraction of polarization of the total intensity plotted as a function
of frequency. 
Bottom: polarization angle estimates plotted as a function of frequency.
Fits to the data are shown with continuous dark
lines (see text for details).
}
\label{fig:sed-pol-ldn1527}
\end{center}
\end{figure}

For comparison with the TMC, we conducted an analysis 
of the SED in polarization of L1527. The
background region used to make the calculations is the same one
as before. The flux densities in polarization are listed
in Table~\ref{tab:sed-pol-ldn1527-results} and plotted as a function
of frequency in Figure~\ref{fig:sed-pol-ldn1527}.
Assuming a uniform fraction of thermal dust emission polarization
in the frequency range 20--353$\,$GHz we find a polarization fraction
slightly higher than in the TMC, but the two estimates are consistent 
with each other within the uncertainties, in agreement with what one
might expect from the level of polarization shown in the 353$\,$GHz 
polarization map (see Figure~\ref{fig:pol_maps}).
The power law index describing the polarized synchrotron variations
with frequency ($\alpha_{\rm synch, pol}= -1.02 \pm 0.20 $) 
now gives a consistent picture to within the uncertainties with
the power law indices from the intensity analysis in temperature and
 flux, and is in agreement with the value $\beta_{\rm synch, obs}=-2.7$ 
found by \citet{wolleben04} from their
analysis of the polarized temperatures in large
scale in regions north of the Taurus molecular cloud complex. 

In order to check for possible Faraday screens that could produce
Faraday depolarization towards L1527, and therefore a rotation of the polarization
angles towards low frequencies, we repeated the fit analysis on the distribution
of polarization angles. The results of the fit,
$RM= 5.92 \pm 1.04\,$rad m$^{-2}$ and 
$\gamma_{0}=1.31^{\circ} \pm 0.05^{\circ}$, are very similar to the
results obtained for the TMC, but the fit is slightly less constrained 
($\chi^2_{\rm red}=1.55$). The quality of the fit is lower for L1527
than for the TMC because the value of the polarization angles at 
frequencies 28.4$\,$GHz and 40.6$\,$GHz departs from the trend shown 
by the other angles at frequencies greater than 10$\,$GHz.
For comparison with the results obtained in the TMC, a look 
at the distribution of the polarization angles (see Figure~\ref{fig:pol_maps}) 
displayed over the area covered by L1527 indeed shows overall uniform distributions 
of the pseudo-vectors at each frequency but with different 
orientations in the frequency range 10--50$\,$GHz. 
The perturbation of the polarization angle values could come from
AME components having different polarization orientations
as a function of frequency but, all in all, the
uncertainties in the polarization angles listed in 
Table~\ref{tab:sed-pol-ldn1527-results} are still high and do not
permit any firm conclusion.

Finally, as was discussed for the TMC, we assume 
the best data to put constraints on the fraction of AME polarization 
are the those for 28.4$\,$GHz. If none of the other components is polarized and
of negligible intensity, we find an upper limit $\pi_{\rm AME} < 4.5\,\%$ 
(95$\,\%$ C.L.). On the other hand, if the other component intensities are not negligible and
their modelled estimates are removed from the total intensity, and if 
the polarization is produced only by the AME component, a worse constraint 
$\pi_{\rm AME} < 5.3\,\% $ (95$\,\%$ C.L.) is obtained.


\section{Discussion} \label{sec:discussion}

In our analysis we have shown that towards and around the TMC 
in the maps at frequencies 0.408$\,$GHz and 1.420$\,$GHz 
the emission of the synchrotron component predominates
over the free--free emission independently of the dust mixing
fraction of the H$\alpha$ with the thermal dust grains. This result
is shown at a frequency of 0.408$\,$GHz in Figure~\ref{fig:map_ff_synch408}, 
where one can see that the emission of the free--free is expected to
be only about 1$\,\%$ or lower of the total emission of the maps.
From a TT-plot analysis of the same two maps we have parameterized
the variation of the synchrotron emission with frequency towards the
TMC by a single power law of power law index 
$\beta_{\rm synch} = 2.80 \pm 0.26$ (see Table~\ref{tab:beta}). 
This result is consistent with that obtained by \citet{wolleben04},  
who showed from the analysis of observations at 1.408$\,$GHz, 1.660$\,$GHz ,
and 1.713$\,$GHz that the diffuse synchrotron temperature component in polarization 
is well described by a power law relation of index $\beta_{\rm synch} \approx -2.7$.

In order to put constraints on the level of polarization of the AME
component we have made an analysis of the SED in intensity
(Figure~\ref{fig:sed-int}) 
and polarization (Figure~\ref{fig:sed-pol}) of
the TMC. The area of the TMC and the background regions defined and used for this are
shown in Figure~\ref{fig:ccs_bg_cloud}. Thanks to the QUIJOTE data
maps at 11$\,$GHz, 13$\,$GHz, 17$\,$GHz, and 19$\,$GHz the multi-fit component of the
SED in intensity shows a clear detection of AME of total intensity
$S_{\rm AME, peak} = 43.0 \pm 7.9\,$Jy peaking at a central
frequency of about 19$\,$GHz (Table~\ref{tab:sed-parameters}). 
A clear detection of a synchrotron
component is not possible in the SED in intensity but a clear
detection is seen in the SED in polarization, indicating that a
synchrotron component is indeed associated with the TMC. From the analysis of
the SED in polarization at a frequency of 28.4$\,$GHz we have estimated 
an upper limit $\pi_{\rm AME}<$4.2$\,\%$ with 95$\,\%$ confidence level
(C.L.) (worst case scenario) and $\pi_{\rm AME}<$3.8$\,\%$ with (95$\,\%$
 C.L.) if the intensity of all the other components is negligible at
this frequency. 

A look at the polarization maps at 1.420$\,$GHz and 353$\,$GHz 
(Figure~\ref{fig:pol_maps}) shows very uniform polarization
patterns that indicate the presence of uniform magnetic fields on a large
scale towards the TMC (these were the main motivation for observing the TMC with
the MFI). A look at the polarization maps in the frequency
range 10--40$\,$GHz, though, indicates variations of the polarization angles on smaller
scales. These variations are also seen in the variation of the polarization
angles $\gamma$ estimated from the analysis of the SED in polarization
(Figure~\ref{fig:sed-pol}), but the uncertainties in the polarization 
angles are generally too large to make any strong conclusions 
(Table~\ref{tab:sed-pol-results}). In addition, such variations of the polarization angles on a 
smaller scale towards the TMC could result in reducing the net
fraction of polarization estimated after integration over the total
area of the TMC and thus mislead the interpretation of the
constraints put on $\pi_{\rm AME}$ in the TMC. 
In order to address this problem and to understand what could be the origin
of these variations of $\gamma$ in the frequency range 10--50$\,$GHz 
we repeated our analysis towards the L1527 dark cloud nebula.  
On the one hand the analysis of the SED in intensity
(Figure~\ref{fig:sed-int-ldn1527}) gives a power law description of the
synchrotron component as a function of frequency consistent with that observed on a 
large scale \citep[this work and ][]{wolleben04}. In addition, the
overall trend of the distribution of the polarization angles 
in the frequency range at 1.42--353$\,$GHz are very similar.
This is confirmed by the fits to the distributions of
polarization angles in the TMC and L1527 respectively.
The $RM$ obtained from the two fits are low and of order $\sim
6.6\,$rad m$^{-2}$ in
the TMC and  $\sim 5.9\,$rad m$^{-2}$ in L1527. For comparison 
\citet{wolleben04} derived towards regions north of the TMC $RM$
of $ -18\,$rad m$^{-2}$ and $ -30\,$rad m$^{-2}$ to model their
observations at frequencies below 2$\,$GHz. 
These results suggest that if there are Faraday screens towards the TMC and L1527 their
effects are quite low at the resolution of the observations.  
On the other hand, the SED in polarization of L1527 (Figure~\ref{fig:sed-pol-ldn1527})
confirms the trends observed in the variations of the polarization
angles $\gamma$ in the
frequency range 10--50$\,$GHz. Since these variations are not expected 
to come from Faraday depolarization, an
alternative explanation is that another component is polarized in a
direction different from the direction of the magnetic fields inferred
from the polarization maps at 1.420$\,$GHz and 353$\,$GHz, but the
uncertainties in the estimates of such a component are too high to
derive any strong conclusion. In any case, if this polarized
component is emitted by the carriers at the origin of the AME we expect its
level of polarization at 28.4$\,$GHz in L1527 to be  
$\pi_{\rm AME}<$5.3$\,\%$ (95$\% $C.L.), or $\pi_{\rm AME}<$4.5$\,\%$
(95$\,\%$C.L.) if the intensities of all the other components 
are negligible at this frequency.


\begin{figure}
\begin{center}
\vspace*{2mm}
\centering
\includegraphics[width=85mm, angle =0]{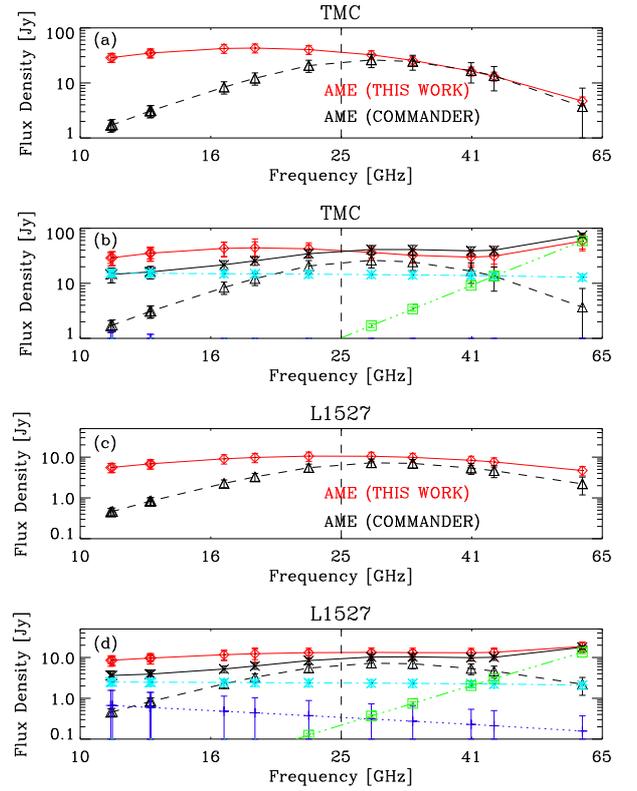}
\caption{\small SEDs of the TMC and L1527 in the frequency range 
10--65$\,$GHz.
Panels (a) and (c): AME flux densities from this analysis and from Commander.
The dark curves show the SEDs of the total flux densities of the
two AME components from Commander and the red curves show the SED of
the AME from this analysis. Panels (b) and (d):
the total flux densities obtained with our multi-fit components are
shown with red curves and diamond symbols. The total flux densities
obtained with Commander are shown with the black curves and
star symbols. The Commander AME flux densities are shown with
black dashed curves and triangle symbols. The Commander thermal dust flux
densities are displayed with the green dashed double dotted lines with
square symbols. The Commander synchrotron flux densities calculated by assuming
a power law in temperature of index $\beta_{\rm synch}=-2.8$ are shown
with dark blue dotted lines and plus symbols. The Commander free--free flux densities are
shown with pale blue dotted lines and star symbols. 
}
\label{fig:sed-ame}
\end{center}
\end{figure}


\begin{table}
\begin{center}
\begin{tabular}{ccc}
\hline\hline
\noalign{\smallskip}
& TMC &  L1527\\
\noalign{\smallskip}
\hline 
Frequency & $\frac{S(\rm Fit)}{S(\rm Commander)}$&  $\frac{S(\rm Fit)}{S(\rm Commander)}$ \\
$[$GHz$]$ & [AME] & [AME] \\
\noalign{\smallskip}
\hline
\noalign{\smallskip}
      11.2    &   17.0 & 12.3 \\
      11.2     &  16.7 & 12.1 \\
      12.8    &   11.3 & 8.3 \\
      12.9   &    11.2 & 8.2 \\
      16.7   &    5.0 &  4.0 \\
      18.7   &    3.5 & 3.0 \\
      22.7   &    2.0 & 1.9 \\
      28.4   &    1.2 & 1.5 \\
      32.9   &    1.1 & 1.4 \\
      40.6   &   1.0  & 1.6 \\
      44.1   &   1.0  & 1.6 \\
      60.5   &    1.3 & 2.1 \\
\noalign{\smallskip}
\hline\hline
\end{tabular}
\end{center}
\normalsize
\medskip
\caption{\small Ratio between the AME intensity flux densities obtained in
  this work and those estimated with Commander for the TMC and
  L1527 as displayed in the plots shown in Figure~\ref{fig:sed-ame}.
}
\label{tab:sed-ratio}
\end{table}

In section~\ref{sec:planck_products} we discuss qualitatively how the 13 GHz
MFI-QUIJOTE SMICA-CMB subtracted map compares morphologically 
with the map obtained by combining the Commander 
separation component templates (synchrotron, free--free, thermal dust, and AME) 
at a reference frequency of 13$\,$GHz. As a further test we now discuss
quantitatively how the Commander AME components compare to the AME component
derived from the SED analysis in intensity. To do this we calculated
the various template maps obtained by combining the two Commander AME
templates at the reference frequencies of the observations 
in the range 10--65$\,$GHz. We then calculated the intensity fluxes of
the Commander AME by using the same masks as those used to
calculate the SED in the TMC and L1527. 
We also calculated the total flux densities of the
synchrotron, free--free, thermal dust, and AME components by 
combining the template maps of all the components and calculated the
flux densities for each component.
The results are shown in Figure~\ref{fig:sed-ame} for both regions. 
In the plots showing only the fits of the AME components the dark curves 
show the AME flux densities obtained with the Commander templates in the
frequency range 10--65$\,$GHz panels a and c.
The red curves show the SEDs of the AME fit components obtained in this analysis. 
In both regions we find the AME flux densities estimated with each method
\citep[our component separation including the QUIJOTE-MFI data and the
component separation with the Commander algorithm][]{cpp2015-25} 
are similar within a factor two at frequencies higher than $\sim$25$\,$GHz. 
On the other hand the lack of information in the
frequency range 10--20$\,$GHz clearly shows the AME components to be
underestimated by the Commander method. Such results 
are understood as originating from cross talk between the synchrotron,
free--free, and AME with the consequence of underestimating the AME
intensities in the TMC and L1527 regions. A look at the flux
densities of the other Commander components is shown 
in panels b and d, Figure~\ref{fig:sed-ame}. 
The Commander AME density fluxes are shown with
black dashed curves and triangle symbols. The Commander thermal dust flux
densities are displayed with the green dashed double dotted lines with
square symbols. The Commander synchrotron flux densities calculated by assuming
a power law in temperature of index $\beta_{\rm synch}=-2.8$ are shown
with dark blue dotted lines and plus symbols. 
The Commander free--free flux densities are
shown with pale blue dotted lines and star symbols.
A comparison of these plots with those in 
panels a and c indeed clearly
shows that the AME Commander flux densities are underestimated 
at the cost of overestimating the Commander free--free flux densities
in both regions since, in the particular case of the Taurus molecular cloud
complex, the Commander synchrotron component associated with the cloud structures
is found to be very low. For that reason one can see that consistent fits
to the total AME components are obtained at frequencies higher than
$\sim$25$\,$GHz by each method and divergence of the fits at frequencies
lower than $\sim$25$\,$GHz. This limit is shown in all the plots
with a dashed vertical line centred at a frequency of 25.5$\,$GHz for illustration.
To summarize these results, the ratio of the flux
densities of the total AME components obtained by each method 
are listed in Table~\ref{tab:sed-ratio}. The high values of the ratios estimated 
in the frequency range 10--20$\,$GHz clearly demonstrate the
relevance and importance of the QUIJOTE-MFI data for sampling the 
electromagnetic spectrum at radio frequencies lower than 20$\,$GHz. 
Without the QUIJOTE-MFI information, the Commander component separation
methods underestimates the AME density fluxes. With the QUIJOTE-MFI
data the location of the AME peaks and the width of the AME features
can be characterized more accurately.

\begin{figure}
\begin{center}
\vspace*{2mm}
\centering
\includegraphics[width=85mm, angle =0]{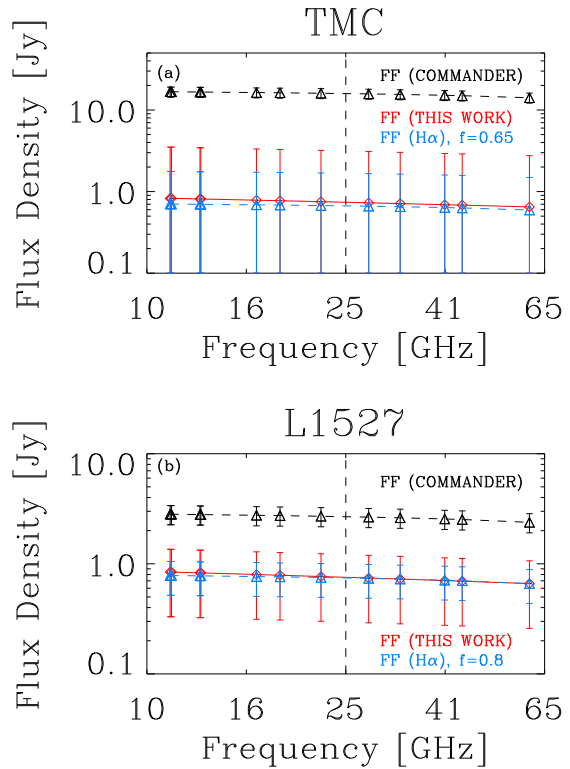}
\caption{\small Free--free SEDs of the TMC (panel a) and L1527 (panel b) in the frequency range 10--65$\,$GHz.
In each panel, the free--free flux densities from this analysis are
shown in red. The free--free flux densities from Commander
are shown in black, and the free--free flux densities
calculated with the H$\alpha$ map are shown in blue.
The free--free flux densities obtained from our analysis
can be reproduced assuming a spatially uniform dust mixing factor $f=0.65$ for the whole
TMC and a dust mixing factor $f=0.8$ for L1527. 
}
\label{fig:sed-ff}
\end{center}
\end{figure}

In section \ref{sec:quijote_morphology} we discuss the similarity between the
  morphology of the Commander free--free map and the
  free--free map using the H$\alpha$ map, and calculated, assuming a dust
mixing fraction, $f=0.8$. In order to quantify how the H$\alpha$
free--free maps could reproduce the free--free SED profiles obtained from the
multi-component fit analysis of the SEDs in intensity we have calculated
the SED spectra obtained in the TMC and in L1527 for various values of
the dust mixing factor. Our best fits to the data are shown in
Figure \ref{fig:sed-ff}. As discussed above, for each region the SED obtained with the
Commander free--free map is overestimated with respect to our fits. On
the other hand, the free--free flux densities obtained from our analysis
can be reproduced assuming a spatially uniform dust mixing factor $f=0.65$ for the whole
TMC and a dust mixing factor $f=0.8$ for L1527. We point out that for
the TMC the consistency between our fit and the free--free H$\alpha$
calculated SED is quite sensitive to the tuning of the dust mixing
fraction. We interpret this as an indication that the dust mixing
fraction is not necessarily uniform over the whole TMC. The extinction
model we use to estimate the H$\alpha$ optical depth 
also assumes a linear relation with the thermal dust optical depth
$\tau_{353}$, but variations are expected towards high and low column
density regions \citep[][]{pir11}. All in all, though, and despite the
above caveats, 
the comparisons of our SED multi-components fit analyses including new data from the
QUIJOTE-MFI with the free--free maps
calculated with the H$\alpha$ maps give quite a consistent picture of the free--free
component assuming an average dust
mixing fraction $f \approx 0.80$. From this analysis
we drive the conclusions that, towards the TMC and L1527, 
the Commander component separation method tends to overestimate the free--free component 
at the cost of underestimating the AME component.


\section{Conclusions} \label{sec:conclusions}

We acquired new linear polarimetry observations of the Taurus molecular
cloud region observed in the frequency 10-20$\,$GHz with the Multi-Frequency
Instrument mounted on the first telescope of the Q-U-I-JOint
TEnerife (QUIJOTE) experiment.  
The data were obtained over a total integration time of about 423 hours
towards an area of $\approx 17^{\circ} \times 17^{\circ}$ during 2015 March--July. 

From the combination of the QUIJOTE data with the \textit{WMAP} 9-yr data
release, the \textit{Planck} second data release, the DIRBE maps, and ancillary
data we detect an AME component of total emission 
$S_{\rm AME, peak} = 43.0 \pm 7.9\,$Jy (5.4 $\sigma$ detection) in the TMC and 
$S_{\rm AME, peak} = 10.7 \pm 2.7\,$Jy (4.0 $\sigma$ detection) in L1527. 
In the TMC the AME peaks around a frequency of 19$\,$GHz and around a frequency of 25$\,$GHz in L1527.
In the TMC, the analysis of the SED in polarization at {the \textit{Planck} channel of
28.4$\,$GHz put constraints on the level of polarization of the AME with 
an upper limit $\pi_{\rm AME}<$4.2$\,\%$ (95$\,\%$ C. L.),
and $\pi_{\rm AME}<$3.8$\,\%$ (95$\,\%$
 C.L.) if the intensity of all the other components is negligible at
this frequency. The same analysis in L1527 leads to 
$\pi_{\rm AME}<$5.3$\,\%$ (95$\,\%$ C.L.) or $\pi_{\rm AME}<$4.5$\,\%$ 
(95$\,\%$ C.L.) if the intensities of all the other components are 
negligible at this frequency. 

From a comparison of the Commander free--free map with the
  free--free maps calculated with the H$\alpha$ map for several dust
  mixing fraction, and from a comparison of the free--free flux
  densities obtained from our analysis including the QUIJOTE-MFI data
  with the Commander free--free flux densities and with the H$\alpha$
  free--free flux densities, we reach the conclusion that in the TMC and
  L1527 the dust gas mixing fraction should be of order $80\%$.
		
In the TMC and L1527 the flux densities associated with the AME
component are consistent to within a factor 2 with those obtained 
from the analysis of the Commander template maps at frequencies 
$\ge 22.7\,$GHz. On the other hand, because the Commander 
component separation products were produced without data in the 
frequency range 10--20 GHz we find that the flux densities of the AME 
are drastically underestimated by the Commander analysis in
this frequency range, and that the Commander free--free flux
  densities are overestimated at all frequencies. 
Our analysis shows the importance of the QUIJOTE-MFI data
to accurately separate the synchrotron, free--free, and
AME components. 

Further investigation including additional QUIJOTE data in the frequency range
10--42$\,$GHz, as well as supplementary information provided by C-BASS 5$\,$GHz
data \citep[e.g.][]{king2014,irfan2015}, should be useful in confirming 
our analysis and constraining the polarization components
observed in the frequency range 1--50$\,$GHz in L1527 and the TMC.

\section*{Acknowledgements}
We thank J. P. Leahy for fruitful discussions
about 3C123. We thank Juan Uson and Wolfgang Reich for useful comments.
We thank the anonymous referee whose comments 
helped to improve this work. We thank T. J. 
Mahoney for revising the English of the draft.
FP acknowledges the European Commission under the Marie 
Sklodowska-Curie Actions within the $European$ $Union's$ $Horizon$ 
$2020$ research and innovation programme under Grant Agreement number
658499 (PolAME). This work has been partially funded by the Spanish Ministry
of Economy and Competitiveness (MINECO) under the projects 
AYA2007-68058-C03-01, AYA2010-21766-C03-02, AYA2012-39475-C02-01, 
AYA2014-60438-P: ESP2015-70646.C2-1-R, AYA2015-64508-P and the 
Consolider-Ingenio project CSD2010-00064  (EPI: Exploring the
Physics of Inflation). This project has received funding from the
European Union$^{\prime}$s Horizon 2020 research and innovation programme
under grant agreement number 687312 (RADIOFOREGROUNDS).
CD and SH acknowledge support from an ERC Consolidator grant
(no.~307209). CD also acknowledges support from an STFC Consolidated
grant (ST/L000768/1). 
We acknowledge the use of data from the $Planck$/ESA mission, downloaded
from the $Planck$ Legacy Archive, and of the Legacy Archive for
Microwave Background Data Analysis (LAMBDA). Support for LAMBDA is
provided by the NASA Office of Space Science. Some of the results in
this paper have been derived using the HEALP{\sc ix} \citep{gorski05}
package. This work made use of the Strasbourg Astronomical Data Center
though the CDS digital platform.




\bibliographystyle{mnras}




\appendix



\section{Additional null tests} \label{sec:addjktests}


The full data set has also been divided in two data sets
by considering the maps showing Back-End Module (BEM) 
averaged temperature values of $\langle T_{\rm BEM}$$\rangle$ lower and higher than
the median value $\langle T_{\rm BEM}$$\rangle$$_{ \rm MEDIAN}$ of the full data set. The
results from this null test analysis are given in
Table~\ref{tab:tbem}. Similarly the null test results obtained by
considering the data sets obtained by splitting the data set into two
groups as a function of azimuth are shown in Table~\ref{tab:azel}. One
can see good consistency with the results obtained from the null test
analysis obtained when considering the first half of the maps
observed with the second half as displayed in Table~\ref{tab:time}.

\begin{table*}
\begin{center}
\begin{tabular}{ccccccccccccc}
\hline\hline
\noalign{\smallskip}
Horn & Freq.   && \multicolumn{2}{c}{$\sigma_{\rm I}$ ($\mu$K beam$^{-1}$)}
  && \multicolumn{2}{c}{$\sigma_{\rm Q}$ ($\mu$K beam$^{-1}$)}    &&
                                                              \multicolumn{2}{c}{$\sigma_{\rm
                                                                     U}$
                                                                     ($\mu$K
                                                                     beam$^{-1}$)} && $\sigma_{\rm Q,U}$ (mK~s$^{1/2}$) \\

\noalign{\smallskip}
\cline{4-5}\cline{7-8}\cline{10-11}\cline{13-13}
\noalign{\smallskip}
& (GHz) && Map & NT && Map & NT && Map & NT && NT \\
\noalign{\smallskip}
\hline
\noalign{\smallskip}
     1&11 &&       45.9 &       29.8 &&       ...&       ...&&       ...&       ... &&       ...\\
     1&13 &&       38.5 &       24.7 &&       ...&       ...&&       ...&       ... &&       ...\\
     2&17 &&       98.0 &       62.4 &&       16.1 &       11.1 &&      15.1 &       10.5 &&       2.5\\
     2&19 &&       115.0 &       75.3 &&       21.9 &       15.6 &&       16.6 &       12.3 &&       2.6\\
     3&11 &&       70.9 &       44.5 &&       21.4 &       13.7 &&       15.1 &       12.7 &&       2.5\\
     3&13 &&       58.2 &       36.0 &&       18.7 &       12.2 &&       16.5 &       11.4 &&       2.3\\
     4&17 &&       103.4 &       71.3 &&       18.1 &       12.7 &&       14.4 &       11.4 &&       2.3\\
     4&19 &&       122.0 &       92.9 &&       21.5 &       14.9 &&       19.3 &       15.6 &&       2.9\\
\noalign{\smallskip}
\hline\hline
\end{tabular}
\end{center}
\normalsize
\medskip
\caption{Same as in Table~\ref{tab:time} by splitting the full data
  set in two groups of maps
  obtained by considering observations with a mean temperature of the
  Back End Module (BEM), $T_{\rm BEM}$, higher or lower than
  the BEM median temperature value, $\langle T_{\rm BEM}$$\rangle$$_{ \rm MEDIAN}$, of the full
  set of observations.
 }
\label{tab:tbem}
\end{table*}

\begin{table*}
\begin{center}
\begin{tabular}{ccccccccccccc}
\hline\hline
\noalign{\smallskip}
Horn & Freq.   && \multicolumn{2}{c}{$\sigma_{\rm I}$ ($\mu$K beam$^{-1}$)}
  && \multicolumn{2}{c}{$\sigma_{\rm Q}$ ($\mu$K beam$^{-1}$)}    &&
                                                              \multicolumn{2}{c}{$\sigma_{\rm
                                                                     U}$
                                                                     ($\mu$K
                                                                     beam$^{-1}$)} && $\sigma_{\rm Q,U}$ (mK~s$^{1/2}$) \\

\noalign{\smallskip}
\cline{4-5}\cline{7-8}\cline{10-11}\cline{13-13}
\noalign{\smallskip}
& (GHz) && Map & NT && Map & NT && Map & NT && NT \\
\noalign{\smallskip}
\hline
\noalign{\smallskip}
     1&11 &&       45.9 &       34.1 &&       ... &       ... &&       ... &       ... &&       ...\\
     1&13 &&       38.5 &       29.1 &&       ... &       ... &&       ... &       ... &&       ...\\
     2&17 &&       98.0 &       63.7 &&       16.1 &       11.6 &&       15.1 &       11.0 &&       2.6\\
     2&19 &&       115.0 &       66.3 &&       21.9 &       14.3 &&       16.6 &       12.4 &&      2.5\\
     3&11 &&       70.9 &       44.1 &&       21.4 &       14.1 &&       15.1 &       12.9 &&       2.6\\
     3&13 &&       58.2 &       35.9 &&       18.7 &       12.8 &&       16.5 &       12.7 &&       2.4\\
     4&17 &&       103.4 &       71.6 &&       18.1 &       12.6 &&       14.4 &       12.0 &&       2.3\\
     4&19 &&       122.0 &       86.6 &&       21.5 &       16.6 &&       19.3 &       13.7 &&       2.8\\
\noalign{\smallskip}
\hline\hline
\end{tabular}
\end{center}
\normalsize
\medskip
\caption{Same as in Table~\ref{tab:time} with the data set split into
  two equal group sampling about the same azimuth and elevation ranges.
 }
\label{tab:azel}
\end{table*}

\section{Background Assessment} \label{sec:assess_bg}

We conducted several tests to assess whether the area used as
the background region captures as accurately as possible an
average picture of the ISM properties in the vicinity of the TMC. 
 The level of the variations of the SED fluxes as a function of the areas
covered by the TMC and background regions  
was assessed by checking the effect of changing the limits on
$A_{V}$ used to define the area of the TMC and the background
regions shown in Figure~\ref{fig:ccs_bg_cloud}. 
Using $A_{V} = 4.0\,$magnitudes and $A_{V} = 5.0\,$magnitudes 
produces, to within the uncertainties, fluxes similar to within
a few Jy to those that are obtained with a limit
value of $A_{V} = 4.5\,$magnitudes at frequencies higher than
10$\,$GHz. Another test to assess the relevance of using the background area displayed
in Figure~\ref{fig:ccs_bg_cloud}  was done by calculating SEDs of the TMC for
various sub-regions of the background area.
To do this we calculated the contribution of the background 
towards apertures of diameter 1.5$^{\circ}$ on a sample of spatially independent 
positions displayed over the background region area.
These apertures are shown in the $A_{V}$ maps 
in Figure~\ref{fig:quijote_assess_bg}. 
For each aperture an SED of the TMC was calculated;
we then estimated the median and the variance of the fluxes at each frequency. 
These estimates are plotted as a function of
frequency in Figure~\ref{fig:quijote_assess_bg2} (top). The relative flux
variations obtained from a comparison of the two methods (fluxes
calculated by considering the full background area versus the median of
the fluxes obtained with the background sampled in the circular
apertures) are plotted as a function of frequency in
Figure~\ref{fig:quijote_assess_bg2} (bottom). One can see 
generally good agreement with variations of order 5$\,\%$ or
better at the QUIJOTE frequencies (except for one point at 13$\,$GHz) confirming that, 
at each frequency, the fluxes obtained when the full background area is sampled by circular
apertures are statistically consistent with the flux
obtained with the full background region. 
As a final test, we found that the results obtained with SED modelling when 
using only the QUIJOTE maps from horn 3 are consistent with 
the SED modelling results obtained when horn 1 and 2 are also included. 
All in all, we believe that the background region as defined
in Figure~\ref{fig:ccs_bg_cloud} and also shown in
Figure~\ref{fig:quijote_assess_bg} (bottom) is good enough to capture
the diffuse ISM properties allowing us to calculate the SED
of the molecular cloud region with accuracy. 


\begin{figure}
\begin{center}
\vspace*{2mm}
\centering
\includegraphics[width=65mm, angle =0]{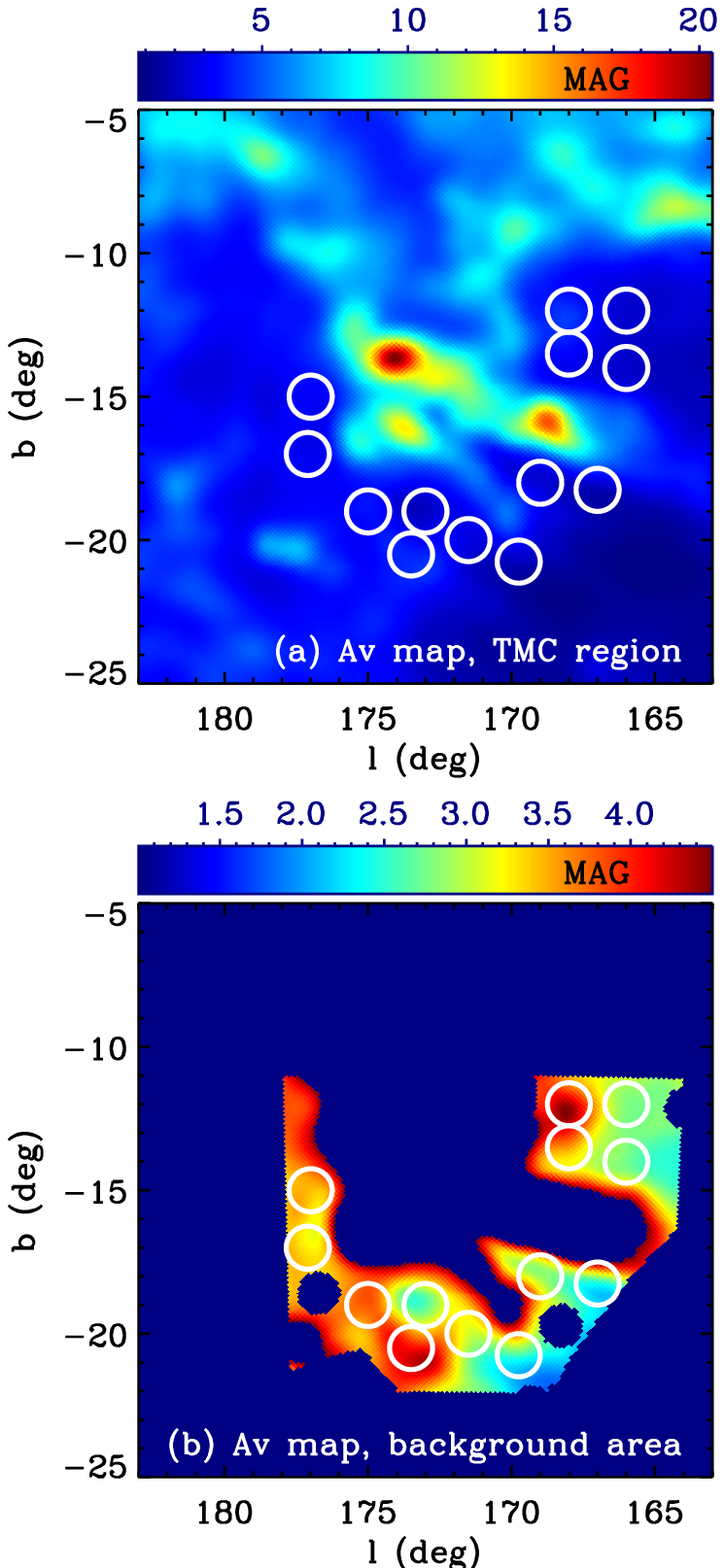}
\caption{\small a) \textit{Planck} $A_{V}$ map of the TMC and its vicinity. The
  location of the sample of apertures is shown with white
  circles. b) Full background region area used to calculate the
  fluxes of the TMC displayed in Table~\ref{tab:sed-results}. The
  sample of 13 circular apertures used to study the variations of the full
  background region are delineated by the white circles.
}
\label{fig:quijote_assess_bg}
\end{center}
\end{figure}


\begin{figure}
\begin{center}
\vspace*{2mm}
\centering
\includegraphics[width=75mm, angle =0]{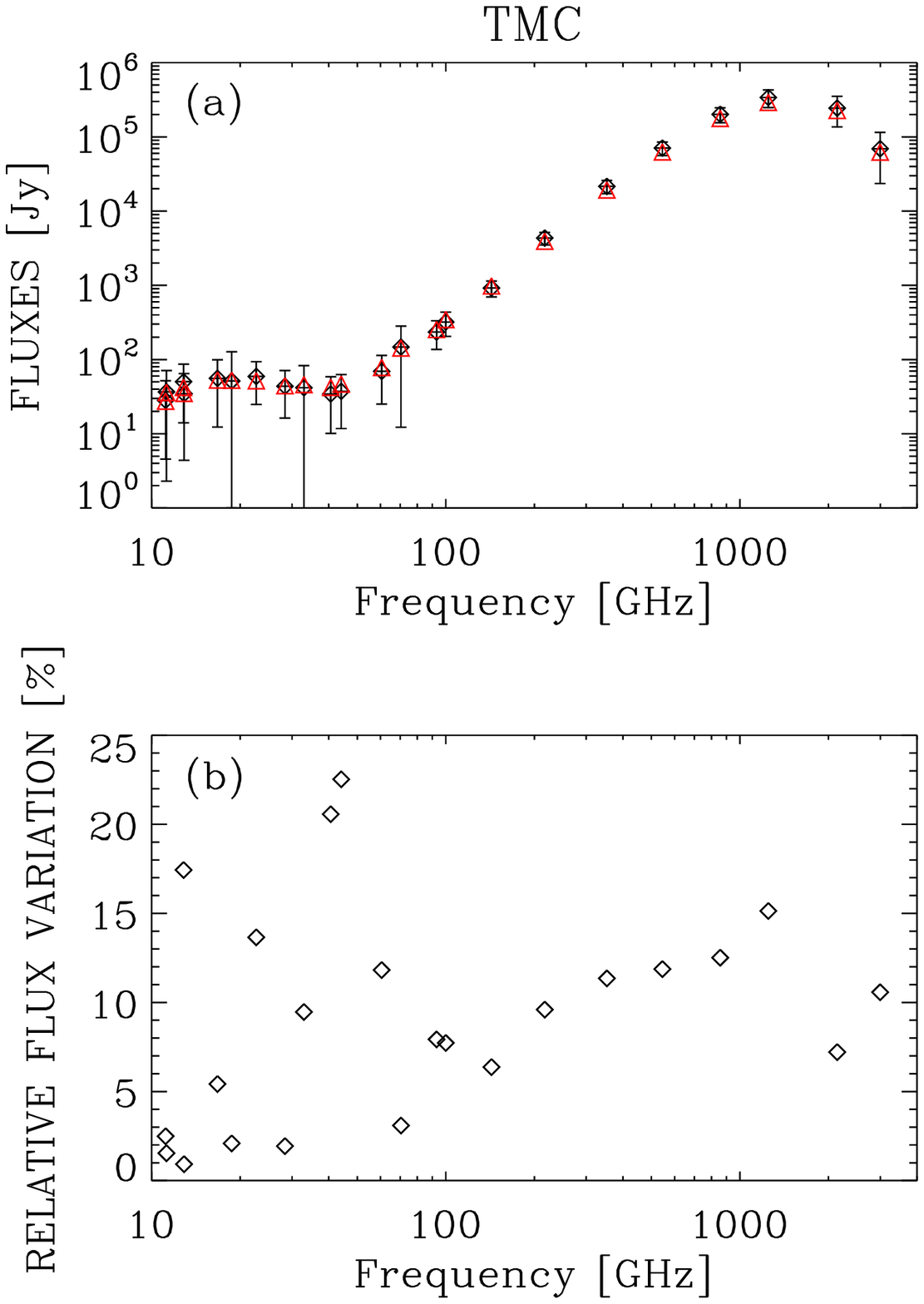}
\caption{\small a) Median values and dispersion of the TMC fluxes obtained when the
  background is sampled with the 13 circular apertures shown with diamond symbols in Figure~\ref{fig:quijote_assess_bg}
  plotted as a function of frequencies. The fluxes obtained towards the TMC with
  the full background area (see Table~\ref{tab:sed-results}) are
  plotted for comparison and shown with red
  triangles. b) Relative flux variation between the
  fluxes displayed in panel (a) as a function of frequency.
}
\label{fig:quijote_assess_bg2}
\end{center}
\end{figure}


\bsp	
\label{lastpage}
\end{document}